\documentstyle[10pt,epsfig,dp_delphititle,float,hangcaption,xspace,amssymb,
amsfonts,amsmath,amsthm,cite,graphicx]{dp_delphi}
\makeindex
\def\DpPaperGroup{EP-PH}
\def\DpPaperRef{2005-019}
\def\DpDate{14 May 2005}
\def\DpAuthors{DELPHI Collaboration}
\def\DpSubmit{(Accepted by Euro. Phys. Journ. )}
\def\DpTitle{{ Charged Particle Multiplicity\\ in Three-Jet Events\\ and Two-Gluon Systems }}

\newcommand{\eref}[1]{\mbox{{Eqn.~}\ref{#1}}}
\newcommand{\tref}[1]{\mbox{{Tab.~}\ref{#1}}}
\newcommand{\fref}[1]{\mbox{{Fig.~}\ref{#1}}}

\newcommand{\sref}[1]{\mbox{{Sect.~}\ref{#1}}}
\newlength{\wi}   \wi 8cm
\newlength{\fwi} \fwi 0.95\wi

\newcommand{\cacf} {\ifmmode{C_A/C_F}       \else{$C_A/C_F$}       \fi}
\newcommand{\epem} {\ifmmode{{\rm e^+e^-}}        \else{$\rm e^+e^-$}        \fi}
\newcommand{\rar} {\ifmmode{\rightarrow}     \else{$\rightarrow$}  \fi}
\newcommand{\bbbar} {\ifmmode{{\rm b\bar{b}}}     \else{$\rm b\bar{b}$}      \fi}
\newcommand{\qqbar} {\ifmmode{{\rm q\bar{q}}}     \else{$\rm q\bar{q}$}      \fi}
\newcommand{\qqg}   {\ifmmode{{\rm q\bar{q}g}}     \else{$\rm q\bar{q}g$}      \fi}
\newcommand{\qqga}   {\ifmmode{{\rm q\bar{q}\gamma}} \else{$\rm q\bar{q}\gamma$}      \fi}
\newcommand{\qbar}  {\ifmmode{{\rm \bar{q}}}      \else{$\rm  \bar{q}$}      \fi}
\newcommand{\bq}     {\ifmmode{{\rm b}}            \else{$\rm  b$}            \fi}
\newcommand{\q}     {\ifmmode{{\rm q}}            \else{$\rm  q$}            \fi}
\newcommand{\g}     {\ifmmode{{\rm g}}            \else{$\rm  g$}            \fi}
\newcommand{\gq}    {\ifmmode{{\rm gq}}           \else{$\rm  gq$}           \fi}
\newcommand{\qg}    {\ifmmode{{\rm qg}}           \else{$\rm  qg$}           \fi}
\newcommand{\ww} {\ifmmode{{\rm W^+W^-}}          \else{$\rm W^+W^-$}        \fi}
\newcommand{\Z}  {\ifmmode{{\rm Z}}               \else{$\rm Z$}             \fi}
\newcommand{\gev}{{\ifmmode   \mbox{Ge\kern-0.2exV}  
                              \else Ge\kern-0.2exV\nolinebreak\fi}}
\newcommand{\mev}{{\ifmmode   \mbox{Me\kern-0.2exV}  
                              \else Me\kern-0.2exV\nolinebreak\fi}}
\newcommand{\msbar} {\ifmmode{\overline{MS}}             \else{$\overline{MS}$}            \fi}
\newcommand{\CF} {\ifmmode{C_F}             \else{$C_F$}            \fi}
\newcommand{\CA} {\ifmmode{C_A}             \else{$C_A$}            \fi}
\newcommand{\nf} {\ifmmode{n_f}             \else{$n_f$}            \fi}
\newcommand{\TR} {\ifmmode{T_R}             \else{$T_R$}            \fi}
\newcommand{\TF} {\ifmmode{T_F}             \else{$T_F$}            \fi}
\newcommand{\as} {\ifmmode{\alpha_s}        \else{$\alpha_s$}       \fi}

\newcommand{\mean}[1]{\ifmmode{{\langle{#1}\rangle}} 
                      \else{$\langle{#1}\rangle$} \fi}
\newcommand{\eps}{{\ifmmode \varepsilon        \else $\varepsilon$\fi}}
\renewcommand{\th} {{\ifmmode \vartheta          \else $\vartheta$\fi}}
\newcommand{\into} {{\ifmmode \rightarrow       \else $\rightarrow$\fi}}
\newcommand{\be}{\begin{equation}}
\newcommand{\ee}{\end{equation}}

\def\bea{\begin{eqnarray}}
\def\eea{\end{eqnarray}}
\def\bal{\begin{align}}
\def\eal{\end{align}}
\def\bsp{\begin{split}}
\def\esp{\end{split}}
\newcommand{\bals}{\begin{align}\begin{split}}
\newcommand{\eals}{\end{split}\end{align}}

\def\bit{\begin{itemize}}
\def\eit{\end{itemize}}

\def\Z0{\mbox{Z}}
\def\epem{$\mbox{e}^+\mbox{e}^-$}
\def\qqbar{$\mbox{q}\bar{\mbox{q}}$}
\def\qqg{$\mbox{q}\bar{\mbox{q}}\mbox{g}$}

\def\sw
{
  \ifmmode  \sin^2 \Theta_W^{eff}
  \else    $\sin^2 \Theta_W^{eff}$
  \fi
}
\def\as
{
  \ifmmode  \alpha_s
  \else    $\alpha_s$
  \fi
}

\def\la
{
  \ifmmode  \Lambda
  \else    $\Lambda$
  \fi
}

\def\asZ
{
  \ifmmode  \alpha_s(m_{\mbox{\small Z}}^2)
  \else    $\alpha_s(m_{\mbox{\small Z}}^2)$
  \fi
}

\def\oasq
{
  \ifmmode  {\mathcal O}(\alpha_s^2)
  \else    ${\mathcal O}(\alpha_s^2)$
  \fi
}
\def\oas
{
  \ifmmode  {\mathcal O}(\alpha_s)
  \else    ${\mathcal O}(\alpha_s)$
  \fi
}

\newcommand{\kalu}{\kappa_{\mathrm{Lu}}}
\newcommand{\kale}{\kappa_{\mathrm{Le}}}

\newcommand{\haho}[1]{\raisebox{1.5ex}[-1.5ex]{#1}}
\begin{document}
%%%%%%%%%%%%%%%%%%%%%%%%%% They are a problem with Coll.Sty ?
\makeatletter
%\input{dp_system:coll.sty}
% Collapse citation numbers to ranges.  Non-numeric and undefined labels
% are handled.  No sorting is done.  E.g., 1,3,2,3,4,5,foo,1,2,3,?,4,5
% gives 1,3,2-5,foo,1-3,?,4,5
\newcount\@tempcntc
\def\@citex[#1]#2{\if@filesw\immediate\write\@auxout{\string\citation{#2}}\fi
  \@tempcnta\z@\@tempcntb\m@ne\def\@citea{}\@cite{\@for\@citeb:=#2\do
    {\@ifundefined
       {b@\@citeb}{\@citeo\@tempcntb\m@ne\@citea\def\@citea{,}{\bf ?}\@warning
       {Citation `\@citeb' on page \thepage \space undefined}}%
    {\setbox\z@\hbox{\global\@tempcntc0\csname b@\@citeb\endcsname\relax}%
     \ifnum\@tempcntc=\z@ \@citeo\@tempcntb\m@ne
       \@citea\def\@citea{,}\hbox{\csname b@\@citeb\endcsname}%
     \else
      \advance\@tempcntb\@ne
      \ifnum\@tempcntb=\@tempcntc
      \else\advance\@tempcntb\m@ne\@citeo
      \@tempcnta\@tempcntc\@tempcntb\@tempcntc\fi\fi}}\@citeo}{#1}}
\def\@citeo{\ifnum\@tempcnta>\@tempcntb\else\@citea\def\@citea{,}%
  \ifnum\@tempcnta=\@tempcntb\the\@tempcnta\else
   {\advance\@tempcnta\@ne\ifnum\@tempcnta=\@tempcntb \else \def\@citea{--}\fi
    \advance\@tempcnta\m@ne\the\@tempcnta\@citea\the\@tempcntb}\fi\fi}
 
\makeatother
%%%%%%%%%%%%%%%%%%%%%%%%%% ??????????????????????????????????
% Generate the title page
\begin{titlepage}
\pagenumbering{roman}
\CERNpreprint{\DpPaperGroup}{\DpPaperRef} % Reference of the paper
\date{{\small\DpDate}} % Date of the paper
\title{\DpTitle} % Title of the paper
\address{\DpAuthors} % General name of the author(s)
\begin{shortabs} % Start the abstract
\noindent
The charged particle multiplicity in hadronic three-jet
events from $Z$ decays is investigated. The topology dependence of the event
multiplicity is found to be well described by a modified leading 
logarithmic prediction. A parameter
fit of the prediction to the data yields a measurement of the colour factor
ratio $C_A/C_F$ with the result
%\begin{displaymath}
%C_A/C_F=2.261 \pm 0.014_{\mathrm{stat.}}
%\pm 0.036_{\mathrm{syst.}}
%\pm 0.052_{\mathrm{theor.}} \pm 0.041_{\mathrm{clus.}} 
%\end{displaymath}
\begin{displaymath}
C_A/C_F=2.261 \pm 0.014_{\mathrm{stat.}}
\pm 0.036_{\mathrm{exp.}}
\pm 0.066_{\mathrm{theo.}} 
\end{displaymath}
in agreement with 
the SU(3) expectation of QCD. The quark-related contribution to
the event multiplicity is subtracted from the three-jet event multiplicity 
resulting in 
a measurement of the multiplicity of two-gluon colour-singlet states 
over a wide energy range.
The ratios $r=N_{gg}(s)/N_{q\bar q}(s)$ of the gluon and quark multiplicities 
and $r^{(1)}=N_{gg}'(s)/N_{q\bar q}'(s)$ of their derivatives are 
compared
with perturbative calculations. While a good agreement between calculations
and data is observed for $r^{(1)}$, larger deviations are found for $r$
indicating that non-perturbative effects
are more important for $r$ than for $r^{(1)}$.

%%% Local Variables: 
%%% mode: latex
%%% TeX-master: "draft2"
%%% End: 
\end{shortabs}
\vfill
\begin{center}
\DpSubmit \ \\ % Horrible hack to allow to have DpSubmit empty
%CW \DpComment \ \\
%CW \DpEMail \ \\
\end{center}
\vfill
\clearpage
\headsep 10.0pt
\addtolength{\textheight}{10mm}
\addtolength{\footskip}{-5mm}
\begingroup
% Commands to process the author names
%
\newcommand{\DpName}[2]{\hbox{#1$^{\ref{#2}}$},\hfill}
\newcommand{\DpNameTwo}[3]{\hbox{#1$^{\ref{#2},\ref{#3}}$},\hfill}
\newcommand{\DpNameThree}[4]{\hbox{#1$^{\ref{#2},\ref{#3},\ref{#4}}$},\hfill}
\newskip\Bigfill \Bigfill = 0pt plus 1000fill
\newcommand{\DpNameLast}[2]{\hbox{#1$^{\ref{#2}}$}\hspace{\Bigfill}}
%
%\small
\footnotesize
\noindent
\DpName{J.Abdallah}{LPNHE}
\DpName{P.Abreu}{LIP}
\DpName{W.Adam}{VIENNA}
\DpName{P.Adzic}{DEMOKRITOS}
\DpName{T.Albrecht}{KARLSRUHE}
\DpName{T.Alderweireld}{AIM}
\DpName{R.Alemany-Fernandez}{CERN}
\DpName{T.Allmendinger}{KARLSRUHE}
\DpName{P.P.Allport}{LIVERPOOL}
\DpName{U.Amaldi}{MILANO2}
% ...
\normalsize
\endgroup
\titlefoot{Department of Physics and Astronomy, Iowa State
     University, Ames IA 50011-3160, USA
    \label{AMES}}
\titlefoot{Physics Department, Universiteit Antwerpen,
     Universiteitsplein 1, B-2610 Antwerpen, Belgium \\
     \indent~~and IIHE, ULB-VUB,
     Pleinlaan 2, B-1050 Brussels, Belgium \\
     \indent~~and Facult\'e des Sciences,
     Univ. de l'Etat Mons, Av. Maistriau 19, B-7000 Mons, Belgium
    \label{AIM}}
\titlefoot{Physics Laboratory, University of Athens, Solonos Str.
     104, GR-10680 Athens, Greece
    \label{ATHENS}}
\titlefoot{Department of Physics, University of Bergen,
     All\'egaten 55, NO-5007 Bergen, Norway
    \label{BERGEN}}
\titlefoot{Dipartimento di Fisica, Universit\`a di Bologna and INFN,
     Via Irnerio 46, IT-40126 Bologna, Italy
    \label{BOLOGNA}}
\titlefoot{Centro Brasileiro de Pesquisas F\'{\i}sicas, rua Xavier Sigaud 150,
     BR-22290 Rio de Janeiro, Brazil \\
     \indent~~and Depto. de F\'{\i}sica, Pont. Univ. Cat\'olica,
     C.P. 38071 BR-22453 Rio de Janeiro, Brazil \\
     \indent~~and Inst. de F\'{\i}sica, Univ. Estadual do Rio de Janeiro,
     rua S\~{a}o Francisco Xavier 524, Rio de Janeiro, Brazil
    \label{BRASIL}}
\titlefoot{Coll\`ege de France, Lab. de Physique Corpusculaire, IN2P3-CNRS,
     FR-75231 Paris Cedex 05, France
    \label{CDF}}
\titlefoot{CERN, CH-1211 Geneva 23, Switzerland
    \label{CERN}}
\titlefoot{Institut de Recherches Subatomiques, IN2P3 - CNRS/ULP - BP20,
     FR-67037 Strasbourg Cedex, France
    \label{CRN}}
\titlefoot{Now at DESY-Zeuthen, Platanenallee 6, D-15735 Zeuthen, Germany
    \label{DESY}}
\titlefoot{Institute of Nuclear Physics, N.C.S.R. Demokritos,
     P.O. Box 60228, GR-15310 Athens, Greece
    \label{DEMOKRITOS}}
\titlefoot{FZU, Inst. of Phys. of the C.A.S. High Energy Physics Division,
     Na Slovance 2, CZ-180 40, Praha 8, Czech Republic
    \label{FZU}}
\titlefoot{Dipartimento di Fisica, Universit\`a di Genova and INFN,
     Via Dodecaneso 33, IT-16146 Genova, Italy
    \label{GENOVA}}
\titlefoot{Institut des Sciences Nucl\'eaires, IN2P3-CNRS, Universit\'e
     de Grenoble 1, FR-38026 Grenoble Cedex, France
    \label{GRENOBLE}}
\titlefoot{Helsinki Institute of Physics and Department of Physical Sciences,
     P.O. Box 64, FIN-00014 University of Helsinki, 
     \indent~~Finland
    \label{HELSINKI}}
\titlefoot{Joint Institute for Nuclear Research, Dubna, Head Post
     Office, P.O. Box 79, RU-101 000 Moscow, Russian Federation
    \label{JINR}}
\titlefoot{Institut f\"ur Experimentelle Kernphysik,
     Universit\"at Karlsruhe, Postfach 6980, DE-76128 Karlsruhe,
     Germany
    \label{KARLSRUHE}}
\titlefoot{Institute of Nuclear Physics PAN,Ul. Radzikowskiego 152,
     PL-31142 Krakow, Poland
    \label{KRAKOW1}}
\titlefoot{Faculty of Physics and Nuclear Techniques, University of Mining
     and Metallurgy, PL-30055 Krakow, Poland
    \label{KRAKOW2}}
\titlefoot{Universit\'e de Paris-Sud, Lab. de l'Acc\'el\'erateur
     Lin\'eaire, IN2P3-CNRS, B\^{a}t. 200, FR-91405 Orsay Cedex, France
    \label{LAL}}
\titlefoot{School of Physics and Chemistry, University of Lancaster,
     Lancaster LA1 4YB, UK
    \label{LANCASTER}}
\titlefoot{LIP, IST, FCUL - Av. Elias Garcia, 14-$1^{o}$,
     PT-1000 Lisboa Codex, Portugal
    \label{LIP}}
\titlefoot{Department of Physics, University of Liverpool, P.O.
     Box 147, Liverpool L69 3BX, UK
    \label{LIVERPOOL}}
\titlefoot{Dept. of Physics and Astronomy, Kelvin Building,
     University of Glasgow, Glasgow G12 8QQ
    \label{GLASGOW}}
\titlefoot{LPNHE, IN2P3-CNRS, Univ.~Paris VI et VII, Tour 33 (RdC),
     4 place Jussieu, FR-75252 Paris Cedex 05, France
    \label{LPNHE}}
\titlefoot{Department of Physics, University of Lund,
     S\"olvegatan 14, SE-223 63 Lund, Sweden
    \label{LUND}}
\titlefoot{Universit\'e Claude Bernard de Lyon, IPNL, IN2P3-CNRS,
     FR-69622 Villeurbanne Cedex, France
    \label{LYON}}
\titlefoot{Dipartimento di Fisica, Universit\`a di Milano and INFN-MILANO,
     Via Celoria 16, IT-20133 Milan, Italy
    \label{MILANO}}
\titlefoot{Dipartimento di Fisica, Univ. di Milano-Bicocca and
     INFN-MILANO, Piazza della Scienza 2, IT-20126 Milan, Italy
    \label{MILANO2}}
\titlefoot{IPNP of MFF, Charles Univ., Areal MFF,
     V Holesovickach 2, CZ-180 00, Praha 8, Czech Republic
    \label{NC}}
\titlefoot{NIKHEF, Postbus 41882, NL-1009 DB
     Amsterdam, The Netherlands
    \label{NIKHEF}}
\titlefoot{National Technical University, Physics Department,
     Zografou Campus, GR-15773 Athens, Greece
    \label{NTU-ATHENS}}
\titlefoot{Physics Department, University of Oslo, Blindern,
     NO-0316 Oslo, Norway
    \label{OSLO}}
\titlefoot{Dpto. Fisica, Univ. Oviedo, Avda. Calvo Sotelo
     s/n, ES-33007 Oviedo, Spain
    \label{OVIEDO}}
\titlefoot{Department of Physics, University of Oxford,
     Keble Road, Oxford OX1 3RH, UK
    \label{OXFORD}}
\titlefoot{Dipartimento di Fisica, Universit\`a di Padova and
     INFN, Via Marzolo 8, IT-35131 Padua, Italy
    \label{PADOVA}}
\titlefoot{Rutherford Appleton Laboratory, Chilton, Didcot
     OX11 OQX, UK
    \label{RAL}}
\titlefoot{Dipartimento di Fisica, Universit\`a di Roma II and
     INFN, Tor Vergata, IT-00173 Rome, Italy
    \label{ROMA2}}
\titlefoot{Dipartimento di Fisica, Universit\`a di Roma III and
     INFN, Via della Vasca Navale 84, IT-00146 Rome, Italy
    \label{ROMA3}}
\titlefoot{DAPNIA/Service de Physique des Particules,
     CEA-Saclay, FR-91191 Gif-sur-Yvette Cedex, France
    \label{SACLAY}}
\titlefoot{Instituto de Fisica de Cantabria (CSIC-UC), Avda.
     los Castros s/n, ES-39006 Santander, Spain
    \label{SANTANDER}}
\titlefoot{Inst. for High Energy Physics, Serpukov
     P.O. Box 35, Protvino, (Moscow Region), Russian Federation
    \label{SERPUKHOV}}
\titlefoot{J. Stefan Institute, Jamova 39, SI-1000 Ljubljana, Slovenia
     and Laboratory for Astroparticle Physics,\\
     \indent~~Nova Gorica Polytechnic, Kostanjeviska 16a, SI-5000 Nova Gorica, Slovenia, \\
     \indent~~and Department of Physics, University of Ljubljana,
     SI-1000 Ljubljana, Slovenia
    \label{SLOVENIJA}}
\titlefoot{Fysikum, Stockholm University,
     Box 6730, SE-113 85 Stockholm, Sweden
    \label{STOCKHOLM}}
\titlefoot{Dipartimento di Fisica Sperimentale, Universit\`a di
     Torino and INFN, Via P. Giuria 1, IT-10125 Turin, Italy
    \label{TORINO}}
%\titlefoot{INFN,Sezione di Torino, and Dipartimento di Fisica Teorica,
%     Universit\`a di Torino, Via P. Giuria 1,\\
%     \indent~~IT-10125 Turin, Italy
\titlefoot{INFN,Sezione di Torino and Dipartimento di Fisica Teorica,
     Universit\`a di Torino, Via Giuria 1,
     IT-10125 Turin, Italy
    \label{TORINOTH}}
\titlefoot{Dipartimento di Fisica, Universit\`a di Trieste and
     INFN, Via A. Valerio 2, IT-34127 Trieste, Italy \\
     \indent~~and Istituto di Fisica, Universit\`a di Udine,
     IT-33100 Udine, Italy
    \label{TU}}
\titlefoot{Univ. Federal do Rio de Janeiro, C.P. 68528
     Cidade Univ., Ilha do Fund\~ao
     BR-21945-970 Rio de Janeiro, Brazil
    \label{UFRJ}}
\titlefoot{Department of Radiation Sciences, University of
     Uppsala, P.O. Box 535, SE-751 21 Uppsala, Sweden
    \label{UPPSALA}}
\titlefoot{IFIC, Valencia-CSIC, and D.F.A.M.N., U. de Valencia,
     Avda. Dr. Moliner 50, ES-46100 Burjassot (Valencia), Spain
    \label{VALENCIA}}
\titlefoot{Institut f\"ur Hochenergiephysik, \"Osterr. Akad.
     d. Wissensch., Nikolsdorfergasse 18, AT-1050 Vienna, Austria
    \label{VIENNA}}
\titlefoot{Inst. Nuclear Studies and University of Warsaw, Ul.
     Hoza 69, PL-00681 Warsaw, Poland
    \label{WARSZAWA}}
\titlefoot{Fachbereich Physik, University of Wuppertal, Postfach
     100 127, DE-42097 Wuppertal, Germany \\
\noindent
{$^\dagger$~deceased}
    \label{WUPPERTAL}}
\addtolength{\textheight}{-10mm}
\addtolength{\footskip}{5mm}
\clearpage
\headsep 30.0pt
\end{titlepage}
%%%%%%%%%%%%%%%%%%%%%%%%%
%
% Change for the document body
%%\pagestyle{heading} % for page numbering
\pagenumbering{arabic} % page numbering in number
\setcounter{footnote}{0} %
\large
\section{Introduction}
Abundant particle multiplicity 
is one of the most obvious properties of hadronic
final states in $e^+e^-$ annihilation. 
The reason for the large multiplicity of particles 
produced in hadronic events
is directly rooted to one of 
the fundamental properties of the strong interaction:
the confinement of quarks and gluons into hadrons. Unlike leptons, quarks
produced in a high energy reaction are not able just to  separate without
further interactions in which the colour charges are balanced 
and free colour singlets are formed.
Ironically, it is the very
same property of QCD which makes
it impossible to predict the average number of hadrons produced in such an
event. However, the hypothesis of Local Parton-Hadron Duality (LPHD) allows
the assumption to be made that the hadronic multiplicity of an event is related only via
a normalisation factor to the partonic multiplicity at a given virtuality
cut-off $Q_0$, which then is a perturbatively calculable quantity.
\par
The multiplicity of $q\bar q$ colour-singlet systems is well understood in
this way. Theoretical predictions of the energy dependence of the multiplicity
produced in such systems \cite{webbermult,gaffneymult,capella} 
have been confirmed by experimental data to  high
accuracy \cite{multmess,restmult,topaz}. 
It is possible to predict 
the multiplicity of two-gluon colour-singlet
systems perturbatively in a similar fashion, but experimental verification of
this quantity is scarce. Two-gluon systems are experimentally difficult to
produce and are observed so far only in 
decays of  resonances with relatively low mass.
The ratio of the multiplicity of quark and gluon systems, $r$,
has been subject to theoretical studies presented in \sref{s:thpred}. 
It is expected that $r$ at high energies resembles asymptotically the
colour factor ratio $C_A/C_F$. $C_F$ and $C_A$, which are
 the Casimir eigenvalues of
the triplet and octet representation of SU(3), 
can be interpreted as effective
colour charges of the triplet quarks and the octet gluons. However, $r$ is
known to have large corrections in higher orders of the 
perturbative expansion and
moreover, is
supposed to be affected by non-perturbative effects. Therefore, the ratio of
the derivatives of the quark and gluon multiplicities with respect to energy, 
$r^{(1)}$, has been
suggested in \cite{r1_suggest} as a better suited observable.
\par
The next slightly more complex 
class of events which can be studied in $e^+e^-$-annihilation
are events with three
jets, which occur if in a $q\bar q$-event a gluon is radiated with a
sufficiently large transverse momentum.
Three-jet events are unique in the sense
that they provide quark jets as well as gluon jets and therefore provide the
opportunity to study the properties of gluon fragmentation once the $q\bar
q$-contributions are understood. There have been numerous experimental
studies where 
%either 
the properties of identified quark and gluon jets are
compared (e.g.~ \cite{qgstudy,alephscales,opal_ngg,scaling,opal_recoil}). 
%or the properties of gluons recoiling against two in the ideal case
%collinear  quark jets (so-called 'unbiased' gluon jets) are studied
%\cite{opal_recoil}. 
A crucial
question arising in this context is how the environment of a three-jet event
affects the properties of a quark or gluon jet compared to the unrestricted
jets of a $q\bar q$- or two-gluon event in terms of phase space restriction or
coherent radiation. Studies have been made to
verify experimentally several proper energy scale like variables which
account for these differences by applying the appropriate evaluation
scales for the jets in
three-jet events (e.g.~in \cite{alephscales,scaling}). 
Coherence effects in the particle production in three-jet
events have been subject to a recent study \cite{coherence}. In parallel with 
the publication of an analysis of the
charged multiplicity of symmetric three-jet events applying a rather 
phenomenological approach \cite{multpaper}, a theoretical prediction of the
multiplicity of three-jet events has been published
\cite{Eden:1998ig,Eden:1999vc}, which is derived in the
colour-dipole picture of the Modified Leading Logarithmic Approximation (MLLA)
and takes into account the effects of colour
coherence  and phase-space restriction in a stringent way. This prediction has
been applied to study the topology-dependence of the event
 multiplicity of symmetric three-jet events in
\cite{multnote} and later in \cite{opalmult}. 
\par
In this paper this prediction is
 not only applied to the charged multiplicity of three-jet events of symmetric 
topologies, but events with more general topologies are also  considered.
%symmetric event topologies are considered for cross check reasons. 
In the
following section the experiment, data selection and the classification of
three-jet event topologies is laid out. \sref{s:thpred} deals with the
relevant theoretical predictions, before in \sref{s_evtmult} the
measured multiplicities are discussed. The following section deals with the
preparation of the predictions which are then  
compared to the data in \sref{s_fit}. 
A parameter fit to determine the colour-factor
ratio $C_A/C_F$ is performed and the systematic uncertainties of this
measurement are discussed. In \sref{s_gluon} the prediction is used to
extract the multiplicity of two-gluon colour singlet systems
from the three-jet event multiplicity
by subtracting the quark contribution. 
With the measurements
of the charged two-gluon multiplicity 
$N_{gg}$ 
the ratios $r$ and $r^{(1)}$ are evaluated and the possibility of
measuring the ratio of the second derivatives, $r^{(2)}$, 
is studied.
The summary and conclusions are presented in \sref{s_summary}.
%, before the analysis will be
%summarised and concluded in section \ref{s_summary}.
%
%
\section{Data and data analysis\label{s:data}}
\begin{table} [htb]
\begin{minipage}{7.5cm}
\begin{center}
  \begin{tabular}{|c|c|}
    \hline
    variable & cut \\
    \hline \hline
    $p$ & $ \ge 0.4~\mathrm{ GeV}$  \\
      \hline
    $\vartheta_{\mathrm{ polar}}$ &  $20^\circ-160^\circ$ \\
      \hline
    $\epsilon_{xy}$ & $\le 5 ~\mathrm{ cm}$ \\
      \hline
    $\epsilon_z$ &  $\le 10 ~\mathrm{ cm}$  \\
      \hline
    $L_{\mathrm{ track}}$ & $\ge 30~\mathrm{ cm}$  \\
      \hline
    $\Delta p/p$ & $\le 100\%$ \\
    \hline
   $E_{\mathrm{HPC}}$ & $0.5~\mathrm{GeV} - 50~\mathrm{GeV}$  \\ \hline
   $E_{\mathrm{EMF}}$ & $0.5~\mathrm{GeV} - 50~\mathrm{GeV}$  \\ \hline
   $E_{\mathrm{HAC}}$ & $1~\mathrm{GeV} - 50~\mathrm{GeV}$  \\ \hline
  \end{tabular}  
  \caption{\label{t_tcuts} Selection cuts applied to charged-particle tracks
and to calorimeter clusters.}
\end{center} 
\end{minipage}
\hfill
\begin{minipage}{7cm}
\begin{center}
  \begin{tabular}{|c|c|}
    \hline
    variable & cut \\  \hline \hline
 \multicolumn{2}{|c|}{general events} \\ \hline
    $E_{\mathrm{ charged}}^{\mathrm{ hemisph.}}$ & $ \ge 0.03\cdot \sqrt{s}$  \\
      \hline
    $E_{\mathrm{ charged}}^{\mathrm{ total}}$ & $ \ge 0.12\cdot \sqrt{s}$  \\
      \hline
    $N_{\mathrm{ charged}}$ & $\ge 5$ \\
      \hline
    $\vartheta_{\mathrm{ sphericity}}$ &  $30^\circ-150^\circ$  \\
      \hline
    $p_{\mathrm{max}}$ &  $45$ GeV  \\
      \hline\hline
 \multicolumn{2}{|c|}{three-jet events} \\ \hline
    $\sum_{i=1}^{3}\theta_i$ & $ > 355^\circ$ \\
      \hline
    $E_{\mathrm{ visible}}/\mathrm{ jet}$ & $ \ge 5$ GeV  \\
      \hline
    $N_{\mathrm{charged}}/\mathrm{ jet}$ & $\ge 2$ \\
      \hline
    $\vartheta_{\mathrm{ jet}}$
            &  $30^\circ-150^\circ$      \\
      \hline
  \end{tabular}  
  \caption{\label{t_evtcut} Selection cuts applied to general events and 
to three-jet events.}
\end{center} 
\end{minipage}
\end{table}
In this paper the hadronic events from $Z$ decays recorded by the {\sc
  Delphi} experiment in the years 1992-1995 are analysed. The {\sc Delphi}
 detector was a hermetic collider detector with a solenoidal
magnetic field, extensive tracking facilities including a micro-vertex
detector,
electromagnetic and hadronic calorimetry as well as extended particle
identification capabilities. % provided mainly by RICH detectors.
The detector and its performance are described in detail elsewhere
\cite{det,perf}.
\par
In order to select well-measured particles originating from the interaction
point, the cuts shown in  \tref{t_tcuts} 
were applied to the measured 
tracks and electromagnetic or hadronic calorimeter clusters.
Here $p$ and $E$ denote the particle's momentum and energy, 
$\vartheta_{\mathrm{polar}}$ denotes the polar angle with respect to the beam, 
$\epsilon_k$ is the distance of closest approach 
to the interaction point in the plane
perpendicular to ($xy$) or along ($z$) the beam, respectively, 
$L_{\mathrm{track}}$ is the measured track length.
$E_{\mathrm{HPC}}$ ($E_{\mathrm{EMF}}$) denotes the 
energy of a cluster as measured with the barrel (forward) electromagnetic 
calorimeter, 
 and  $E_{\mathrm{HAC}}$ the cluster energy measured by the
hadronic calorimeter.
\par
The general event cuts shown in 
 \tref{t_evtcut} select hadronic decays of the $Z$ and suppress
background from leptonic $Z$ decays, $\gamma\gamma$ interactions
or beam-gas interactions. Further reduction of
background to a negligible level
is achieved by the jet-selection cuts given also in 
\tref{t_evtcut} .
The cut variables are the visible charged energy, 
$E^{\mathrm{total}}_{\mathrm{charged}}$ and $E^{\mathrm{hemisph.}}_{\mathrm{charged}}$,
observed in the
event or in each event hemisphere, respectively. 
Event hemispheres are defined by 
the plane perpendicular to the sphericity axis.
The polar angle of this axis with respect to the beam is 
$\vartheta_{\mathrm{sphericity}}$ and $N_{\mathrm{charged}}$ 
is the observed charged multiplicity.
Events are discarded, if they
contain charged particles with momenta apparently above the kinematic limit.
%
%To provide a high
%  quality of the event reconstruction, the measured tracks of charged
%  particles have to pass the cuts listed in \tref{t_chcut} while the energy
%  depositions of neutral particles have to be within the boundaries given in
%  \tref{t_necut}. The events reconstructed from the accepted tracks and energy
%  depositions have in turn then to fulfil the requirements given in
%  \tref{t_evcut} in order to be accepted as hadronic events. 
\par
In addition to the event selection above, a procedure to tag events with
initial $b$-quarks is applied. Details of the tagging procedure, which is based
on the combined information of several observables, can be found in
\cite{b-tag2}. A cut on the tagging variable $\lambda$ by demanding
$\lambda<1.5$ is used to reject $b$-events leading to a light quark event
purity of $\sim 92\%$. Both, the complete set of accepted hadronic ($udscb$-)
events as well as the set of anti-tagged light ($udsc$-) quark events are 
analysed for cross-check reasons.
\par
In the accepted events three jets are then reconstructed
  using the angular ordered Durham algorithm \cite{cambridge} 
 taking into
  acount reconstructed momenta of charged particles and 
  of neutral particles seen in the
  electromagnetic calorimeter. 
No cut-off variable for the cluster
  algorithm is used, every event is clustered into three jets. For cross-check
  reasons,  
  the Durham \cite{durham}, Cambridge \cite{cambridge} and Luclus
  \cite{luclus}
  jet
  clustering algorithms have been used alternatively\footnote{It is
  in the nature of the Cambridge algorithm, that depending on the structure of
  an event not every number of jets can be resolved. The small number of 
  events where three
  jets could not 
   be resolved are discarded when using the Cambridge algorithm.}. 
As three-jet events have to be planar due to momentum conservation, the three
reconstructed jets are projected into the event-plane, which is defined by the 
first two eigenvectors of the sphericity tensor \cite{sphericity}.  
\par
The three-jet event quality requirements, also shown in
\tref{t_evtcut}, assure well-measured jets. 
Here $\theta_i$ denotes 
the angle between the two jets opposite to jet $i$,
$\vartheta_{\mathrm{jet}}$ the polar angle of a jet
and  $E_{\mathrm{visible}} /\mathrm{jet}$ the total visible energy per jet,
taking into account all tracks and information from 
the electromagnetic calorimeter
assigned to this jet. 
%
%
% After the jet reconstruction the cuts listed in \tref{t_jetcut} are applied
%  to the jet structure of the event. 
Assuming massless jet kinematics
  the topology of an event can be 
  characterised by the three angles $\theta_{1,2,3}$ between the jets. The
  inter-jet angles are ordered according to their size with $\theta_1$ being
  the smallest and $\theta_3$ the largest angle. This reflects the
  wide-spread convention due to which jets are numbered according to their
  energy with jet 1 being the most energetic, while the inter-jet angles are
  numbered according to their opposing jet. With massless jets this implies
  the numbering of the angles given above.
\par 
  The  
  inter-jet angles of the projected jets 
  add up to $360^\circ$. Therefore giving two of the three
  angles is sufficient to fully determine the topology of an event. Here,
  $\theta_2$ and $\theta_3$ are used to define the binning. The
  angular intervals used to define bins of different topology classes are
  listed in \tref{t_binning}. 
% Note that due to momentum conservation
%  no inter-jet angle can exceed $180^\circ$. 
The allowed values
  for $\theta_2$ depend on the value of $\theta_3$, as
  $\theta_2<\theta_3$ and $\theta_2>180^\circ-\theta_3/2$, the latter being a
  consequence of $\theta_2>\theta_1$.
\begin{table}[tb]
\begin{center}
\begin{tabular}{|rr|r|}
\hline
\multicolumn {2}{|c|}{$\theta_2$} & \multicolumn {1}{c|}{$\theta_3$} \\
\hline
$ 98^\circ - 100^\circ$ & $119^\circ - 123^\circ$ & $120^\circ - 130^\circ$\\
$100^\circ - 102^\circ$ & $123^\circ - 128^\circ$ & $130^\circ - 140^\circ$\\
$102^\circ - 104^\circ$ & $128^\circ - 133^\circ$ & $140^\circ - 145^\circ$\\
$104^\circ - 106^\circ$ & $133^\circ - 138^\circ$ & $145^\circ - 150^\circ$\\
$106^\circ - 108^\circ$ & $138^\circ - 143^\circ$ & $150^\circ - 155^\circ$\\
$108^\circ - 111^\circ$ & $143^\circ - 148^\circ$ & $155^\circ - 160^\circ$\\
$111^\circ - 115^\circ$ & $148^\circ - 155^\circ$ & $160^\circ - 165^\circ$\\
$115^\circ - 119^\circ$ &                         &                        \\
\hline
\end{tabular}
\end{center}
\caption{\label{t_binning} The bins in $\theta_2$ and $\theta_3$}
\end{table}
%
%
%% \begin{table}[bt]
%% \hspace*{-1cm}
%% \begin{tabular}{c|cccccccccccccccc}
%% $\theta_2$ &
%% $98 ^\circ$ &
%% $100^\circ$ &
%% $102 ^\circ$ &
%% $104 ^\circ$ &
%% $106 ^\circ$ &
%% $108 ^\circ$ &
%% $111^\circ$ &
%% $115^\circ$ &
%% $119 ^\circ$ &
%% $123 ^\circ$ &
%% $128 ^\circ$ &
%% $133 ^\circ$ &
%% $138 ^\circ$ &
%% $143 ^\circ$ &
%% $148 ^\circ$ &
%% $155^\circ$ \\
%% \hline
%% $\theta_3$  &
%% $120 ^\circ$ &
%% $130 ^\circ$ &
%% $140 ^\circ$ &
%% $145 ^\circ$ &
%% $150^\circ$ &
%% $155 ^\circ$ &
%% $160^\circ$ &
%% $165 ^\circ$ &&&&&&&&
%% \end{tabular}
%% \caption{\label{t_binning} The borders of the bins in $\theta_2$
%%   and $\theta_3$}
%% \end{table}
%% %
%
\par
Besides these  general classes of topologies, classes of {\em symmetric
  topologies} are selected, where symmetric events are defined by the
  requirement of the two larger inter-jet angles of an event, $\theta_2$ and
  $\theta_3$ to be equal within small tolerances, i.e.~
  $\theta_3\le\theta_2+\Delta\theta$. An angular tolerance less than
  $5^\circ$ is 
  not meaningful, 
  as Monte Carlo studies showed that the differences between the
  inter-jet angles in the partonic and in the hadronic state scatter with a
  standard deviation 
of $\sim5^\circ$. Therefore $\Delta\theta=5^\circ$ is chosen
  here. Since $\theta_3$ has to be larger than $120^\circ$, the binning is
  completely determined by the choice of $\Delta\theta$.
  Symmetric events have been used in several previous studies
(e.g.~\cite{qgstudy,alephscales,opal_ngg,scaling,multpaper,multnote,opalmult}),  
  exploiting the special property of these events. With
  jet 2 and jet 3 being 
   a quark and a gluon jet with exactly the same topological
  environment, a similar energy scale is present. Here symmetric
  topologies, which are also contained as a subset in the general event 
  topologies, are studied also separately. Due to the additional constraint
  on symmetric topologies, giving only one inter-jet angle is sufficient to
  fully describe the topology of a symmetric event. For consistency with
  previous studies $\theta_1$ is used for this purpose. The reduction to only 
  one independent variable considerably simplifies most correction procedures.
  Note that the symmetric events are
  not fully contained in the general event topologies listed in
  \tref{t_binning}, as the minimum value for $\theta_1$ allowed by the binning
  in \tref{t_binning} is $40^\circ$. In symmetric events jet 3 is produced
  with maximum energy, therefore smaller opening angles $\theta_1$ can be
  accepted. However, Monte Carlo studies showed that the resolved hadronic jet
  structure does not reflect the partonic structure of an event, if $\theta_1$
  is smaller than $\sim20^\circ$, which therefore gives a lower boundary for
  $\theta_1$. 
\par
  For each $(\theta_2,\theta_3)$-bin of the general topologies, and each
  $\theta_1$-bin of the symmetric event sample, the multiplicity distribution
  is measured. To correct these distributions for detector acceptance, a
  matrix correction is used. 
%% The measured multiplicity distribution
%%   $P_n^{\mathrm{acc}}$ is related to the true multiplicity distribution 
%%   $P_n^{\mathrm{gen}}$ via the acceptance matrix $M_{nm}$:
%% \be
%%   P_n^{\mathrm{acc}}=M_{nm}P_m^{\mathrm{gen}}\quad .
%% \ee
%%  Applying the inverse of $M_{nm}$ to the measured multiplicity distribution
%%  would therefore perform the acceptance correction. However, 
%%   $M_{nm}$ usually contains singularities, for example due to events which
%%   failed the acceptance criteria, and can
%%   not be inverted. 
For each angular bin
  a matrix $M_{mn}$ is calculated
  by tracing in a Monte Carlo simulation
  how many particles $m$ in an accepted event with $n$ detected 
  particles have been generated. In order to obtain the generated number of
  charged particles, particles with lifetimes shorter than
  $10^{-9}$s, like $K_{\mathrm{S}}^0$ and $\Lambda$, have been forced to decay. 
  $M_{mn}$ is then
  applied to the measured distributions. Since generated
  events which fail to fulfil the
  selection criteria are not considered when calculating $M_{mn}$ and these
  events are biased towards low multiplicities, the correction is slightly 
  overestimated.
%
%singularities of $M_{nm}$ have
%  been neglected so far, the application of $\tilde M_{mn}^{-1}$ overestimates
%  the correction. 
In order to correct for this overestimation, the
  multiplicity distributions are multiplied by the ratio of the multiplicity
  distributions of all generated events with the multiplicity distributions of
  all generated and not rejected events for the respective angular bin. 
  While the application of 
  $M_{mn}$ increases 
  the mean of the distributions by $\sim30\%$, the
  multiplicative second correction  results in a reduction
  by only $\sim4\%$. 
%The decrease in the second correction reflects the fact,
%  that events with a small multiplicity are less likely to be accepted. 
\par
  When correcting the multiplicity distributions of $udsc$-events, the matrix
  corrected multiplicity distributions still contain some contributions due to
mistagged $b$-events, which are not considered in the second multiplicative
  correction. Therefore the remaining contribution due to $b$-events is
  taken from the Monte Carlo simulation and 
   subtracted from the matrix corrected distribution. This correction is
  small, the mean of the multiplicity distributions is changed by only
  $\sim0.6\%$. 
\par
% To get the central results 
The central values for the results of
the mean multiplicities as a function of the event
 topology are taken to be the arithmetic mean of the respective multiplicity
 distributions. However, in order to estimate the systematic effects of the
 correction procedure, two variations of the correction procedure are 
 also applied:
\begin{itemize}
\item 
The mean multiplicity is determined from the expectation value of
a negative binomial distribution
fitted to every multiplicity distribution.
%The mean multiplicity is not estimated by the arithmetic mean of a
%  multiplicity distribution, but a negative binomial is fitted to every
%  multiplicity distribution and the mean is taken from the fit parameters.
\item Instead of correcting the distributions, the means of the uncorrected
  distributions are taken and corrected with a multiplicative factor taken
  from Monte Carlo simulation.
\end{itemize}
The corrected mean multiplicities obtained by using all three correction procedures
agree well, the small variation is considered in the systematic errors of the
analysis. 
\par
This procedure provides fully corrected mean multiplicities of hadronic 
three-jet events with initial $udscb$-quark -- and $udsc$-quarks for cross-check
reasons --
as a function of the event topology for general and symmetric topologies.
Additionally, the multiplicity distribution and mean multiplicity of all
accepted events  regardless of angular restrictions is treated in the same way
to compare this measurement with previous measurements of this quantity. The
events entering this overall multiplicity distribution are not required to
fulfil the cuts on the jet structure given in the lower part of
\tref{t_evtcut}. 
%These cuts are
%not necessary for the overall hadronic event multiplicity, as this quantity is
%anyway dominated by two-jet like events. 
%Since these cuts are defined to provide well
%resolved three-jet events, the application of the cuts in
%\tref{t_evtcut} would introduce a bias of the order of 3\% towards higher
%multiplicities. 
%
%
%
\section{Theoretical Predictions \label{s:thpred}}
The ratio of gluon and quark multiplicities $r$ has been subject to theoretical
studies for some time. The naive expectation that this ratio reflects the ratio
of the effective colour charges of quarks and gluons given by the colour
factor ratio $C_A/C_F$ is subject to large corrections. The multiplicity
ratio 
\be
\label{e_r}
r=\frac{N_{gg}}{N_{q\bar q}}=r_0[1-r_1\gamma_0-r_2\gamma_0^2-r_3\gamma_0^3]
\ee
is given as an expansion in the anomalous dimension $\gamma_0$, which can be
expressed in terms of the strong coupling $\alpha_s$ as
\be
\gamma_0 = \sqrt{ \frac{2 C_A \alpha_s}{\pi}} \quad .
\ee
Here $r_0$ denotes the colour factor ratio $C_A/C_F$, the coefficients $r_i$
of the correction terms have been calculated in Leading Order (LO) ($r_1$) 
\cite{mue}, Next-to-Leading Order (NLO)
($r_1$, $r_2$) \cite{gaffneymult} and, using a different approach in which
energy conservation is considered, in Next-to-Next-to-Next-to-Leading Order
(3NLO) 
($r_1$, $r_2$ and $r_3$) \cite{dremin}.  As non-perturbative effects of the
hadronisation process or energy conservation lead to large corrections
for the ratio $r$, the suggestion has been made rather to study the ratio of
the derivatives of the multiplicities with respect to the energy scale
$r^{(1)}$ \cite{r1_suggest} which should be less affected by non-perturbative
effects. The ratio has been calculated in 3NLO \cite{capella} as
\be
\label{e_r1cap}
r^{(1)}=\frac{d<N_{gg}>/ds}{d<N_{q\bar q}>/ds}= \frac{r}{\rho_1}
\ee
with
\be
\begin{split}
\rho_1&=1\quad-\quad \frac{\beta_0}{8C_A}r_1\gamma_0^2
\quad\left[\quad   1\quad +\quad \left(a_1+r_1+\frac{2r_2}{r_1}\right)\,
\gamma_0 \quad+
\right. \\
 &\left. \left(
\frac{2r_2a_1}{r_1}+a_1r_1+3r_2+\frac{3r_3}{r_1}+a_2 + a_1^2+r_1^2 +
\frac{\beta_1}{4C_A\beta_0}\right)
%%%% ACHTUNG !!!! Faktor 2 vor beta1 eingef"ugt wegen Definition aus kap. 2
\,\gamma_0^2 \quad\right] 
\end{split}
\ee
and $r$ from \eref{e_r}. The coefficients $r_i$ and $a_i$ have been calculated
in \cite{capella} and are given in \tref{t_as} and \tref{t_rs}, $\beta_0$ and
$\beta_1$ are the first to coefficients of the QCD $\beta$-function.
\par
In another study \cite{Eden:1998ig} the ratio $r^{(1)}$ has been derived in MLLA
within the framework of the colour dipole model. It is implicitly expressed
in the energy evolution of the gluon multiplicity in relation to that 
of the quark
multiplicity: 
%This relation is given by
%
\begin{equation}
\label{eden_gluon}
\left.\frac{dN_{gg}(L')}{dL'}\right|_{L'=L+c_g-c_q} = \frac{C_A}{C_F}
                        \left( 1 - \frac{\alpha_0 c_r}{L}\right)
                        \frac{d}{dL}N_{q\bar{q}} (L) 
\end{equation}
with
\begin{displaymath}
%\begin{xalignat*}{7}
L = \ln\left(\frac{s}{\Lambda^2}\right) \quad,\quad
\alpha_0=\frac{6}{11-2N_F/C_A}
\quad,\quad c_g=\frac{11}{6}
\quad,\quad c_q=\frac{3}{2}\quad,\quad c_r=\frac{10}{27}\pi^2-\frac{3}{2}
\qquad .
%\nonumber
\end{displaymath}
%\end{xalignat*}
%
The constants $c_q$ and $c_g$ are corrections to the phase space available for
the quark and gluon evolution, $c_r$ determines a MLLA correction which is
calculated with an estimated uncertainty of 10\% and $\Lambda$ is the QCD scale
parameter.
The solution of this differential equation implies a constant of integration.
%representing a non-perturbative term.
Extrapolating the solution of \eref{eden_gluon} to small scales,
neglecting the constant of integration, would
imply that the multiplicity in a gg-system would still be significantly larger
than in a $\rm q\bar{q}$-system. 
At very small scales, however, the hadronic multiplicity of both systems
should mainly be determined  by hadronic phase space and thus should become
almost equal \cite{Eden:1998ig}:
\begin{equation}
\label{eden_n0}
N_{gg} (L_0) \approx N_{q\bar{q}} (L_0) = N(L_0) \qquad .
\end{equation}
Thus a non-perturbative constant term appears in the solution for the gluon
multiplicity as expected from the behaviour of the fragmentation 
functions. 
In \cite{Eden:1998ig} it is suggested to determine $N(L_0)$ from data on charmonium
or bottomium states.
\par
The multiplicity of three-jet events is then given by the two
alternative formulations \cite{Eden:1999vc}:
\begin{subequations}
\begin{equation}
%\tag{{\it Eden a}}
\tag{{7A}}
\label{eden_3jet_a}
N_{q\bar q g}=N_{q\bar q}(L_{q\bar q},\kappa_{\mathrm Lu}) +
\frac{1}{2}N_{gg}(\kappa_{\mathrm Le})~~,   
\end{equation}
\begin{equation}
%\tag{{\it Eden b}}
\tag{{7B}}
\label{eden_3jet_b}
N_{q\bar q g}=N_{q\bar q}(L,\kappa_{\mathrm Lu}) +
\frac{1}{2}N_{gg}(\kappa_{\mathrm Lu})\quad ,
\end{equation}
\end{subequations}
\setcounter{equation}{6}
henceforth referred to as predictions Eden A and Eden B,
with
\addtocounter{equation}{1}
\begin{displaymath}
L=\ln\left(\frac{s}{\Lambda^2}\right)\quad,\quad
L_{q \bar q}=\ln\left(\frac{s_{q \bar q}}{\Lambda^2}\right) \quad,\quad
\kappa_{\mathrm Lu}= \ln\left(\frac{p_{t, \mathrm{Lu}}^2}{\Lambda^2}\right)
\quad,\quad
\kappa_{\mathrm Le}= \ln\left(\frac{p_{t, \mathrm{Le}}^2}{\Lambda^2}\right)
\end{displaymath}
and
\begin{displaymath}
p_{t, \mathrm{Lu}}^2=\frac{s_{qg}s_{\bar qg}}{s} \quad,\quad
p_{t, \mathrm{Le}}^2=\frac{s_{qg}s_{\bar qg}}{s_{q \bar q}}
\quad,\quad
s_{ij}=(p_i+p_j)^2\quad.
\end{displaymath}
\noindent  
The predictions \eref{eden_3jet_a} and \eref{eden_3jet_b}  
differ in the
definition of the gluon contribution to the event multiplicity.
In  \eref{eden_3jet_a} the $ q \bar q$-contribution is
determined mainly by the invariant mass of the \qqbar-system which is
also the relevant scale in an \qqbar-event of the same topology
with the gluon being replaced by a hard photon. In \eref{eden_3jet_b} 
the \qqbar-contribution is
given by the centre-of-mass energy of the whole event reflecting the 
phase-space available to the $ q\bar q$-pair if no hard 
gluon had been emitted.
%and therefore independent of the event topology.
%
\par
The expression $N_{q \bar q}(L,\kappa)$ for the quark contribution to the 
three-jet multiplicity takes into account that the resolution of a gluon jet
at a given $p_t$ implies restrictions on the phase space of the quark
system. This restricted multiplicity is linked to the multiplicity of an
unrestricted  \qqbar-system $N_{q \bar q}(L)$ via \cite{Eden:1998ig}:
\begin{equation}
\label{eden_nqq}
N_{q \bar q}(L,\kappa_{\mathrm{cut}}) = 
N_{q \bar
q}(\kappa_{\mathrm{cut}}+c_q)+(L-\kappa_{\mathrm{cut}}-c_q)
\left. \frac{dN_{q \bar q}(L')}{dL'}\right|_{L'=\kappa_{\mathrm{cut}}+c_q}.
\end{equation}
It is important to
 note that the unrestricted multiplicity is always a function of only
one logarithmic energy scale, while the restricted multiplicity demands two
logarithmic arguments.
The equivalent 
restriction for the gluon contribution to the event multiplicity
can be neglected, as the evolution scale of the gluon is the $p_t$-like
variable $\kappa$ and therefore coincides with the
cut-off scale in \eref{eden_nqq}. Note also, that
%Both predictions \eref{eden_3jet_a} and \eref{eden_3jet_b} use
% different scales for this effect,
 the topology dependence of the \qqbar-term in \eref{eden_3jet_b}
enters only due to this phase space restriction. 
\par
\begin{table}[tb]
\begin{center}
\begin{minipage}[tb]{6cm}
\begin{center}
\begin{tabular}{|c|r@{.}l|r@{.}l|r@{.}l|}
\hline 
$N_F$ & \multicolumn{2}{c|}{$a_1$}   & \multicolumn{2}{c|}{$a_2$}   & 
\multicolumn{2}{c|}{$a_3$} \\
\hline 
  3   & 0&280   & -0&379    & 0&209  \\
\hline 
  4   & 0&297   & -0&339    & 0&162  \\
\hline 
  5   & 0&314   & -0&301    & 0&112\\
\hline 
\end{tabular}
\end{center}
\caption{\label{t_as} The coefficients $a_i$ from \cite{capella}}
\end{minipage}
\hspace{1.5cm}
\begin{minipage}[tb]{6cm}
\begin{center}
\begin{tabular}{|c|r@{.}l|r@{.}l|r@{.}l|}
\hline 
$N_F$ & \multicolumn{2}{c|}{$r_1$}   & \multicolumn{2}{c|}{$r_2$}   & 
\multicolumn{2}{c|}{$r_3$} \\
\hline 
  3   & 0&185   & 0&426    & 0&189  \\
\hline 
  4   & 0&191   & 0&468    & 0&080  \\
\hline 
  5   & 0&198   & 0&510    & -0&041\\
\hline 
\end{tabular}
\end{center}
\caption{\label{t_rs} The coefficients $r_i$
from \cite{capella}}
\end{minipage}
\end{center}
\end{table}
The explicit energy dependence of the multiplicity of unrestricted
(``unbiased'') $ q\bar q$ events has already been calculated in 
\cite{webbermult}:
\be \label{e_nwebber}
\left\langle N (Q^2)\right\rangle = a \cdot \alpha_s^b(Q^2) \cdot 
\exp\left( \frac{c}{\sqrt{\alpha_s(Q^2)}}\right) \cdot 
\Bigg[ 1 + {\cal O}\left(\sqrt{\alpha_s(Q^2)}\right)\Bigg]
\ee
with
\begin{align}
b&=  \frac{1}{4}+
      \frac{2}{3}\frac{N_F}{\beta_0}\left(1-\frac{C_F}{C_A}\right)
      \qquad \mbox{and} \\
c&=  \sqrt{\frac{32\pi\cdot C_A}{\beta_0^2}}\quad.
\end{align}
In a more recent publication \cite{capella} the multiplicity of $q\bar q$ 
or $gg$ colour singlet systems has been
calculated in 3NLO as an 
expansion of $r$ and $\gamma=\langle N_g\rangle'/\langle
N_g\rangle$, where the prime denotes a derivative with respect to the
logarithmic energy scale $y=\log(p\Theta/Q_0)$ with $\Theta$ being the
opening angle of the first splitting and $Q_0$ the cut-off of the perturbative
expansion. The mean multiplicities are then
given by
\be
\label{e_capellang}
\langle N_g(y)\rangle = k_g \cdot y^{-a_1 C^2}\cdot
 \exp\left[2C\sqrt{y}+\delta_g(y)\right]
\ee
and
\be
\label{e_capellanq}
\langle N_q(y)\rangle = \frac{k_q}{r_0} \cdot y^{-a_1 C^2}\cdot
 \exp\left[2C\sqrt{y}+\delta_q(y)\right]\qquad.
\ee
$k_g$ and $k_q$ denote free normalisations, the quantities
$r_i$ and $a_i$ are calculated
 in \cite{capella} and given in \tref{t_as} and \tref{t_rs}, 
$C=\sqrt{4N_C/\beta_0}$ and the
 additional contributions to the exponent are given as
\be
\delta_g(y)=
\frac{C}{\sqrt{y}}\left[2a_2C^2+\frac{\beta_1}{\beta_0^2}
\left\{ \log(2y)+2\right\}\right]+
\frac{C^2}{y}\left[a_3C^2-\frac{a_1\beta_1}{\beta_0^2}
\left\{ \log(2y)+1\right\}\right]
\ee
and
\be
\delta_q(y)=\delta_g(y) +\frac{C}{\sqrt{y}}r_1
+\frac{C^2}{y}\left( r_2 + \frac{r_1^2}{2}\right) \quad.
\ee
Due to the definitions of the evolution variables, which are not symmetric
 with respect to quarks and gluons, the multiplicity of the two-gluon system
 is given in a slightly higher order than the multiplicity of the
 quark-antiquark system. 
\section{Mean multiplicity of hadronic events\label{s_evtmult}}
%
%
%\begin{figure}[tb]
%\begin{center}
%\begin{minipage}[h]{10cm}
%\mbox{\epsfig{file=mult_distr_event.eps,width=10cm}}
%\end{minipage} 
%\end{center}
%\caption
%{\label{f_mudis_evt} The overall multiplicity distribution for 
%$udscb$- and $udsc$-events. The lines represent a fit of negative binomials.
%}
%\end{figure}
\begin{figure}[tb]
\begin{center}
\begin{minipage}[h]{14cm}
\mbox{\epsfig{file=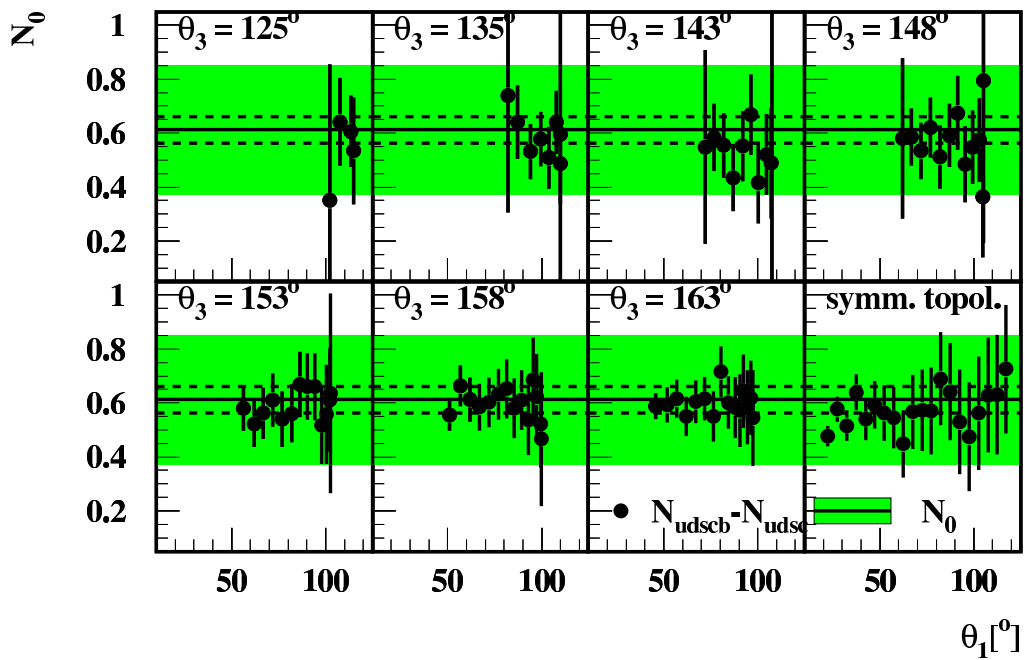,width=14cm}}
\end{minipage} 
\end{center}
\caption
{\label{f_n0const} The multiplicity difference between $udsc$- and
  $udscb$-events. The shaded area reflects the value for $N_0$ from
  \eref{e_n0}, the dashed lines indicate statistical errors only.}
\end{figure}
%
%%%%%%%%%%%%%%%%%%%%%%%%%%%%%%%%%%%%%%%%%%%%%%%%%%%%%%%%%%%%%%%%%%%%%
%Multiplicity is a relatively difficult quantity to measure, since many
%tracks
%especially of soft particles are incorrectly reconstructed, which leads to
%acceptance corrections of the order of 30\%.
%In order to estimate biases in this analysis, the measured overall hadronic
%multiplicity is compared with previous measurements of this quantity. 
%\par
%%%%%%%%%%%%%%%%%%%%%%%%%%%%%%%%%%%%%%%%%%%%%%%%%%%%%%%%%%%%%%%%%5
%In \fref{f_mudis_evt} the multiplicity distribution for all hadronic $udsc$-
%(open markers) and $udscb$-events (solid markers) regardless of event topology
%is shown. It can clearly be observed that the multiplicity distribution for
%$udscb$-events is slightly shifted to higher multiplicity with respect to
%$udsc$-events due to the additional
%multiplicity contribution from the decay of the heavy $b$-quarks. The
%plotted lines indicate negative binomial distributions which have been fitted
%to the data. 
%The mean values of these distributions are $20.963\pm0.004$ for $udscb$- and
%$20.353\pm0.004$ for $udsc$-events with statistical errors only. 
The mean multiplicity of hadronic events  at $\sqrt{s}=m_Z$, 
without consideration of the event
topology, is found in this analysis 
to be $20.963\pm0.004$ for $udscb$- and
$20.353\pm0.004$ for $udsc$-events with statistical errors only. The
multiplicity of $udscb$-events is expected to be slightly larger than the
multiplicity of $udsc$-events due to the additional multiplicity produced in
$B$ decays.
These values are in good agreement with previous measurements of this quantity
\cite{multmess}. From this measurement the difference in multiplicity between
$udscb$- and $udsc$-events, which proves to be crucial for this analysis, has
been found to be
\be
N_0\equiv \delta_{udscb-udsc}= N_{udscb} - N_{udsc} = 
0.610\pm 0.002_{\mathrm{stat.}} \quad,
\ee
where the error of the difference has been determined assuming a correlation
of $\sqrt{1-R_b}\sim0.9$ between the two measurements. $R_b$ denotes the
fraction of hadronic events with initial $b$-quarks, which has been measured
on the $Z$ resonance \cite{rb}: 
\be
\label{e_rb}
R_b(m_Z)=0.21638\pm0.00066 \qquad.
\ee
 From previous
measurements of the multiplicity of $udscb$- and $b$-events
\cite{n0mes_d,n0mes_o}, 
$N_0$ can be calculated using the relation
\be
N_0=2\frac{R_b}{1-R_b} \cdot
\Bigl(\langle n_h\rangle_b -\langle n_h\rangle\Bigr) \quad,
\ee
where $\langle n_h\rangle_b$ and $\langle n_h\rangle$ represent the
multiplicity per hemisphere in a $b$- or a general hadronic event,
respectively.  
The value obtained from the results in \cite{n0mes_d} is 
%\be
$N_0=0.60\pm0.06$,
%\ee
while the value obtained from the results in \cite{n0mes_o} is
$N_0=0.62\pm0.07$, both in very good agreement with the value of this
analysis. 
The given statistical errors are larger than the statistical error
of $N_0$ from this analysis, as in \cite{n0mes_d} and \cite{n0mes_o} $N_b$ is
measured instead of $N_{udsc}$, so the determination of $N_0$ is less direct
and based on a smaller event sample. 
The systematic errors of
$N_b$ are $\pm 0.22$ in \cite{n0mes_d} and $\pm 0.24$ in \cite{n0mes_o}.
The average of these two values for $N_0$ is 
\be
\label{e_n0}
N_0=0.61\pm 0.24
\ee
where also the given
systematic errors on $N_b$ have been considered.  
%As
%the treatment of the event multiplicity without consideration to the event
%topology is more thorough in a way that would go beyond the scope of this
%analysis, the averaged value \eref{e_n0}, which agrees completely with the
%value obtained here, is used further on.
\par
\begin{figure}[tb]
\begin{center}
\begin{minipage}[h]{15cm}
\mbox{\epsfig{file=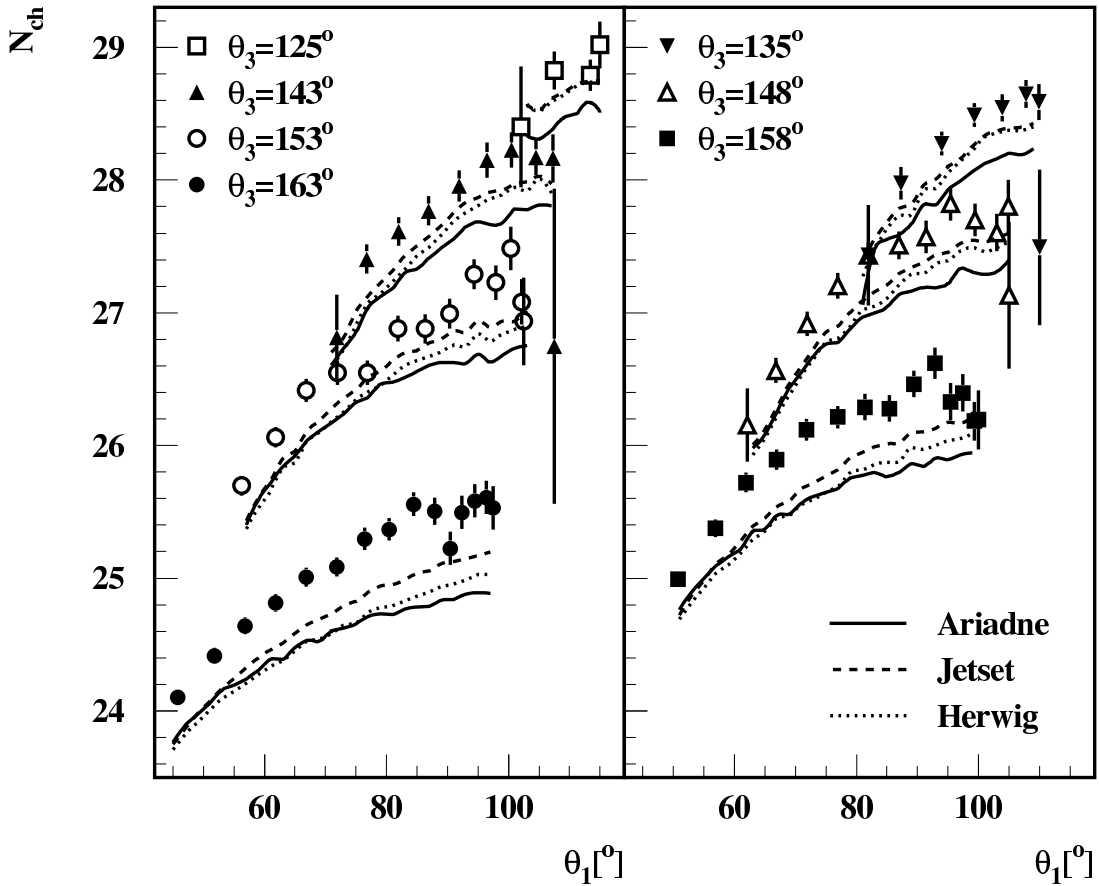,width=15cm}}
\end{minipage} 
\end{center}
\caption
{\label{f_mcmult_gen}
The mean charged
multiplicity of $udscb$-events with general topologies as a function
of the opening angle $\theta_1$ for different values of $\theta_3$ in
comparison with the predictions of Monte Carlo models. The data points have
been plotted in two diagrams with alternating $\theta_3$-bins, 
as the neighbouring curves would otherwise overlap. 
The errors shown 
are  statistical only.}
\end{figure}
\begin{figure}[tb]
\begin{center}
\begin{minipage}[h]{12cm}
\mbox{\epsfig{file=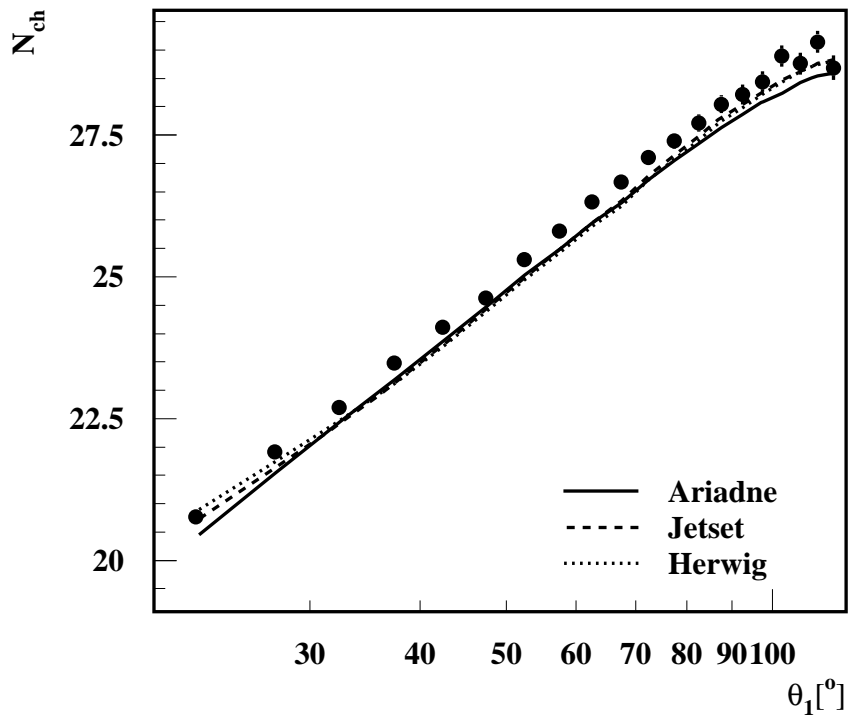,width=12cm}}
\end{minipage} 
\end{center}
\caption
{\label{f_mcmult_sym} The mean charged multiplicity of $udscb$-events with
  symmetric topology as a function of the opening angle $\theta_1$ in
  comparison with the predictions of Monte Carlo models
 The errors shown 
are  statistical only.}
\end{figure}
\begin{table}[p]
\begin{center}
 \begin{tabular}{|c|c|r@{~$\pm$}l@{~$\pm$}l||c|c|r@{~$\pm$}l@{~$\pm$}l|}
 \hline
 $\vartheta_2$&$\vartheta_3$&
 \multicolumn{3}{c|}{$N^{udscb}_{\mathrm{ch}}$}&
 $\vartheta_2$&$\vartheta_3$&
 \multicolumn{3}{c|}{$N^{udscb}_{\mathrm{ch}}$}
 \\ \hline
 $ 99^\circ$ &$ 163^\circ$ & 25.53 &0.16 &   0.12 &
 $ 121^\circ$ &$ 153^\circ$ & 26.88 &0.11 &  0.06 \\
 $ 101^\circ$ &$ 158^\circ$ & 26.19 &0.22 &   0.14 &
 $ 121^\circ$ &$ 158^\circ$ & 26.29 &0.10 &  0.02 \\
 $ 101^\circ$ &$ 163^\circ$ & 25.61 &0.12 &   0.06 &
 $ 121^\circ$ &$ 163^\circ$ & 25.30 &0.08 &  0.05 \\
 $ 103^\circ$ &$ 153^\circ$ & 26.94 &0.33 &   0.11 &
 $ 126^\circ$ &$ 125^\circ$ & 28.83 &0.15 &  0.12 \\
 $ 103^\circ$ &$ 158^\circ$ & 26.18 &0.15 &   0.10 &
 $ 126^\circ$ &$ 135^\circ$ & 28.49 &0.09 &  0.09 \\
 $ 103^\circ$ &$ 163^\circ$ & 25.58 &0.12 &   0.10 &
 $ 126^\circ$ &$ 143^\circ$ & 27.96 &0.12 &  0.09 \\
 $ 105^\circ$ &$ 148^\circ$ & 27.13 &0.55 &   0.42 &
 $ 126^\circ$ &$ 148^\circ$ & 27.51 &0.10 &  0.06 \\
 $ 105^\circ$ &$ 153^\circ$ & 27.08 &0.17 &   0.09 &
 $ 126^\circ$ &$ 153^\circ$ & 26.88 &0.10 &  0.05 \\
 $ 105^\circ$ &$ 158^\circ$ & 26.40 &0.14 &   0.09 &
 $ 126^\circ$ &$ 158^\circ$ & 26.21 &0.09 &  0.04 \\
 $ 105^\circ$ &$ 163^\circ$ & 25.49 &0.13 &   0.11 &
 $ 126^\circ$ &$ 163^\circ$ & 25.08 &0.07 &  0.02 \\
 $ 107^\circ$ &$ 143^\circ$ & 26.75 &1.19 &   0.79 &
 $ 131^\circ$ &$ 125^\circ$ & 28.40 &0.46 &  0.22 \\
 $ 107^\circ$ &$ 148^\circ$ & 27.80 &0.20 &   0.13 &
 $ 131^\circ$ &$ 135^\circ$ & 28.28 &0.09 &  0.07 \\
 $ 107^\circ$ &$ 153^\circ$ & 27.49 &0.16 &   0.07 &
 $ 131^\circ$ &$ 143^\circ$ & 27.77 &0.11 &  0.05 \\
 $ 107^\circ$ &$ 158^\circ$ & 26.33 &0.14 &   0.07 &
 $ 131^\circ$ &$ 148^\circ$ & 27.43 &0.10 &  0.08 \\
 $ 107^\circ$ &$ 163^\circ$ & 25.23 &0.12 &   0.05 &
 $ 131^\circ$ &$ 153^\circ$ & 26.55 &0.09 &  0.08 \\
 $ 110^\circ$ &$ 135^\circ$ & 27.49 &0.59 &   0.35 &
 $ 131^\circ$ &$ 158^\circ$ & 26.12 &0.08 &  0.04 \\
 $ 110^\circ$ &$ 143^\circ$ & 28.16 &0.18 &   0.12 &
 $ 131^\circ$ &$ 163^\circ$ & 25.01 &0.07 &  0.07 \\
 $ 110^\circ$ &$ 148^\circ$ & 27.60 &0.14 &   0.08 &
 $ 136^\circ$ &$ 135^\circ$ & 27.98 &0.12 &  0.07 \\
 $ 110^\circ$ &$ 153^\circ$ & 27.23 &0.13 &   0.04 &
 $ 136^\circ$ &$ 143^\circ$ & 27.61 &0.11 &  0.08 \\
 $ 110^\circ$ &$ 158^\circ$ & 26.62 &0.12 &   0.06 &
 $ 136^\circ$ &$ 148^\circ$ & 27.20 &0.10 &  0.06 \\
 $ 110^\circ$ &$ 163^\circ$ & 25.50 &0.10 &   0.09 &
 $ 136^\circ$ &$ 153^\circ$ & 26.55 &0.09 &  0.06 \\
 $ 113^\circ$ &$ 135^\circ$ & 28.59 &0.13 &   0.06 &
 $ 136^\circ$ &$ 158^\circ$ & 25.89 &0.08 &  0.08 \\
 $ 113^\circ$ &$ 143^\circ$ & 28.17 &0.13 &   0.09 &
 $ 136^\circ$ &$ 163^\circ$ & 24.81 &0.07 &  0.07 \\
 $ 113^\circ$ &$ 148^\circ$ & 27.70 &0.12 &   0.11 &
 $ 141^\circ$ &$ 135^\circ$ & 27.43 &0.38 &  0.24 \\
 $ 113^\circ$ &$ 153^\circ$ & 27.29 &0.11 &   0.07 &
 $ 141^\circ$ &$ 143^\circ$ & 27.41 &0.11 &  0.07 \\
 $ 113^\circ$ &$ 158^\circ$ & 26.46 &0.10 &   0.05 &
 $ 141^\circ$ &$ 148^\circ$ & 26.92 &0.09 &  0.09 \\
 $ 113^\circ$ &$ 163^\circ$ & 25.56 &0.09 &   0.09 &
 $ 141^\circ$ &$ 153^\circ$ & 26.41 &0.09 &  0.06 \\
 $ 117^\circ$ &$ 125^\circ$ & 29.02 &0.18 &   0.12 &
 $ 141^\circ$ &$ 158^\circ$ & 25.72 &0.07 &  0.05 \\
 $ 117^\circ$ &$ 135^\circ$ & 28.65 &0.10 &   0.06 &
 $ 141^\circ$ &$ 163^\circ$ & 24.64 &0.06 &  0.05 \\
 $ 117^\circ$ &$ 143^\circ$ & 28.23 &0.13 &   0.06 &
 $ 146^\circ$ &$ 143^\circ$ & 26.81 &0.32 &  0.31 \\
 $ 117^\circ$ &$ 148^\circ$ & 27.82 &0.12 &   0.07 &
 $ 146^\circ$ &$ 148^\circ$ & 26.56 &0.09 &  0.07 \\
 $ 117^\circ$ &$ 153^\circ$ & 26.99 &0.11 &   0.06 &
 $ 146^\circ$ &$ 153^\circ$ & 26.06 &0.08 &  0.09 \\
 $ 117^\circ$ &$ 158^\circ$ & 26.28 &0.10 &   0.03 &
 $ 146^\circ$ &$ 158^\circ$ & 25.38 &0.07 &  0.07 \\
 $ 117^\circ$ &$ 163^\circ$ & 25.37 &0.09 &   0.05 &
 $ 146^\circ$ &$ 163^\circ$ & 24.41 &0.06 &  0.07 \\
 $ 121^\circ$ &$ 125^\circ$ & 28.79 &0.12 &   0.08 &
 $ 152^\circ$ &$ 148^\circ$ & 26.16 &0.28 &  0.14 \\
 $ 121^\circ$ &$ 135^\circ$ & 28.54 &0.10 &   0.10 &
 $ 152^\circ$ &$ 153^\circ$ & 25.70 &0.07 &  0.07 \\
 $ 121^\circ$ &$ 143^\circ$ & 28.15 &0.13 &   0.08 &
 $ 152^\circ$ &$ 158^\circ$ & 24.99 &0.05 &  0.08 \\
 $ 121^\circ$ &$ 148^\circ$ & 27.57 &0.12 &   0.08 &
 $ 152^\circ$ &$ 163^\circ$ & 24.10 &0.04 &  0.08 \\
 \hline
 \end{tabular}
\end{center}
\caption{\label{t_mult_gen} The multiplicity of $udscb$-events in dependence
 of the event topology for general topologies. The values for $\theta_2$ and
 $\theta_3$  represent the centre of the bins, not averages, i.e. due to bin
 overlap it can
 happen that the given value for $\theta_2$ is larger than the corresponding
 value for $\theta_3$.  
 The first errors are statistical, the second errors
 are systematic errors due to the choice of the cluster algorithm and
 variations of track and event cuts. 
% Note that no error on the effect of the
% correction procedure on the absolute value of the multiplicities is 
%included.
} 
\end{table}
\begin{table}[tb]
\begin{center}
 \begin{tabular}{|c|r@{~$\pm$}l@{~$\pm$}l||c|r@{~$\pm$}l@{~$\pm$}l|}
 \hline
 $\vartheta_1$&\multicolumn{3}{c|}{$N^{udscb}_{\mathrm{ch}}$}&
 $\vartheta_1$&\multicolumn{3}{c|}{$N^{udscb}_{\mathrm{ch}}$}
 \\ \hline
 $ 22^\circ$ &  20.77 &0.03 &   0.24 &
 $ 72^\circ$ &  27.11 &0.13&   0.10 \\
 $ 27^\circ$ &  21.91 &0.04 &   0.18 &
 $ 77^\circ$ &  27.40 &0.14&   0.08 \\
 $ 32^\circ$ &  22.70 &0.05 &   0.13 &
 $ 82^\circ$ &  27.71 &0.15&   0.11 \\
 $ 37^\circ$ &  23.48 &0.06 &   0.10 &
 $ 87^\circ$ &  28.05 &0.16&   0.06 \\
 $ 42^\circ$ &  24.11 &0.07 &   0.09 &
 $ 92^\circ$ &  28.21 &0.17&   0.13 \\
 $ 47^\circ$ &  24.63 &0.08 &   0.09 &
 $ 97^\circ$ &  28.44 &0.18&   0.10 \\
 $ 52^\circ$ &  25.30 &0.09 &   0.06 &
 $ 102^\circ$ &  28.89 &0.19&   0.14 \\
 $ 57^\circ$ &  25.81 &0.10 &   0.08 &
 $ 107^\circ$ &  28.76 &0.19&   0.07 \\
 $ 62^\circ$ &  26.32 &0.11 &   0.10 &
 $ 112^\circ$ &  29.14 &0.20&   0.07 \\
 $ 67^\circ$ &  26.67 &0.12 &   0.05 &
 $ 117^\circ$ &  28.68 &0.22&   0.14 \\
 \hline
 \end{tabular}
\end{center}
\caption{\label{t_mult_sym} The multiplicity of $udscb$-events in dependence
 of the event topology for symmetric topologies. 
 The first errors are statistical, the second errors
 are systematic errors due to the choice of the cluster algorithm and
 variations of track and event cuts. 
%Note that no error on the effect of the
% correction procedure on the absolute value of the multiplicities is 
%included.
} 
\end{table}
The additional multiplicity due to initial $b$-quarks is in MLLA predicted to
be independent of the centre-of-mass energy due to energy conservation and
coherence effects in the gluon-radiation from heavy quarks
\cite{mlla_n0const}. This energy independence has been experimentally verified
(see e.g.~\cite{kha}). 
In \fref{f_n0const} the multiplicity difference 
between $udscb$- and $udsc$-events $N_0$ is shown as a function of the event
topology. The shaded area reflects the value given in 
\eref{e_n0}, the dashed lines
indicate the purely statistical error on $N_0$. No significant tendency can
be observed, $N_0$ can be considered as independent of the event topology
within the angular regions studied. 
\par
The dependence of the  multiplicity of $udscb$-events on
the event topology is shown in
\fref{f_mcmult_gen} for general and in \fref{f_mcmult_sym} for symmetric
event topologies. For symmetric topologies a nearly logarithmic 
increase of multiplicity with increasing $\theta_1$ can be observed. Also for
general topologies the event multiplicity increases with $\theta_1$ for a
fixed value of $\theta_3$. For
general event topologies also an increase of multiplicity with decreasing
$\theta_3$ can be observed. This dependence is more pronounced than the
$\theta_1$-dependence as a change in $\theta_1$ of $50^\circ$ results in a
multiplicity change of roughly 2 units, while the same change in $\theta_3$
leads to a change in multiplicity of roughly 4 units. The Monte Carlo models,
although mutually agreeing well, underestimate the multiplicity by $\sim0.4$
units. In \fref{f_mcmult_gen} it can be seen that the deviation is stronger
for large values of $\theta_3$ than for smaller values.
 \par
%However, 
%for extremely small $\theta_1$ a
%topology dependence would be expected due to the so-called dead-cone
%effect, which would result in a depletion of $b$-events in these topology
%classes leading to an effective reduction of $N_0$.
%
%
\section{Preparation of the prediction \label{s_predprep}}
\begin{figure}[p]
\begin{center}
\begin{minipage}[h]{12cm}
\mbox{\epsfig{file=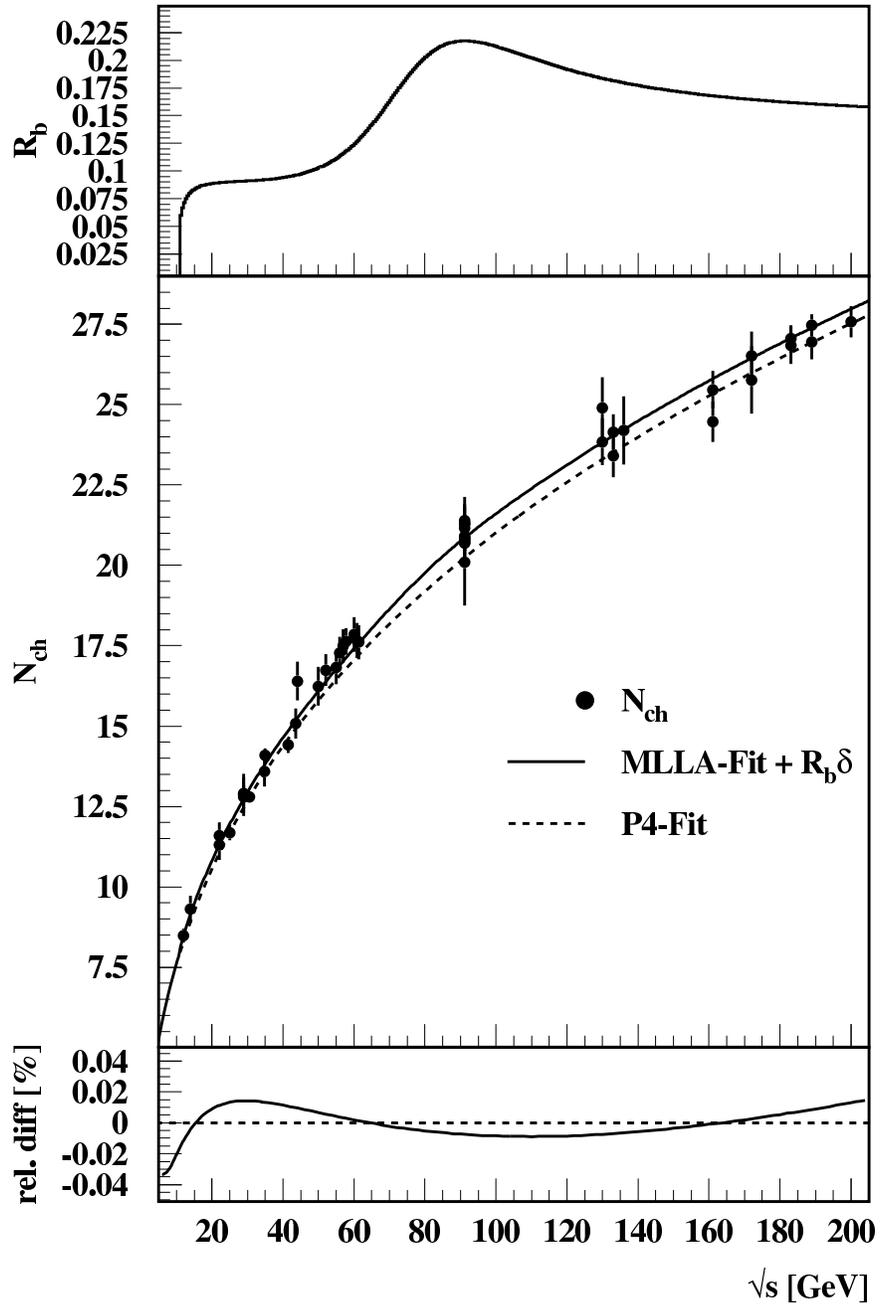,width=12cm}}
\end{minipage} 
\end{center}
\caption{\label{f_fitepem} 
 {\bf Top:} The branching ratio $R_b$ as a function of $\sqrt{s}$~
 {\bf Middle:} The measurements of $N_{q\bar q}$
 \protect{\cite{multmess,restmult}} with the fitted
 parametrisation \eref{e_fitepem}. 
Note that the 4th order polynomial P4 fit does not account for the additional
 multiplicity due to $b$-events.
 {\bf Bottom:}
The relative deviation of the fitted polynomial P4
from the parametrisation \eref{e_nwebber}}
\end{figure}
\eref{eden_3jet_a} and \eref{eden_3jet_b} give 
the average multiplicity of a
three-jet event in terms of the restricted $q\bar q$-multiplicity
and the multiplicity of a two-gluon system. Both contributions can be derived
from the unrestricted $q\bar q$-multiplicity using \eref{eden_nqq} and
\eref{eden_gluon}, 
%and both have
which has
 been measured in numerous $e^+e^-$ experiments at
different centre-of-mass energies. The aim of this section is to provide a closed
parametrisation of $N_{q\bar q g}$ using a parametrisation of the measured
unrestricted multiplicity $N_{q\bar q}(\sqrt{s})$. 
\par
The measurements of $N_{q\bar
  q}(\sqrt{s})$ used \cite{multmess} are shown in the central part of
  \fref{f_fitepem}. 
%In the middle part of \fref{f_fitepem} the measurements of $N_{q\bar
%  q}(\sqrt{s})$ which are used 
%are shown \cite{multmess}. 
Not included are older
  measurements of the {\sc Jade} and {\sc Pluto} collaboration in which the
  pions from the decay $K^0\to\pi\pi$ have not been added to the
  multiplicity \cite{multmess_notused}. Recent results of a re-analysis of the
  {\sc Jade} data \cite{jade_reana} are, taken into account.
  In these measurements events with initial $b$-quarks are included. Due to the
  decay products of the $b$-containing mesons these events have an increased
  multiplicity. As the fraction of events with initial $b$-quarks varies with
  the centre-of-mass energy, this additional multiplicity introduces an
  additional energy dependence of the multiplicity which is not due to strong
  interactions and which  has to be corrected for. The upper plot of 
\fref{f_fitepem} shows the branching ratio
\be
R_b=\frac{\sigma(e^+e^-\to b\bar b)}{\sigma(e^+e^-\to \mathrm{hadrons})}
\ee
as a function of the centre-of-mass energy. The curve has been obtained
using the LUXTOT routine of the {\sc Jetset}/{\sc Pythia} package \cite{luclus}.
%It has been determined using the
%  LUXTOT routine of the {\sc Jetset}/{\sc Pythia} package \cite{pythia}. Beginning at
%  $\sqrt{s}>10$GeV it raises rapidly to the value of $R_b\sim\frac{1}{11}$
%  which is given by the electric charge of the $b$-quark and stays nearly
%  constant where the photon production is the dominant $s$-channel process. 
%  In the vicinity of the $Z$-resonance the $b$-fraction is increasing as the
%  decay-width of the $Z$ into $b$'s is larger than the one of the photon.
%  In the energy region above the $Z$-resonance an average value is
%  asymptotically reached.
%\par
The average additional multiplicity due to $b$-events can be expressed in
terms of $N_0$:
\be
\delta_{b-udsc}=2\cdot\Bigl(\langle n_h\rangle_b -
\langle n_h\rangle_{udsc}\Bigr)=\frac{N_0}{R_b(m_Z)}\quad.
\ee
With the values from \eref{e_rb} and \eref{e_n0}, the multiplicity difference
can be determined giving $\delta_{b-udsc}=2.83\pm0.23$. 
\par
The $q\bar q$-multiplicities are then fitted with the function
\be
\label{e_fitepem}
N_{udscb}(\sqrt{s})\,=\,N_{udsc}(\sqrt{s})\,+\,\delta_{b-udsc}\cdot 
R_b(\sqrt{s})
\ee
 where \eref{e_nwebber} is used to parametrise $N_{udsc}(\sqrt{s})$. The
 description of the data is very good with a $\chi^2$ per degree of freedom of
 $33/48$. The values found for the fit parameters are
\begin{align}
a&= 0.10252 \pm 0.0025 \nonumber\\
\Lambda&= (0.243\pm 0.012)~\mathrm{GeV} \quad .\nonumber
\end{align}
The fitted function is indicated in \fref{f_fitepem} as a solid line. To
simplify the integration of
$N_{udsc}(\sqrt{s})$ required due to \eref{eden_gluon}, a polynomial of
order four in $L=\log(s/\Lambda^2)$ is fitted to the parametrisation of
$N_{udsc}(\sqrt{s})$. The fit is performed with $\Lambda=0.25$ GeV, and the
values obtained for the coefficients are given
in \tref{t_p4coeff}. The polynomial is indicated as a dashed line in
the middle plot of \fref{f_fitepem}. The difference between the dashed
polynomial and the solid line indication of \eref{e_fitepem} is due to the
omission of the $\delta_{b-udsc}\cdot 
R_b(\sqrt{s})$ term. In the lower part of \fref{f_fitepem} the relative
deviation of the polynomial from the fitted  $N_{udsc}(\sqrt{s})$ is
shown. The deviations are smaller than $2\cdot10^{-4}$ 
over most of the fitted region  
  and nowhere exceed $4\cdot10^{-4}$,
indicating an excellent description of $N_{udsc}(\sqrt{s})$ by the
polynomial. 
\begin{table}[tb]
\begin{center}
\begin{tabular}{ll}
\hline
$p_0$&$ 0.18981 $\\
$p_1$&$ 0.40950 $\\
$p_2$&$ 0.57358\cdot10^{-1}$ \\
$p_3$&$ 0.10349\cdot10^{-2}$ \\
$p_4$&$ 0.28640\cdot10^{-3}$\\
\hline
\end{tabular}
\caption{\label{t_p4coeff} The coefficients of the polynomial fitted to
$N_{ch}^{udsc}(\sqrt{s})$}
\end{center}
\end{table}
The value of the polynomial at $\sqrt{s}=m_Z$ is 20.2561. This value is in
good agreement with the mean multiplicity of $udsc$-events of $20.353\pm0.004$
measured in this
analysis. The small difference is taken into account when considering the
systematic errors of this analysis.
\par
Inserting the polynomial into \eref{eden_gluon} and \eref{eden_nqq} leads to a
closed form of the predictions of \eref{eden_3jet_a} and \eref{eden_3jet_b}. To
fix the constant of integration, left free in \eref{eden_gluon}, a measurement
of $N_{gg}$ from  $\chi'$ decays by the {\sc Cleo} collaboration is used 
\cite{cleomult}. 
The  value of $N_{gg}(9.9132~\mathrm{GeV})=9.339\pm 0.090\pm 0.045$
using $\Lambda=250$ MeV 
results in $L_0=5.86$, which corresponds to a centre-of-mass energy of 4.68 GeV
at which quark and gluon multiplicities are equal.
\par
Assuming massless jet kinematics, the scale variables $\kalu$, $\kale$
and $L_{q\bar q}$ can directly be expressed only
in terms of the inter-jet angles
$\theta_i$ (and the constant centre-of-mass energy). This yields 
$N_{q\bar qg}$ 
as a function of the inter-jet angles under the assumption that a certain jet
is the gluon jet. As no explicit identification of the gluon jet is made,
$N_{q\bar qg}$  is calculated for all three possible gluon jet hypotheses and
the weighted mean of these values is taken, where each hypothesis is weighted
with the corresponding value of the three-jet matrix-element
which can be calculated from the inter-jet angles assuming again massless jet
kinematics. 
%\be
%\frac{x_q^2+x_{\bar q}^2}{(1-x_q)(1-x_{\bar q})}
%\ee
%which again can be expressed in terms of the inter jet angles via
%\be
%x_i=2\cdot\frac{\sin\theta_i}{\sin\theta_1+\sin\theta_2+\sin\theta_3}\quad.
%\ee
\par
However, it is a known feature of the hadronisation process that
due to coherent
production of inter-jet particles
%, especially between close-by jets, 
%these 
close-by 
jets are pulled even closer together.
 The calculations leading to \eref{eden_3jet_a} and
\eref{eden_3jet_b}  refer to the partonic structure of an event. While
for the change in multiplicity due to the hadronisation process in accordance
with the LPHD hypothesis an overall normalisation constant can be found, for
the change in the event topology, i.e.~the inter-jet angles, this is not the
case. Therefore,
 a topology dependent hadronisation correction has to be applied. 
The effect can be most easily understood for symmetric event topologies. During
the hadronisation process jets two and three are pulled closer together
resulting in a smaller opening angle $\theta_1$ at the hadronic than at the 
partonic
level of the event. As the multiplicity  increases with $\theta_1$, 
the reduced $\theta_1$ found at the hadronic level thus 
results in an underestimation of the multiplicity
compared to 
the prediction which is based on the value of $\theta_1$ at parton level.  
\par
\begin{figure}[tb]
\begin{center}
\begin{minipage}[h]{14cm}
\mbox{\epsfig{file=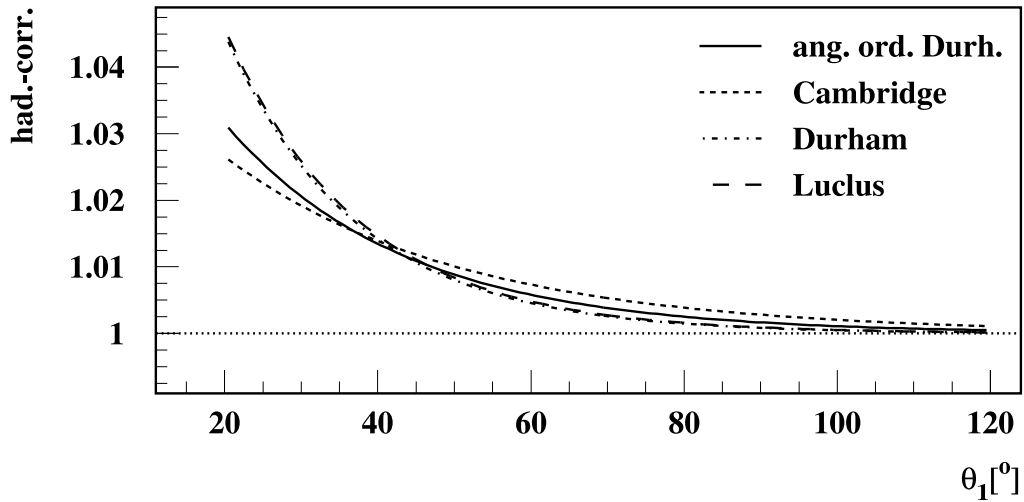,width=14cm}}
\end{minipage} 
\end{center}
\caption
{\label{f_hadcor_sym} 
The hadronisation correction factor for symmetric topologies from the {\sc
  Ariadne }
Monte Carlo. Plotted is the
ratio of multiplicities with $\theta_1$ determined at the hadronic level and
the multiplicities with $\theta_1$ determined at the partonic level. 
}
\end{figure}
\begin{table}[tb] \begin{center}
\begin{tabular}{|l|c|r@{.}l|r@{.}l|r@{.}l|r@{.}l|}
\hline
\multicolumn{2}{|c|}{ }&\multicolumn{2}{c|}{a.o.D.}&\multicolumn{2}{c|}{Camb.}
&\multicolumn{2}{c|}{Durh.}&\multicolumn{2}{c|}{Lucl.}\\
\hline
& $a_1$ & -0&0417134 & -0&0317331 & -0&0564096 & -0&0554275 \\
\cline{2-10}
\haho{\sc{Ariadne}}
& $a_2$ & 0&07052 &0&0487817&  0&133435&  0&132929 \\ 
\hline
& $a_1$ & -0&0591038 &  -0&0455694 & -0&0750201  &  -0&0564708 \\
\cline{2-10}
\haho{\sc{Jetset}}
& $a_2$ &   0&111077  &   0&0804121 & 0&240021 &  0&149093 \\
\hline
& $a_1$ &    -0&0843598 & -0&147902& -0&0788673 &  -0&0680853 \\
\cline{2-10}
\haho{\sc{Herwig}} & $a_2$ &    0&549409   &0&869146&  0&727511 &0&521014 \\
\hline
\end{tabular}
\end{center}
\caption
{\label{t_hadcor_sym} The parameters of the hadronisation correction factors
  for symmetric event topologies}
\end{table}
\begin{figure}[p]
\begin{center}
\begin{minipage}[h]{14cm}
\mbox{\epsfig{file=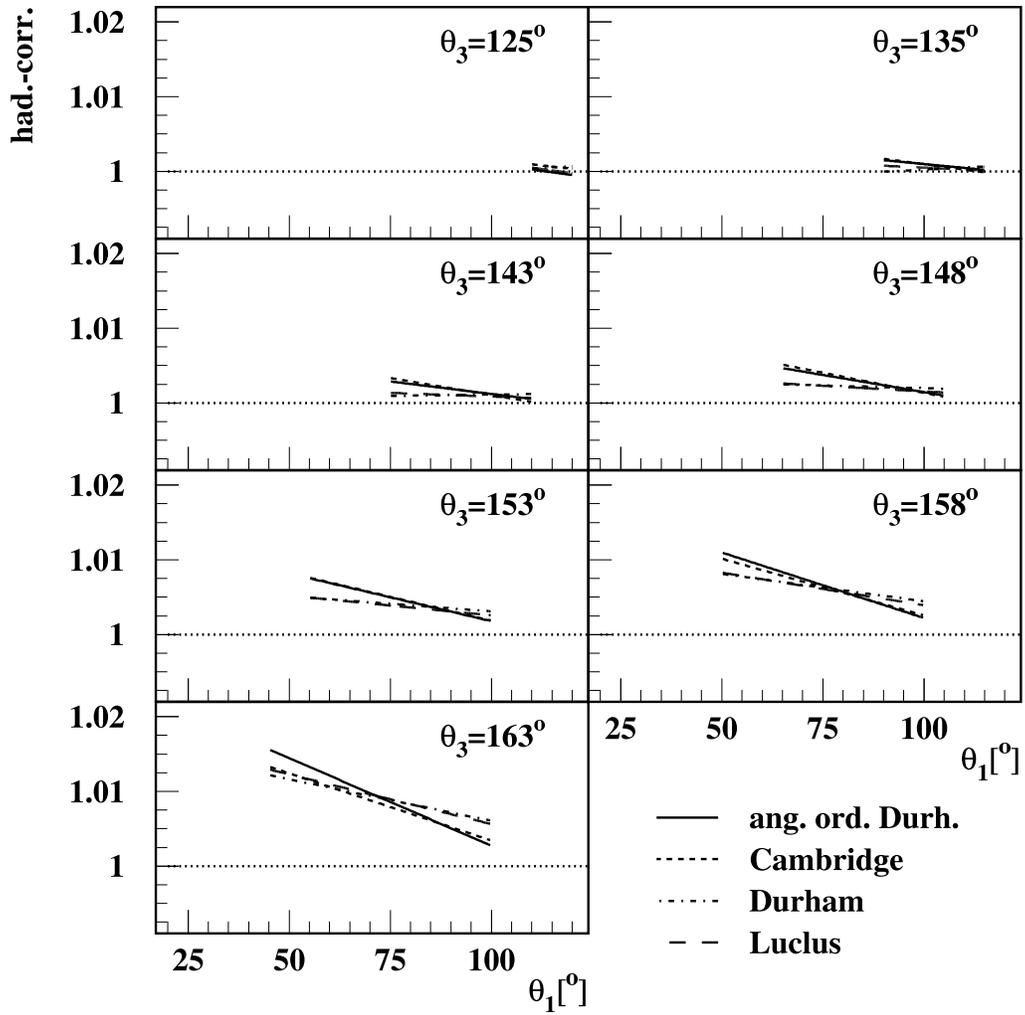,width=14cm}}
\end{minipage} 
\end{center}
\caption
{\label{f_hadcor_all} 
The hadronisation correction factors for general topologies.
Plotted is the
ratio of multiplicities with the inter-jet angles
 determined at the hadronic level over
the multiplicities with the inter-jet angles determined at the partonic level. 
}
\end{figure}
\begin{table}[tb] \begin{center}
\begin{tabular}{|l|c|r@{.}l|r@{.}l|r@{.}l|r@{.}l|}
\hline
\multicolumn{2}{|c|}{ }
&\multicolumn{2}{c|}{a.o.D.}&\multicolumn{2}{c|}{Camb.}
&\multicolumn{2}{c|}{Durh.}&\multicolumn{2}{c|}{Lucl.}\\
\hline
& $a_1$          & 1&45743  &   1&19004  & 1&26382   & 1&58878 \\    
\cline{2-10}
& $a_2\cdot10^3$ & -6&7041  &   -2&85481  & -4&1925  & -8&40213 \\
\cline{2-10}
& $a_3\cdot10^3$ & -4&27398 &   -1&15495  & -1&83942 & -4&62464 \\
\cline{2-10}
\haho{{\sc Ariadne}}
& $a_4\cdot10^5$ & 6&25913 &   1&78559 & 2&94159 & 6&54661 \\
\cline{2-10}
& $a_5\cdot10^5$ & 2&48986 &   1&11675  & 1&64381 & 3&00994 \\
\cline{2-10}
& $a_6\cdot10^7$ & -2&31969&   -0&728499 & -1&15423& -2&32587\\
\hline
& $a_1$ & 0&925643     & 1&34605       & 1&47339      &  1&33754 \\    
\cline{2-10}
& $a_2\cdot10^3$ & 0&604669  & -4&93063   & -6&55149  &  -4&90623 \\
\cline{2-10}
& $a_3\cdot10^3$ & 1&5079    & -2&33696   & -3&10136  &  -2&0464  \\
\cline{2-10}
\haho{{\sc Jetset}}
& $a_4\cdot10^5$ & -1&75222 & 3&32752   & 4&17083  &  2&84691 \\
\cline{2-10}
& $a_5\cdot10^5$ & -0&019105  & 1&80813   & 2&29184  &  1&81215 \\
\cline{2-10}
& $a_6\cdot10^7$ & 0&449687  & -1&23386  & -1&42457 &  -1&01679\\
\hline
& $a_1$ & 0&910054     & 1&22464      & 1&41922      &  0&820707 \\      
\cline{2-10}
& $a_2\cdot10^3$ & 0&994144  & -3&11472  & -6&18148  &  1&78567    \\
\cline{2-10}
& $a_3\cdot10^3$ & 2&14597   & -0&3132   & -2&81892  &  2&43204    \\
\cline{2-10}
\haho{{\sc Herwig}}
& $a_4\cdot10^5$ & -2&94057 & 0&232189  & 4&03189  &  -3&00669 \\
\cline{2-10}
& $a_5\cdot10^5$ & -0&208228 & 1&12039  & 2&28406  &  -0&351293  \\
\cline{2-10}
& $a_6\cdot10^7$ & 0&947642  & -0&0622476 & -1&46606 &  0&888795   \\
\hline
\end{tabular}
\end{center}
\caption
{\label{t_hadcor_all} 
The parameters for the hadronisation correction factors for general
topologies for the angular ordered Durham (a.o.D.), Cambridge (Camb.), Durham
(Durh.) and Luclus (Lucl.) algorithm} 
\end{table}
This effect is corrected for by a topology dependent correction factor which
is applied to the prediction. In \fref{f_hadcor_sym} the correction factors
obtained for symmetric event topologies are shown. The correction factors are
calculated by dividing the mean hadron
multiplicity obtained for a certain $\theta_1$
bin (with $\theta_1$ measured at the hadronic level of the simulated event) by
the mean hadron
multiplicity obtained with $\theta_1$ measured at the partonic level of
the event. The correction factors shown are taken from the {\sc Ariadne}
 Monte Carlo
simulation for each of the four cluster algorithms used. The correction factor
is around unity for large opening angles and increases for small
$\theta_1$. This follows the expectation that jets which are close
together are pulled even closer together than jets with a larger angle between
them. In order to get a smooth correction the function
\be
\label{e_hadcor_sym}
c=\frac{1}{1-a_2e^{a_1\theta_1}}
\ee
is fitted to the correction factors. The correction is described very well by
the fitted function,  the corrections obtained for the four studied cluster
algorithms are shown in
%which is indicated as a solid line in
\fref{f_hadcor_sym}. The fitted parameters obtained using the Monte Carlo
generators {\sc Ariadne}, {\sc Herwig} and {\sc Jetset}
 are given in \tref{t_hadcor_sym}
for the four cluster algorithms.
The same procedure is applied also to events with general topologies. In
\fref{f_hadcor_all} the correction factors obtained with {\sc Ariadne}
for the four used cluster
algorithms are shown for several values of $\theta_3$ as a function of
$\theta_1$. In order to describe these corrections smoothly the two-dimensional
function 
\be
\label{e_hadcor_all}
c=(a_1+a_2\theta_3+a_5\theta_3^2)+(a_3+a_4\theta_3+a_6\theta_3^2)
\cdot\theta_1
\ee
is fitted to the correction factors. The description of the correction by the
fit is reasonable, and the parameter values obtained are listed in
\tref{t_hadcor_all}. In general, the correction becomes smaller with
increasing opening angle $\theta_1$, 
as expected for this hadronisation effect.
\section{The fit of $C_A/C_F$ \label{s_fit}}
\begin{figure}[p]
\begin{center}
\begin{minipage}[h]{14cm}
\mbox{\epsfig{file=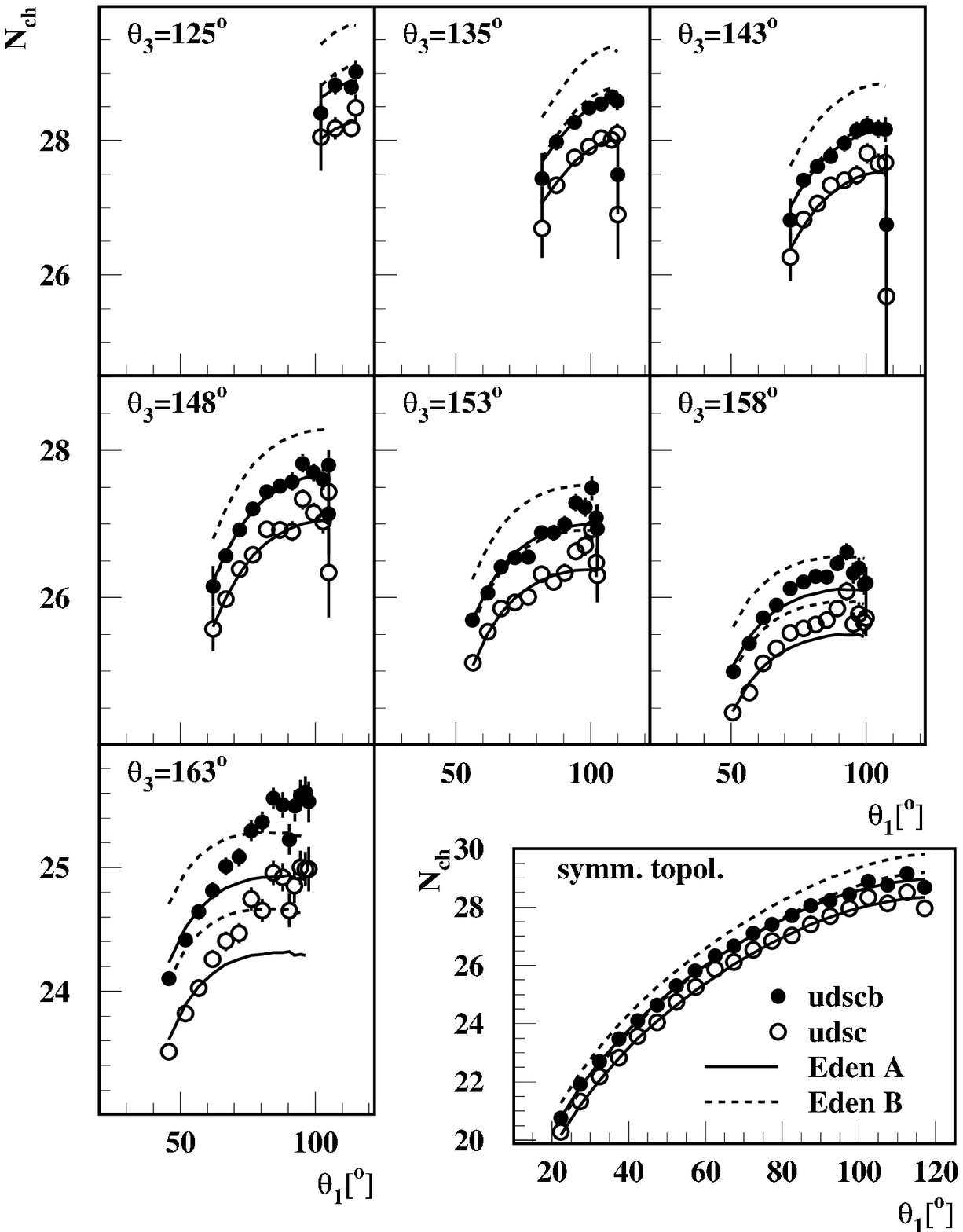,width=14cm}}
\end{minipage} 
\end{center}
\caption
{\label{f_pred} The mean multiplicity of $udscb$- and $udsc$-events with
  symmetric and general topologies in comparison with the predictions
 \eref{eden_3jet_a} and \eref{eden_3jet_b} 
 prepared as described in \sref{s_predprep}. The error bars indicate
  statistical errors only.  } 
\end{figure}
\begin{figure}[p]
\begin{center}
\begin{minipage}[h]{14cm}
\mbox{\epsfig{file=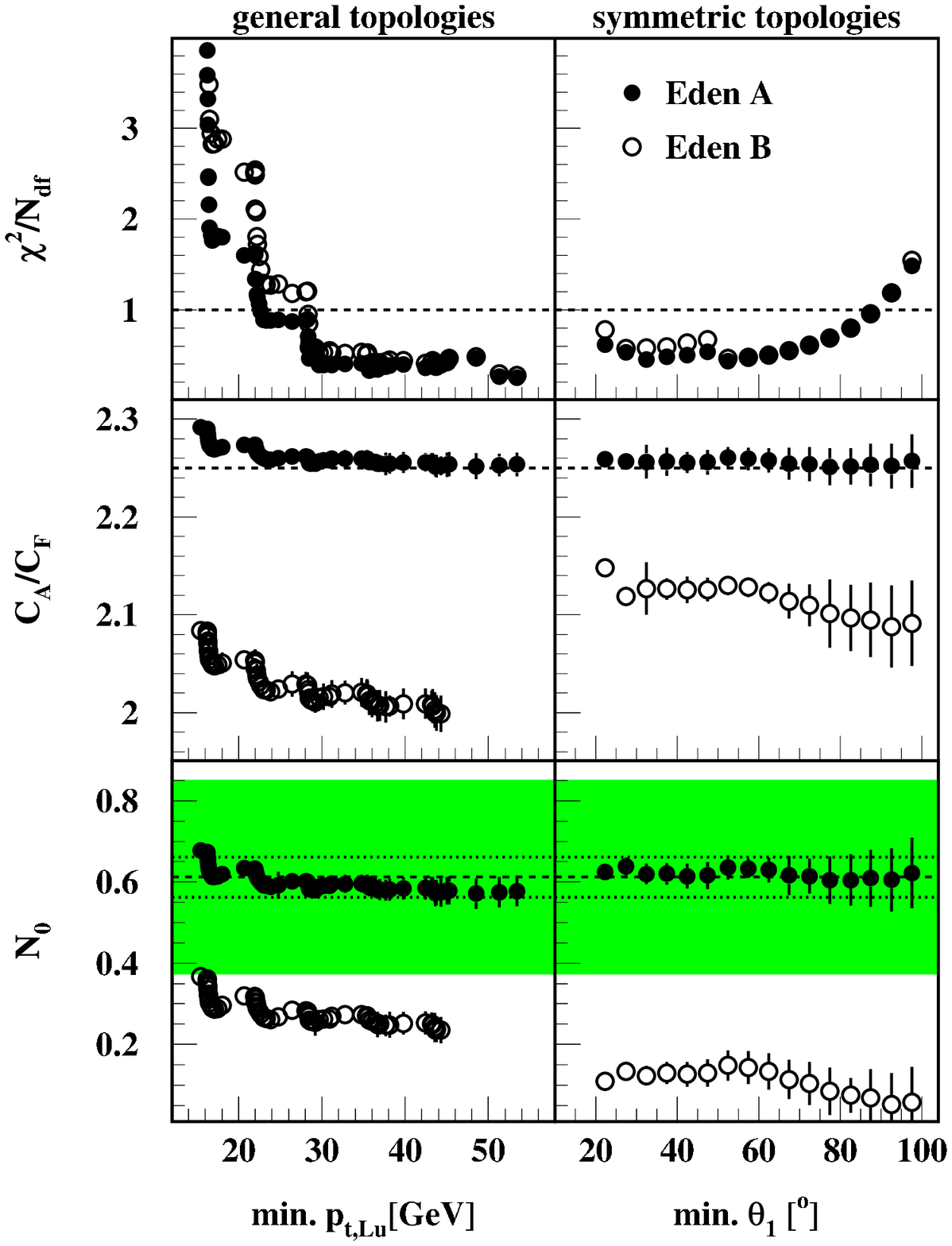,width=14cm}}
\end{minipage} 
\end{center}
\caption
{\label{f_stability} The fit parameters 
$\chi^2/N_{df}$, $C_A/C_F$ and $N_0$ 
as function of the fit range. The shaded area indicates the expected value
of $N_0$ with statistical and systematical uncertainties, the dotted lines
indicate statistical errors only. The dashed lines indicate 1.0 in the top,
$C_A/C_F=2.25$ in the middle and the $N_0$ value from \eref{e_n0} in the bottom
plots.  }
\end{figure}
The closed form of $N_{q\bar qg}$ as a function of the event topology obtained
from \eref{eden_3jet_a} (Eden A) and \eref{eden_3jet_b} (Eden B)
with the procedure described 
in \sref{s_predprep} can now be compared with the measured three-jet event
multiplicities. In \fref{f_pred} the measured multiplicities of $udsc$- (open
markers) and
$udscb$-events (solid markers) are shown for symmetric and general topologies.
The two solid lines indicate the respective predictions of \eref{eden_3jet_a}
where for $udscb$-events the constant $N_0$ of \eref{e_n0} has been added to
the prediction. The dashed lines represent the respective predictions of
\eref{eden_3jet_b}. It can be observed that prediction \eref{eden_3jet_a}
describes the multiplicity of symmetric events in an excellent way, while
\eref{eden_3jet_b} overestimates the multiplicity by $\sim 0.6$, with the
result that the prediction for $udsc$-events almost coincides with the
multiplicities measured in $udscb$-events. Moreover, the slope of
\eref{eden_3jet_b} seems to be larger than the slope of the measured
multiplicities. 
For general topologies an overall
good description of the event multiplicities by \eref{eden_3jet_a} can be
seen. Only for large $\theta_3$ and then especially for large $\theta_1$,
i.e.~in topologies where jet 3 is not strongly pronounced, larger deviations
occur. Again, \eref{eden_3jet_b} overestimates the multiplicities
significantly and shows a larger slope with respect to $\theta_1$.
\par
In order to determine the colour factor ratio $C_A/C_F$ from the event
multiplicities, the predictions \eref{eden_3jet_a} and \eref{eden_3jet_b} are
fitted to the data. To avoid systematic uncertainties entering through the
use of a $b$-tagging procedure, $udscb$-events are used to obtain the central
result,
while the results obtained for $udsc$-events are considered when
estimating systematic uncertainties.
%However, $udsc$-events are fitted for cross-check reasons.
$C_A/C_F$ enters the prediction only via the {\em
  derivative} of the gluon multiplicity in \eref{eden_gluon}, so this parameter
is expected to be sensitive to the change of multiplicity rather than to its
absolute value. Therefore, the additional constant $N_0$ is allowed to
vary freely in
the fit (also when fitting $udsc$-events where $N_0$ is expected to be zero)
in order to avoid an influence of the absolute value on the fit result
for $C_A/C_F$. 
\par
The fit range used is determined by the value of $\chi^2/N_{df}$ obtained with
the fit. While for symmetric event topologies it is straightforward to
determine the data points used in the fit by demanding a minimum opening angle
$\theta_1$, the situation is not as clear for the general event topologies,
where two variables are needed to describe a topology. A reasonable order
criterium is a `three-jet likeliness' of a topology, i.e.~the emphasis which
is put on the third jet. Therefore the $p_t$ of the third jet is a reasonable
ordering variable. It turned out that $p_{t,\mathrm{Lu}}$ is the best suited
of the used $p_t$-like variables, because 
the fits converge fastest, if the fit range
is determined by $p_{t,\mathrm{Lu}}$. 
The fit range is then defined by demanding
a minimum   $p_{t,\mathrm{Lu}}$-value for the event topologies used.
\par
In the upper plots of \fref{f_stability} the values obtained for
$\chi^2/N_{df}$ when fitting $udscb$-events with the predictions of
\eref{eden_3jet_a} and \eref{eden_3jet_b} are shown for both
general and symmetric
topologies. The values of  $\chi^2/N_{df}$ obtained for symmetric events are
always smaller than 1.0, regardless of the chosen minimal $\theta_1$-value
except for extremely small fit-ranges. The curves show a shallow minimum around
$\theta_1^{\mathrm{min}}\sim30^\circ$ which is chosen to determine the
fit-range. For general topologies, and a small
minimal $p_{t,\mathrm{Lu}}$, the $\chi^2/N_{df}$ values are large but decrease
fast with a more restricted fit range. The
$\chi^2/N_{df}$ curve shows a step-like structure, where every step
corresponds to a completely excluded bin in $\theta_3$. From a minimum 
$p_{t,\mathrm{Lu}}$ value of $\sim 22$ GeV on a plateau in the vicinity of 
$\chi^2/N_{df}=1$ can be observed and
a minimum $p_{t,\mathrm{Lu}}$ of 25 GeV is chosen.
%Taking also the convergence of the fitting
%procedure into account, the fit range for general topologies
%is determined by the choice of
%a minimal $p_{t,\mathrm{Lu}}$ of 25 GeV.
\par
\begin{table}[tb]
\begin{center}
\begin{tabular}{|c|c|r@{.}l@{$\pm$}r@{.}l|r@{.}l@{$\pm$}r@{.}l|
r@{.}l@{$\pm$}r@{.}l|r@{.}l@{$\pm$}r@{.}l|}
\hline
\multicolumn{2}{|c|}{ }& \multicolumn{8}{c|}{Eden A} & 
\multicolumn{8}{c|}{Eden B} \\
\multicolumn{2}{|c|}{ }
& \multicolumn{4}{c|}{$C_A/C_F$} &\multicolumn{4}{c|}{$N_0$}
& \multicolumn{4}{c|}{$C_A/C_F$} &\multicolumn{4}{c|}{$N_0$}\\
\hline
&symm. & 2&257 & 0&019 & 0&62 & 0&03 &  2&127 & 0&034 &  0&12 & 0&03  \\
\cline{2-18}
\haho{$udscb$}
&gen. & 2&261 & 0&014 & 0&60 & 0&03 &  2&028 & 0&050 &  0&28 & 0&03\\
\hline
& symm & 2&237 & 0&030 & 0&09 & 0&05 & 2&101  & 0&096 & -0&42 & 0&13\\
\cline{2-18}
\haho{$udsc$}
& gen. & 2&272 & 0&015 & 0&01 & 0&04 & 2&039  & 0&056 & -0&30 & 0&08\\
\hline
\end{tabular}
\end{center}
\caption{\label{t_cafitres} The results of the parameter fit for 
  $C_A/C_F$ and $N_0$}
\end{table}
Fitting the data within these ranges results in the parameter
values given in \tref{t_cafitres}. As can be seen, the prediction of
\eref{eden_3jet_a} results in compatible
values for $C_A/C_F$ when fitting $udsc$- and $udscb$-events of symmetric or
general topologies. Both values for $N_0$ obtained by fitting $udscb$-events
agree with the expectation of $N_0=0.61$ as well as both $N_0$ values for
$udsc$-events agree with the expectation of $N_0=0$.
%On the other hand, the results obtained for $C_A/C_F$ and $N_0$ when fitting 
%\eref{eden_3jet_b} to the data are not compatible
%when fitting
%symmetric and general topologies. 
 The results obtained for $C_A/C_F$ and $N_0$ with
\eref{eden_3jet_b} in
symmetric topologies are not in good agreement with the values obtained in 
 general topologies. 
Also the values for $N_0$ are not
in agreement with the respective expectations. In the lower plots of
\fref{f_stability} the fit results for $C_A/C_F$ and $N_0$ are shown for
$udscb$-events as a function of the fit range. 
The parameters are correlated, but as this plot is only to
show the
influence of the fit range, when plotting the dependence of one
parameter the other parameter is fixed to the respective value of
\tref{t_cafitres}. While \eref{eden_3jet_a} leads to results which are
independent of the fit range, the results of \eref{eden_3jet_b} exhibit a more
pronounced dependence on the fit range. Again, the discrepancy of the results
obtained with \eref{eden_3jet_b} for symmetric and general topologies can be
observed, as well as the disagreement of the obtained values for $N_0$ with the
expectation indicated by the shaded area while
fixing $N_0$ to the expected values leads to an
unacceptably high $\chi^2/N_{df}$.
\par
Since the prediction \eref{eden_3jet_b}, 
which depicts a rather extreme scenario of the phase-space assignment to quark
and gluon jets, 
fails to give consistent results for symmetric and
general topologies and because \eref{eden_3jet_b} overestimates the event
multiplicity in general which results in unphysically low values of $N_0$,
this formulation of the prediction is discarded in favour of
\eref{eden_3jet_a} for the central result.
%To avoid the systematic uncertainties of the $b$-tagging procedure, the results
%obtained with the undiscriminated $udscb$-events are considered as the central
%values, while the results obtained for $udsc$-events are considered when
%estimating of systematic uncertainties. 
The central results of the fits are therefore
\be
\frac{C_A}{C_F}=2.257\pm0.019_{\mathrm{stat.}}
\ee
for symmetric and
\be
\frac{C_A}{C_F}=2.261\pm0.014_{\mathrm{stat.}}
\ee
for general event topologies.
\par
Several sources of systematic uncertainties have been considered. The
systematic error on $C_A/C_F$ is estimated by fitting prediction
\eref{eden_3jet_a} with the variations to the analysis listed below. As the
parameters $C_A/C_F$ and $N_0$ are correlated and this correlation should not
be reflected by the estimated systematic uncertainties, $N_0$ is fixed at
0.61. The values obtained for $C_A/C_F$ with the variations of the analysis
are then compared to the result obtained by the standard analysis but with
$N_0$ fixed at the above value.
\par
\begin{table}[tb]
\begin{minipage}{7cm}
\begin{center}
\begin{tabular}{|c|c|c|}
\hline
variable & tight & loose \\
    \hline
    \hline
$p$ & $\ge 0.5$ GeV & $\ge 0.3$ GeV \\
    \hline
$\vartheta_{\mathrm{polar}}$ & $30^\circ-150^\circ$ & $20^\circ-160^\circ$ \\
    \hline
$\epsilon_{xy}$ & $\le 2 ~\mathrm{ cm}$ & $\le 8 ~\mathrm{ cm}$ \\
      \hline
$\epsilon_z$ &  $\le 8 ~\mathrm{ cm}$  &  $\le 12 ~\mathrm{ cm}$  \\
      \hline
$\Delta p/p$ & $\le 80\%$ & $\le 120\%$ \\
    \hline
\end{tabular}
%\end{center}
\caption{\label{t_tcutsvar} Variations of the track cuts given in \tref{t_tcuts} for
  systematic studies}
%\end{table}
\end{center}
\end{minipage}
\hfill
\begin{minipage}{7cm}
%\begin{table}[tb]
\begin{center}
\begin{tabular}{|c|c|c|}
\hline
variable & tight & loose \\
\hline
\hline
 \multicolumn{3}{|c|}{general events} \\ \hline
$E_{\mathrm{ charged}}^{\mathrm{ hemisph.}}$ & $ \ge 0.05\cdot \sqrt{s}$
& $ \ge 0.02\cdot \sqrt{s}$  \\
      \hline
$E_{\mathrm{ charged}}^{\mathrm{ total}}$ & $ \ge 0.18\cdot \sqrt{s}$  
& $ \ge 0.06\cdot \sqrt{s}$  \\
      \hline
$N_{\mathrm{ charged}}$ & $\ge 6$ & $\ge 4$ \\
      \hline
$\vartheta_{\mathrm{ sphericity}}$ &  $40^\circ-140^\circ$  
& $30^\circ-150^\circ$  \\
      \hline
$p_{\mathrm{max}}$ &  $20$ GeV  &  $70$ GeV  \\
      \hline\hline
 \multicolumn{3}{|c|}{three-jet events} \\ \hline
$\sum_{i=1}^{3}\theta_i$ & $ > 357.5^\circ$ & $ > 350^\circ$ \\
      \hline
$E_{\mathrm{ visible}}/\mathrm{ jet}$ & $ \ge 8$ GeV   & $ \ge 3$ GeV  \\
      \hline
$\vartheta_{\mathrm{ jet}}$
            &  $40^\circ-140^\circ$   &  $30^\circ-150^\circ$      \\
      \hline
\end{tabular}
\caption{\label{t_evtcutvar} Variation of the event cuts given in \tref{t_evtcut}
 for systematic studies}
\end{center}
\end{minipage}
\end{table}
\begin{table}[tb]
\begin{center}
\begin{tabular}{|l|l||r@{.}l|r@{.}l|r@{.}l||r@{.}l|r@{.}l|r@{.}l|}
\hline
\multicolumn{2}{|c||}{}&\multicolumn{6}{c||}{symm. topol.}&
\multicolumn{6}{c|}{general topol.}\\
\hline
& track cuts & 1&2\% &\multicolumn{2}{c|}{ }&\multicolumn{2}{c||}{ }
& 0&4\% &\multicolumn{2}{c|}{ }&\multicolumn{2}{c|}{ }\\
experi-& event cuts&0&8\%&\multicolumn{2}{c|}{ }&\multicolumn{2}{c||}{ }
&0&6\%&\multicolumn{2}{c|}{ }&\multicolumn{2}{c|}{ }\\
\cline{2-4}\cline{9-10}
mental& hadr.-corr. & 1&2\% & \multicolumn{2}{c|}{ } &\multicolumn{2}{c||}{ }
& 0&4\% &\multicolumn{2}{c|}{ }&\multicolumn{2}{c|}{ }\\
& acc.-corr. & 0&4\%  & \multicolumn{2}{c|}{ }&\multicolumn{2}{c||}{ }
& 0&2\%  & \multicolumn{2}{c|}{ }&\multicolumn{2}{c|}{ }\\
& $udsc$- / $udscb$-sample & 1&0\% &2&5\%& \multicolumn{2}{c||}{ }
& 0&8\% &1&6\%& \multicolumn{2}{c|}{ }  \\
& fit range & 0&1\% &\multicolumn{2}{c|}{ }&\multicolumn{2}{c||}{ }
& 0&0\% &\multicolumn{2}{c|}{ }&\multicolumn{2}{c|}{ }\\
& normalisation of $N_{q\bar q}$&0&5\% &\multicolumn{2}{|c|}{ }&4&0\%
&0&5\% &\multicolumn{2}{c|}{ }&3&3\%\\
& $e^+e^-$ fitted data &0&9\% &\multicolumn{2}{c|}{ }&\multicolumn{2}{c||}{ }
&0&7\% &\multicolumn{2}{c|}{ }&\multicolumn{2}{c|}{ }\\
& variation of $\delta_{b-udsc}$&0&8\% &\multicolumn{2}{c|}{
}&\multicolumn{2}{c||}{ }
&0&7\% &\multicolumn{2}{c|}{ }&\multicolumn{2}{c|}{ }\\
\cline{1-6}\cline{9-12}
& variation of $\Lambda$ &1&0\%&\multicolumn{2}{c|}{
}&\multicolumn{2}{c||}{ }
&1&0\%&\multicolumn{2}{c|}{
}&\multicolumn{2}{c|}{ }\\
theo-  & variation of $c_r$ &2&0\%& 3&1\% & \multicolumn{2}{c||}{ }
&2&1\%& 2&9\% & \multicolumn{2}{c|}{ }\\
retical & variation of $L_0$ &0&0\%&\multicolumn{2}{c|}{
}&\multicolumn{2}{c||}{ }
&0&0\%&\multicolumn{2}{c|}{
}&\multicolumn{2}{c|}{ }\\
%\cline{1-6}\cline{9-12}
%\multicolumn{2}{|l||}{clustering algorithm} 
%&\multicolumn{2}{|c}{ }&2&2\%&\multicolumn{2}{c||}{ }&\multicolumn{2}{c}{ }
%&1&8\%&\multicolumn{2}{c|}{ }\\
&clustering algorithm & 2&2\% & \multicolumn{2}{c|}{ }& \multicolumn{2}{c||}{ }& 
1&8\% & \multicolumn{2}{c|}{ }&\multicolumn{2}{c|}{ }\\
\hline
\end{tabular}
\end{center}
\caption{\label{t_syserr} The relative
systematic uncertainties of the measurement of
  $C_A/C_F$}  
\end{table}

The studied experimental sources of uncertainty are:
\begin{itemize}
%\item Variations of the track cuts listed in \tref{t_tcuts} 
%  by $\pm 50\%$. However,
%  the
%  minimal track length is not lowered in order to maintain a contribution of
%  the TPC detector to the track reconstruction;
\item  Variations of the track cuts listed in \tref{t_tcuts} 
      within the ranges given in \tref{t_tcutsvar}.  
  The
  minimal track length is not lowered in order to maintain a contribution of
  the TPC detector to the track reconstruction;

\item Variations of the cuts on the event- and jet-structure listed in
  \tref{t_evtcut} within the ranges given in \tref{t_evtcutvar} ;
\item The hadronisation corrections are calculated from {\sc Jetset} instead of
  {\sc Ariadne} simulation. Alternatively $30\%$ of the hadronisation correction are
  regarded as this correction's uncertainty;
\item Alternatively to the arithmetic mean of the multiplicity distributions,
 the mean multiplicities are determined by parameter fits of negative
 binomials to the distributions. As a further alternative, the matrix
 correction is replaced by a multiplicative correction to the mean
 multiplicities;
\item The $udsc$ sample is used to estimate the uncertainty of $C_A/C_F$ due
  to the $b$-quark
  multiplicity;
\item The lower limit of the
  fit range is varied between $\theta_1>25^\circ$ and $\theta_1>35^\circ$ 
   for symmetric
  events and between $p_{t,\mathrm{Lu}}>22.5$ GeV 
and $p_{t,\mathrm{Lu}}>28$ GeV for general topologies;
\item The difference between the parametrisation of $N_{q\bar q}$ at
  $\sqrt{s}=m_Z$ and the event multiplicity measured in this analysis is 0.101
  units, i.e.~0.5\%. In the worst case this reflects an uncertainty on the
  normalisation of $N_{q\bar q}$ which would directly be reflected in
  $C_A/C_F$. Therefore this deviation is considered conservatively
  as the relative uncertainty on $C_A/C_F$;
\item The data set used to fit the parametrisation of $N_{q\bar q}$ is varied,
   by excluding the results of the
    {\sc Topaz},
       {\sc Jade} and {\sc Mark}-I collaborations or the results of the
  {\sc Amy} collaboration respectively;
\item The correction due to the additional multiplicity in $b$-events is
  varied within the errors of  $\delta_{b-udsc}$ when fitting the
  parametrisation of $N_{q\bar q}$.
\end{itemize}
Additionally, the uncertainty due to the choice of the clustering algorithm
has been estimated by comparing the result obtained with the angular ordered
Durham algorithm with the results obtained using the Durham, Cambridge and
Luclus algorithms. Moreover the following uncertainties of the
theoretical prediction have been studied:
\begin{itemize}
\item The scale variable $\Lambda$ has been varied between 200 MeV
  and 300 MeV;
\item The constant $c_r$ has been varied within the given theoretical
  uncertainty of $\sim10\%$;
\item The measurement of $N_{gg}$ by {\sc Cleo}
 which has been used to determine 
  $L_0$ has been varied within the given errors. As
  $L_0$ is a constant of integration it is supposed to affect mainly
  $N_0$ and not $C_A/C_F$. 
  Therefore $N_0$ is varied freely when estimating the effect of 
  variations of $L_0$. Indeed no dependence of $C_A/C_F$ on the variation of
  $L_0$ is observed.
\end{itemize}
The relative uncertainties caused by the different sources are given in
\tref{t_syserr}.  Alternatively to estimating the systematic uncertainty
  inherent to the prediction \eref{eden_3jet_a} by varying the parameters as
  discussed above, it is suggested in \cite{edenfehler} to study the
  difference between the prediction \eref{eden_gluon} based upon the
  colour-dipole model with predictions based upon parton-shower evolution,
  as e. g. \eref{e_r1cap}. Since this measurement of $C_A/C_F$ is
  sensitive rather to the energy dependence of the multiplicities than to the
  absolute value of multiplicity, the uncertainty of \eref{eden_gluon} is
  relevant here. Both predictions, \eref{e_r1cap} and \eref{eden_gluon} are
  shown in \fref{f_r1}. The difference between the two predictions, evaluated
  at the centre of the data around $\sim 40$GeV is $\sim3.7\%$. This
  corresponds to an one-sigma error-estimate of 2.1\% which is completely
  compatible with the theoretical error given in \tref{t_syserr}.
The result obtained with symmetric topologies is therefore
\be
\frac{C_A}{C_F}=2.257\pm0.019_{\mathrm{stat.}}
\pm0.056_{\mathrm{exp.}}
\pm0.070_{\mathrm{theo.}}
%\pm0.050_{\mathrm{clus.}}
%\frac{C_A}{C_F}=2.257\pm0.019_{\mathrm{stat.}}
%\pm0.056_{\mathrm{exp.}}
%\pm0.050_{\mathrm{theo.}}
%\pm0.050_{\mathrm{clus.}}
\ee
and 
\be
\frac{C_A}{C_F}=2.261\pm0.014_{\mathrm{stat.}}
\pm0.036_{\mathrm{exp.}}
\pm0.066_{\mathrm{theo.}}
%\pm0.041_{\mathrm{clus.}}
%\frac{C_A}{C_F}=2.261\pm0.014_{\mathrm{stat.}}
%\pm0.036_{\mathrm{exp.}}
%\pm0.052_{\mathrm{theo.}}
%\pm0.041_{\mathrm{clus.}}
\ee
with general topologies. Both results are strongly correlated due to a 
large fraction of common events, so an average cannot be made
 here. Instead, the
 more precise result obtained with general topologies is taken as the
central result. This result is the most precise measurement of the colour
factor ratio so far with an overall relative uncertainty of 3.4\%.
\par
\begin{figure}[tbh]
\begin{center}
\begin{minipage}[h]{10cm}
\mbox{\epsfig{file=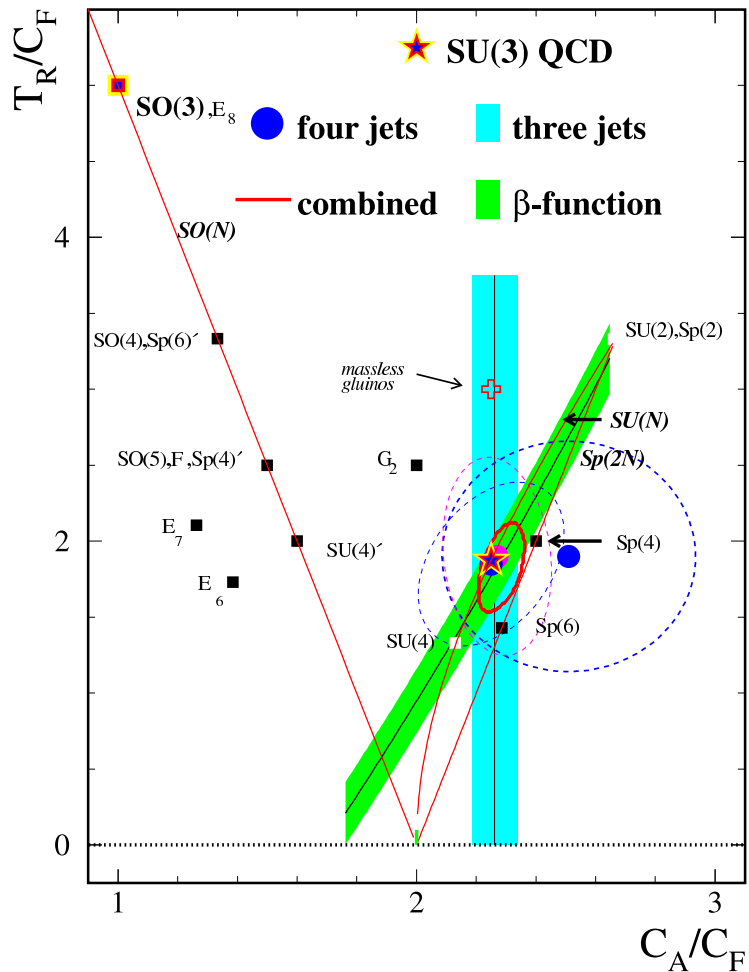,width=10cm}}
\end{minipage} 
\end{center}
\caption
{\label{f_wim} 
%Measurements of the QCD colour factor ratios in comparison with
%the values expected for several groups. 
The plot shows the Casimir eigenvalues expected for the
special orthogonal (SO(N), straight line on the left) and the special unitary
(SU(N), curved line on the right) 
groups as well as for the symplectic groups 
(SP(2N), straight line on the right).
%SO(N) groups are to be found on the straight line left,
%of
%$C_A/C_F=2$, 
%Sp(2N) groups on the straight line right of $C_A/C_F=2$. SU(N) groups are on
%the curve right of %$C_A/C_F=2$ with 
% SU(3)  marked by the large star at $(2.25,2)$. 
$\mathrm{E}_6$, $\mathrm{E}_7$,
 $\mathrm{E}_8$, $\mathrm{G}_2$ and F are the five exceptional Lie groups.
% whose
% algebras are not member of one of the four infinite sets of Lie
% algebras. 
%Massless gluinos increase the number of fermions a gluon can split
% into and therefore increase $T_R$ without affecting $C_A/C_F$. 
The shaded bands and the dashed ellipses indicate the result of this 
and other \cite{betafunction,fourjets}
 experimental analyses, which are combined in the
 solid ellipse.
%diagonal
%band is the result of a measurement of the QCD $\beta$-function
% \cite{betafunction}, the vertical band indicates the result of this analysis.
% The dashed ellipses represent measurements of the
%inter-jet angles in four-jet events \cite{fourjets}. The combination of these measurements
%is indicated by the solid ellipse.
% is in excellent agreement with the QCD
%expectation of SU(3).
} 
\end{figure}
In \fref{f_wim} the colour factor ratios $C_A/C_F$ and $T_R/C_F$ 
are mapped for several symmetry groups where $T_R=N_FT_F$ with $N_F$ as the
number of active quark flavours and $T_F=1/2$ as the normalisation of the
SU(3) representation.  
In this plot the present
analysis result is indicated by the shaded vertical band. The shaded diagonal
band is the result of a measurement of the QCD $\beta$-function correlating
$T_R/C_F$ with $C_A/C_F$
\cite{betafunction}. The dashed ellipses represent measurements of the
four-jet cross section as function of the 
inter-jet angles \cite{fourjets}. 
 The measurements are combined by adding their $\chi^2$-functions. The
  solid contour represents the $\Delta\chi^2=2.4$ limit, corresponding to a
  confidence level of 68\%.
%The combination 
%of these measurements
%indicated by the solid ellipse 
 It is in excellent agreement with the QCD
expectation of SU(3).
%, while nearly all other indicated symmetry groups are
%ruled out thus confirming the group structure of QCD.
%
%
%\clearpage
\section{The extraction of $N_{gg}$ \label{s_gluon}}
\begin{table}[tb]
\begin{center}
\begin{tabular}{|c|c@{$\pm$}c@{$\pm$}c||c|c@{$\pm$}c@{$\pm$}c|}
 \hline
 $\kale$ [GeV] & $N_{gg}$ & (stat.) & (sys.)&
 $\kale$ [GeV] & $N_{gg}$ & (stat.) & (sys.)\\
 \hline
 13.36 &11.21 &  0.14&  0.14 &
 35.26 &19.96 &  0.14&  0.11 \\
 15.57 &12.41 &  0.12&  0.08 &
 38.01 &20.73 &  0.16&  0.13 \\
 17.82 &13.43 &  0.11&  0.09 &
 40.69 &21.56 &  0.16&  0.15 \\
 20.13 &14.32 &  0.11&  0.08 &
 43.37 &22.09 &  0.17&  0.18 \\
 22.50 &15.56 &  0.11&  0.08 &
 45.87 &22.72 &  0.18&  0.18 \\
 24.92 &16.53 &  0.11&  0.13 &
 48.19 &23.81 &  0.19&  0.21 \\
 27.45 &17.57 &  0.12&  0.16 &
 50.19 &23.71 &  0.19&  0.11 \\
 30.01 &18.32 &  0.13&  0.10 &
 51.69 &24.60 &  0.20&  0.15 \\
 32.61 &19.27 &  0.14&  0.15 &
 52.50 &23.75 &  0.22&  0.22 \\
 \hline
 \end{tabular}\end{center}
\caption{\label{t_ngg_sym}
$N_{gg}$ extracted from three-jet events with symmetric topologies}
\end{table}
\begin{table}[tb]
\begin{center}
\begin{tabular}{|c|c@{$\pm$}c@{$\pm$}c||c|c@{$\pm$}c@{$\pm$}c|}
 \hline
 $\kale$ [GeV] & $N_{gg}$ & (stat.) & (sys.)&
 $\kale$ [GeV] & $N_{gg}$ & (stat.) & (sys.)\\
 \hline
 24.29 &16.28 &  0.09&  0.17 &
 36.14 &20.78 &  0.12&  0.17 \\
 26.44 &16.96 &  0.09&  0.15 &
 36.38 &20.56 &  0.12&  0.17 \\
 27.25 &17.23 &  0.28&  0.14 &
 36.50 &20.39 &  0.14&  0.15 \\
 27.79 &17.64 &  0.10&  0.09 &
 37.13 &20.49 &  0.11&  0.15 \\
 28.78 &17.89 &  0.10&  0.12 &
 37.73 &20.17 &  0.38&  0.34 \\
 28.99 &18.64 &  0.33&  0.20 &
 38.75 &20.88 &  0.11&  0.15 \\
 29.44 &17.89 &  0.10&  0.14 &
 39.67 &18.90 &  1.19&  1.12 \\
 29.60 &18.08 &  0.10&  0.10 &
 40.01 &21.33 &  0.12&  0.14 \\
 29.93 &18.55 &  0.10&  0.12 &
 40.52 &21.42 &  0.12&  0.12 \\
 30.27 &18.55 &  0.11&  0.14 &
 40.88 &21.78 &  0.13&  0.13 \\
 30.45 &18.79 &  0.11&  0.20 &
 41.33 &21.84 &  0.18&  0.13 \\
 30.50 &19.00 &  0.17&  0.20 &
 41.35 &21.96 &  0.13&  0.17 \\
 30.59 &19.39 &  0.11&  0.11 &
 41.63 &21.88 &  0.13&  0.17 \\
 30.66 &19.27 &  0.13&  0.28 &
 43.86 &22.25 &  0.09&  0.14 \\
 30.66 &19.80 &  0.16&  0.25 &
 44.60 &20.75 &  0.59&  0.51 \\
 31.69 &18.82 &  0.10&  0.15 &
 45.88 &22.83 &  0.09&  0.15 \\
 32.33 &18.68 &  0.32&  0.44 &
 46.71 &23.09 &  0.13&  0.17 \\
 33.25 &19.43 &  0.10&  0.12 &
 46.97 &23.02 &  0.10&  0.16 \\
 34.42 &19.93 &  0.11&  0.16 &
 47.55 &23.28 &  0.10&  0.13 \\
 34.51 &19.33 &  0.55&  0.51 &
 48.00 &22.82 &  0.46&  0.35 \\
 34.86 &19.96 &  0.11&  0.10 &
 50.14 &23.85 &  0.15&  0.17 \\
 35.27 &20.12 &  0.11&  0.13 &
 51.52 &24.37 &  0.18&  0.16 \\
 35.78 &20.27 &  0.12&  0.13 &
 51.72 &23.93 &  0.12&  0.16 \\
 36.07 &20.76 &  0.20&  0.26 &
 &\multicolumn{3}{c|}{}\\
 \hline
 \end{tabular}\end{center}
\caption{\label{t_ngg_gen}
$N_{gg}$ extracted from three-jet events with general topologies}
\end{table}
Instead of fitting $C_A/C_F$ with the measured three-jet event multiplicities,
the predictions of \eref{eden_3jet_a} and \eref{eden_3jet_b} can be used to
extract the multiplicities of unrestricted two-gluon colour-singlet systems
from the three-jet event multiplicity. As the direct
experimental access to this
quantity is severely limited, this proves to be an interesting option. The
predictions \eref{eden_3jet_a} and \eref{eden_3jet_b} are solved for the 
gluonic contribution yielding
\begin{subequations}
\begin{align}
\tag{29A}
\label{e_ngga}
N_{gg}(\kappa_{\mathrm Le})&=2\cdot
\Bigl(N_{q\bar q g}(\vartheta_2,\vartheta_3)
-N_{q\bar q}(L_{q\bar q},\kappa_{\mathrm Lu})\Bigr)\\
\tag{29B}
\label{e_nggb}
N_{gg}(\kappa_{\mathrm Lu})&=2\cdot
\Bigl(N_{q\bar q g}(\vartheta_2,\vartheta_3)
-N_{q\bar q}(L,\kappa_{\mathrm Lu})\Bigr)\quad .
\end{align}
\end{subequations}
\setcounter{equation}{29}
As discussed already in the previous section, \eref{eden_3jet_b} fails to
describe the data so only \eref{e_ngga}, which is based on \eref{eden_3jet_a},
is used to extract the gluon multiplicity.
The hadronisation correction discussed in \sref{s_predprep} is taken into
account by dividing $N_{q\bar q g}$ by the appropriate correction
factor. $N_{q\bar q}$ is calculated for all three gluon-jet hypotheses and
averaged with the appropriate weights as discussed above before subtracting it
from the three-jet multiplicity. Analogously, the value of $\kale$, at which
$N_{gg}$ is being determined, is the weighted average of the $\kale$-values for
the three possible gluon-jet hypotheses.
\par
\begin{figure}[tb]
\begin{center}
\begin{minipage}[h]{14cm}
\mbox{\epsfig{file=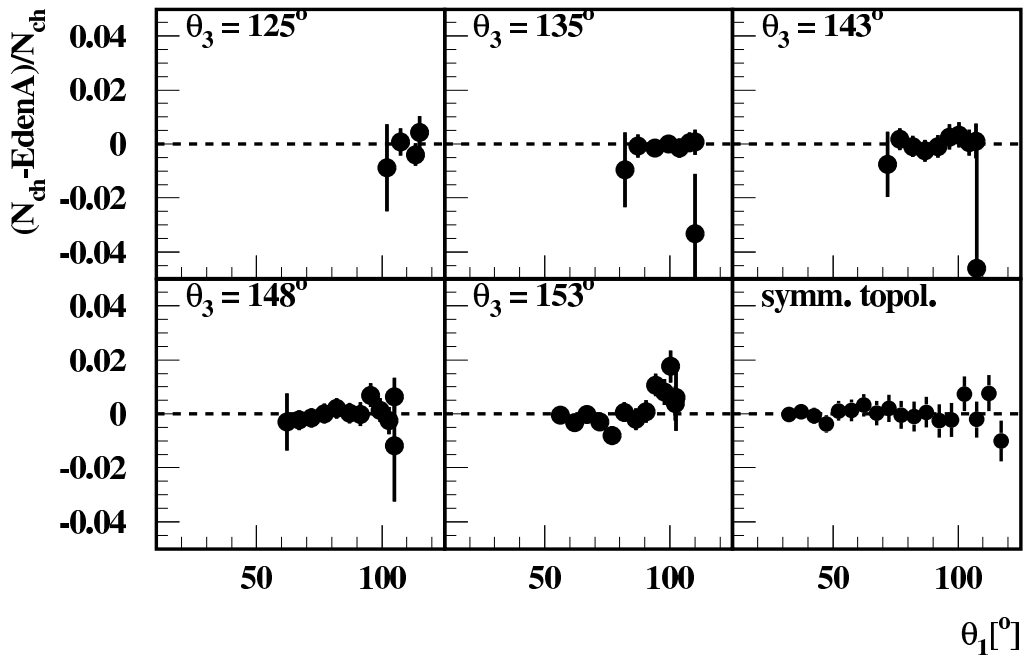,width=14cm}}
\end{minipage} 
\end{center}
\caption
{\label{f_diff_n0_data} Difference between the measured three-jet event
  multiplicity and the prediction \eref{eden_3jet_a}. 
%The open dots represent
%  topologies not used according to the criteria discussed in \sref{s_fit}.}
}
\end{figure}
According to the procedure of \sref{s_fit} the angular ordered Durham
algorithm is applied to obtain the central results. Also, only the topologies
which entered the fit in \sref{s_fit} are considered in order 
to provide a good
description of the three-jet multiplicity by the prediction.  
As the additional multiplicity due to $b$-quark decays has been found to be
topology independent, $udscb$-events are used in order to avoid the
uncertainties inherent in the $b$-tagging procedure and $N_0$ is fitted to the
data yielding  $N_0=0.628\pm0.017$ for general and
$N_0=0.628\pm0.024$ for symmetric topologies 
 with the $\chi^2/N_{df}$ values
of 80.4/74 and 7.3/17 respectively. The differences between the
prediction and the measured three-jet event multiplicity are shown in
\fref{f_diff_n0_data}. The agreement between data and prediction is good,
especially for symmetric topologies.
Anti-$b$-tagged
$udsc$-events, with $N_0$ fixed at 0, 
are taken into account when estimating the
systematic uncertainties of $N_{gg}$.  
\par
\begin{figure}[p]
\begin{center}
\begin{minipage}[h]{16cm}
\mbox{\epsfig{file=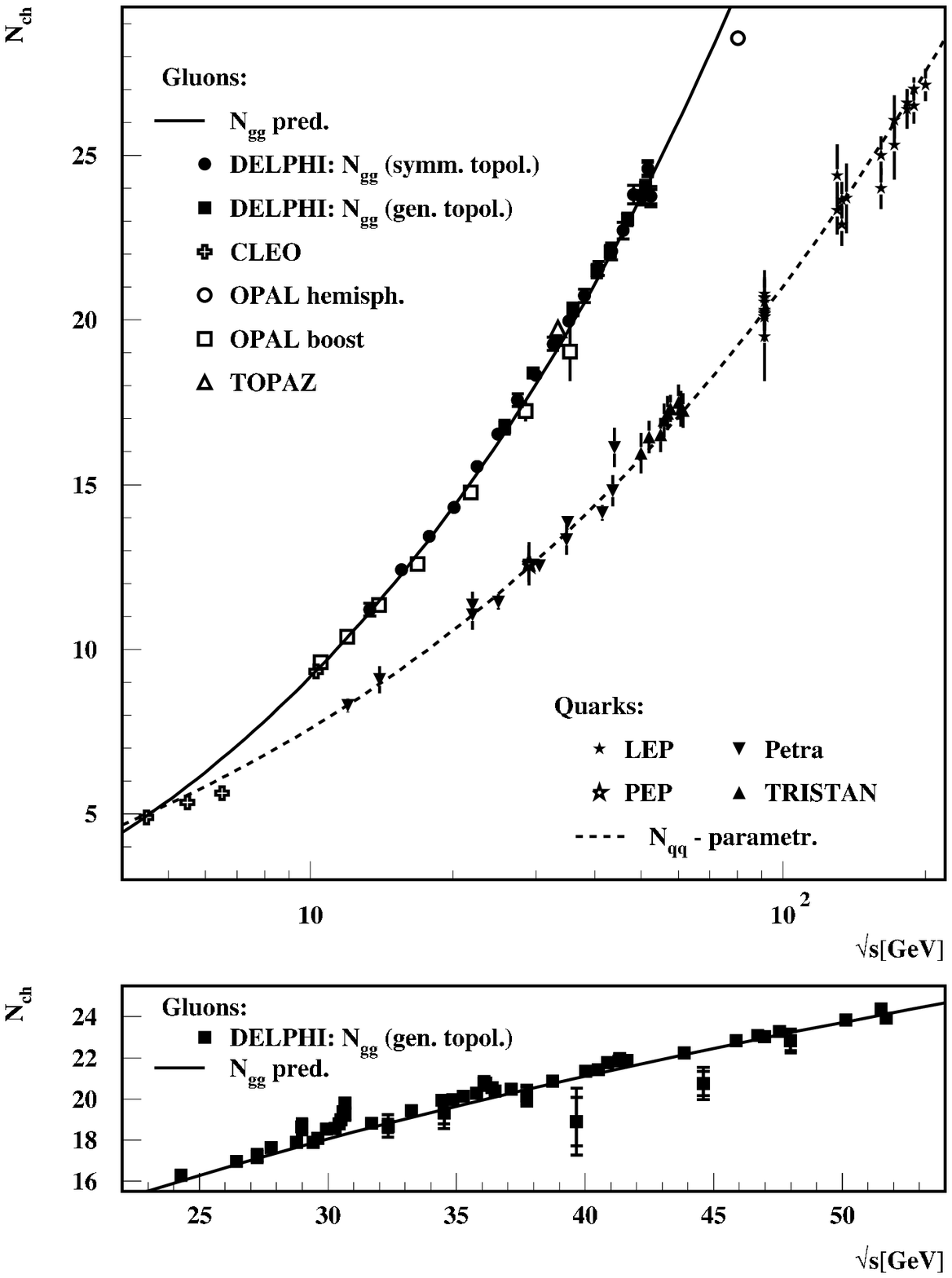,width=16cm}}
%\mbox{\epsfig{file=gluons2.eps,width=14cm}}
\end{minipage} 
\end{center}
\vspace*{-.7cm}
\caption 
{ \label{f_ngg} 
Gluon and quark multiplicities. For clarity, gluon
  multiplicities from general topologies are rebinned in the upper plot. In
  the bottom plot these Multiplicities are shown for each topology bin.}
\end{figure}
The gluon multiplicities obtained are given in \tref{t_ngg_sym} and
\tref{t_ngg_gen}. They are shown in \fref{f_ngg} as a function of
$p_{t,\mathrm{Le}}$ which has been identified with the centre-of-mass energy
of the unrestricted
two-gluon colour-singlet system. 
%Open dots indicate the results from general,
%solid dots from symmetric event topologies. 
Solid circles indicate the results from symmetric, solid squares from general
event topology. For clarity the multiplicities from general topologies have
been rebinned in $p_t$ in the upper plot. The unrebinned results are shown in
the lower plot of \fref{f_ngg}.  
Results of both topology classes
are found to be in good agreement. The good agreement between prediction
and data in symmetric topologies also for small
opening angles allows gluon multiplicities at small values of
$\kale$ to be obtained 
resulting in a larger kinematic range covered by the gluon
multiplicities from symmetric than from general topologies. 
\par
The solid line in
\fref{f_ngg} represents the prediction for the gluon multiplicity which has
been described in \sref{s_predprep}. The agreement between this
prediction and the extracted gluon multiplicities reflects the good
description of the three-jet event multiplicities by \eref{eden_3jet_a}.
The measurement of $N_{gg}$ of the {\sc Cleo} collaboration at $\sim 10$ GeV
shown in
\fref{f_ngg} has been used to fix the constant of integration when setting up
the prediction for $N_{gg}$, therefore it agrees by definition with the
prediction. However, this value does not enter in the determination of
$N_{gg}$.
The measurement at $\sim 5$ GeV is represented with statistical
errors only. 
The {\sc Opal} 
measurement of $N_{gg}$ around 80 GeV is based on the measurement
of gluon jets recoiling against two identified $b$-quark jets \cite{opal_ngg}.
The {\sc Opal} measurements at lower energies used identified gluon
jets from three-jet events and an effective jet energy scale 
\cite{opal_recoil}.  Also included is a measurement of the {\sc Topaz}
collaboration obtained from fully symmetric events \cite{topaz}.
The measurement \cite{opalmult} is not included in \fref{f_ngg}, as it is
based on \eref{eden_3jet_b} which has been found not to describe the 
data satisfactorily.
The overall agreement between the several measurements is good.
\par
After the preliminary
 presentation of the analysis of symmetric three-jet events
in \cite{multnote}, a similar analysis has been published \cite{opalmult},
where different conclusions have been reached. Especially the prediction
\eref{eden_3jet_b} has been preferred over \eref{eden_3jet_a} due to the 
observed extrapolation behaviour of the extracted gluon
multiplicities. 
 However, only symmetric topologies have been studied in
\cite{opalmult}, while the main arguments of this analysis to disfavour
\eref{eden_3jet_b} compared to \eref{eden_3jet_a} are the inconsistent
results for symmetric and general topologies obtained when using
\eref{eden_3jet_b}. Moreover, the extrapolation behaviour of the 
gluon multiplicities extracted with \eref{eden_3jet_a} to direct
 measurements at lower energies  
is satisfying, when a hadronisation
correction is applied.  Note that in \cite{opalmult} no hadronisation
 correction has been used and the extrapolation behaviour of the gluon
 multiplicities 
 extracted using \eref{eden_3jet_a} has been found to be unsatisfying, leading
 to the preference of \eref{eden_3jet_b} over \eref{eden_3jet_a} in
 \cite{opalmult}.  A more extensive attempt to compare both analyses 
can be
 found in \cite{drmartin}. 
% in a way that a satisfying result is reached by using
%\eref{eden_3jet_a}. 
\par
 \begin{figure}[tb]
 \begin{center}
 \begin{minipage}[h]{14cm}
 \mbox{\epsfig{file=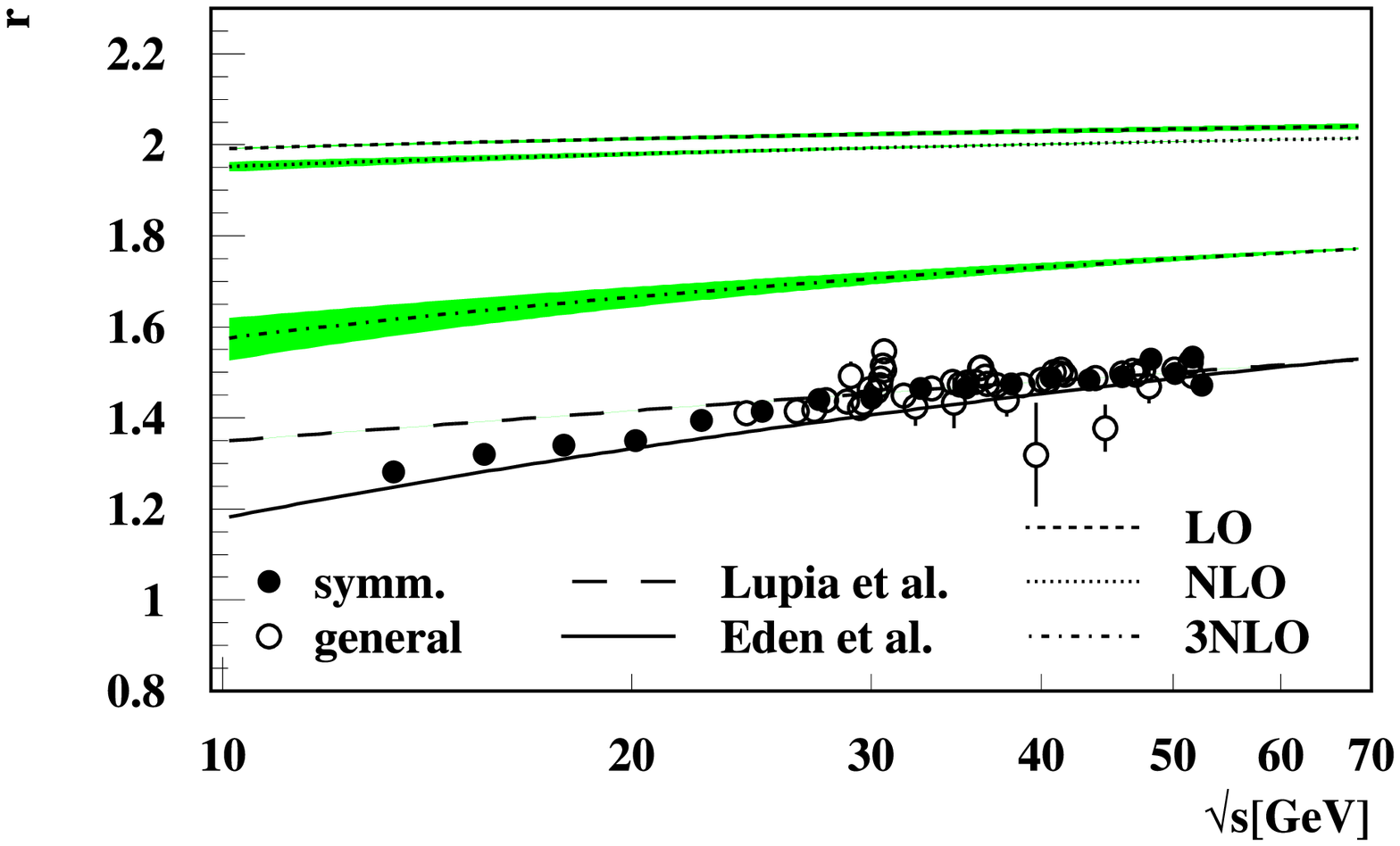,width=14cm}}
 \end{minipage} 
 \end{center}
 \caption
 {\label{f_r0} Measurements of $r$ as defined in \eref{e_r} in comparison 
with theoretical
 predictions. The shaded bands indicate the effects of a variation of $N_F$
 between 3 and 5. }
 \end{figure}
In addition to the gluon multiplicity, in \fref{f_ngg} the unrestricted
$q\bar q$-multiplicity measured by several $e^+e^-$ experiments which have been
used to parametrise the prediction are shown. The measurements have been
corrected for the varying $b$-contributions, so the multiplicity of
$udsc$-events is shown. The line indicates the fitted parametrisation from
\eref{e_nwebber}. The higher multiplicity and the larger slope of the gluon
multiplicity with respect to energy in comparison with the $q\bar
q$ multiplicity can be clearly observed illustrating the observation of the
larger colour charge of gluons. 
\par
In order to obtain the ratio $r$ of
the gluon and quark multiplicity, the extracted gluon multiplicities are
divided by the parametrisation of $N_{q\bar q}$ evaluated at the respective
energy. The obtained values for $r$ are shown in \fref{f_r0}. Open dots
represent the results from general, solid dots from symmetric topologies.
The curves indicate the LO, NLO and 3NLO predictions of $r$ discussed in
\sref{s:thpred} with the shaded area representing the effects of a variation
of $N_F$ between 3 and 5. Additionally, the result of a numeric calculation
\cite{ochs_r} is indicated. 
The predictions clearly overestimate the measured
ratio. The overestimate is larger the lower the order of the calculation is. 
The difference between the
predictions is also rather large. The reason for the poor description
of the measurement by these predictions are non-perturbative effects to which
the absolute value of multiplicity is especially sensitive.
Only the line derived from the
prediction of \eref{eden_gluon} describes the data, but in the determination of
this curve experimental input, especially the measurement of the gluon
multiplicity at $\sqrt{s}=10$ GeV 
which fixes the constant of integration, enters, so this line
%does not represent a purely perturbative calculation. 
includes a non-perturbative correction.
\par
 \begin{figure}[tb]
 \begin{center}
 \begin{minipage}[h]{14cm}
 \mbox{\epsfig{file=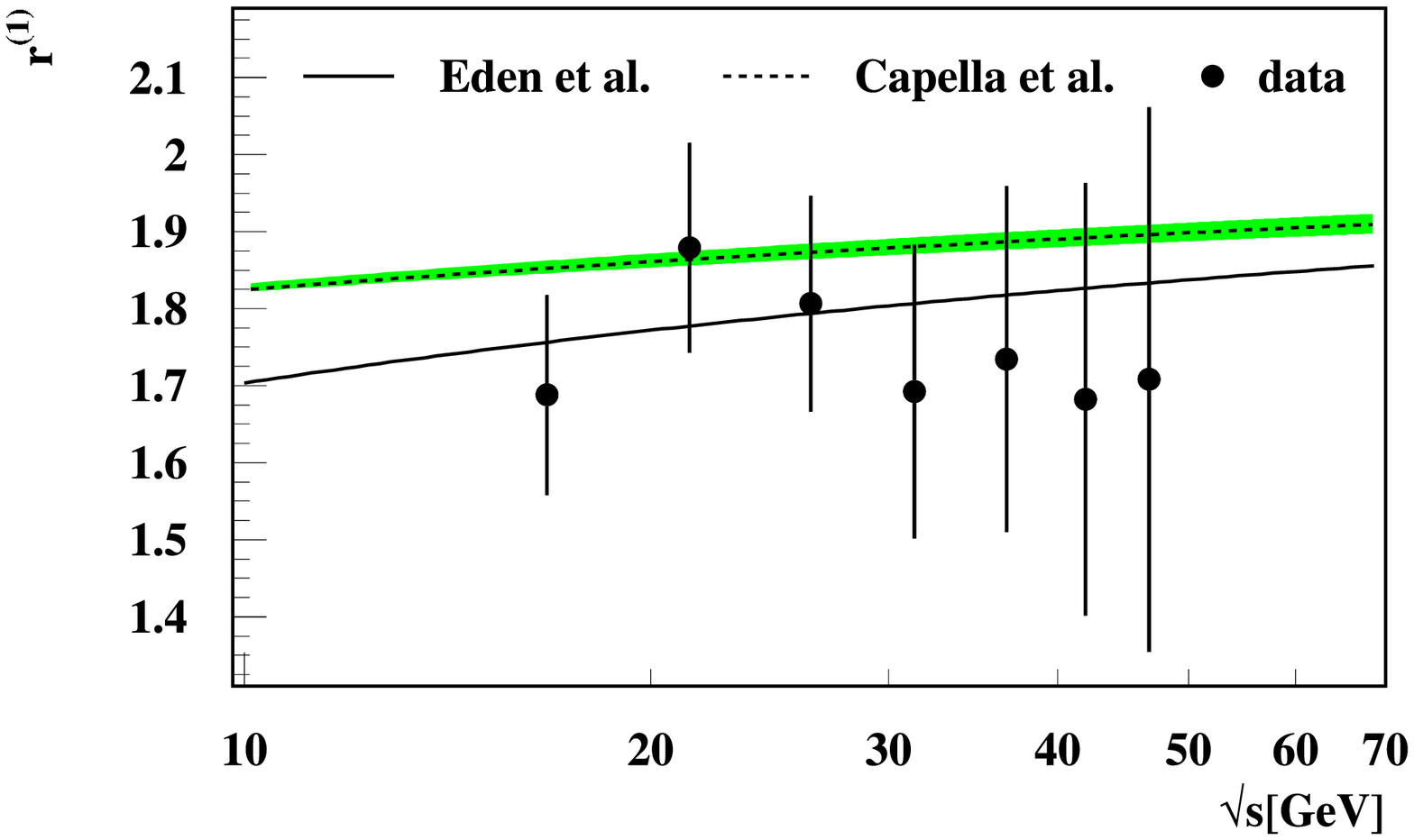,width=14cm}}
 \end{minipage} 
 \end{center}
 \caption
 {\label{f_r1} Measurements of $r^{(1)}$ 
as defined in \eref{e_r1cap} in comparison 
with theoretical
 predictions. The shaded bands indicate the effects of a variation of $N_F$
 between 3 and 5. }
 \end{figure}
The ratio of the derivatives of gluon and quark multiplicity with respect to
the energy,
%\begin{displaymath}
$r^{(1)}$,
%=\frac{d<N_g>/ds}{d<N_q>/ds}
%\end{displaymath}
 is expected to be less sensitive to such non-perturbative
effects.  $r^{(1)}$ is obtained from the measured gluon multiplicities by
taking the linear slope of mutually exclusive 
pairs of measured multiplicities
which is then divided by the derivative of the parametrisation of $N_{q\bar q}$
taken at the respective energy value. The obtained values for $r^{(1)}$ are
shown in \fref{f_r1}. Only symmetric topologies have been considered here, as
they cover a larger kinematic range and are more evenly distributed in energy.
 However, the results obtained from general topologies for $r^{(1)}$ are in
full agreement with the results shown in \fref{f_r1}.
The measured $r^{(1)}$ are constant within their errors, the weighted mean is:
\be
r^{(1)}=1.75\pm 0.07\quad.
\ee
$r^{(1)}$ is therefore significantly higher than $r$ due to the weaker
dependence on non-perturbative effects, which reduce the ratio from the
asymptotic value of $C_A/C_F$. The predictions \eref{e_r1cap} and 
\eref{eden_gluon} are indicated in \fref{f_r1} as curves. 
A reasonable 
agreement between the perturbative calculation and data can be observed
here. Note that \eref{eden_gluon} is a purely perturbative prediction for
the slopes. Also the difference between both predictions is smaller
for $r^{(1)}$ than for $r$. Both observations confirm that the energy dependence of
multiplicity is an observable superior to the absolute value of
multiplicity.
\par
\begin{figure}[p]
\begin{center}
\begin{minipage}[h]{14cm}
\mbox{\epsfig{file=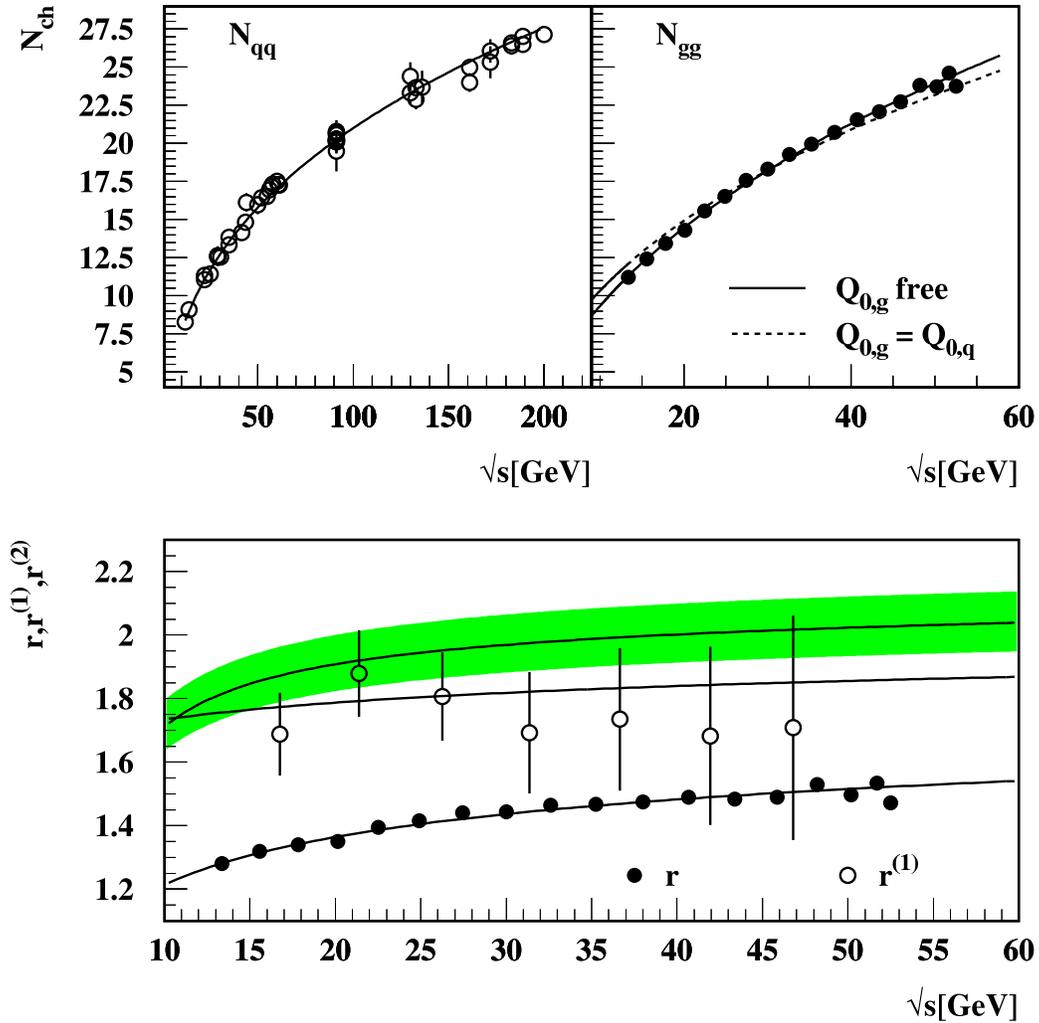,width=14cm}}
\end{minipage} 
\end{center}
\caption
{\label{f_3nlo} The top plots show the
3NLO predictions \eref{e_capellang} and \eref{e_capellanq}
fitted to the measured multiplicities. The lower plot shows from bottom to top
$r$, $r^{(1)}$ and
$r^{(2)}$ which are derived from the fitted parametrisations together with the
measured points for $r$ and $r^{(1)}$.}
\end{figure}
\begin{table}[tb]
\begin{center}
\begin{tabular}{|l|c|c|c|}
\hline
& $k$ & $Q_0$[GeV] & $\chi^2/N_{df}$\\
\hline
$N_{q\bar q}$ & $ 0.243 \pm 0.011$ & $0.454\pm0.051$ & 0.75 \\
$N_{gg}$ & $ 0.1980 \pm 0.0005$ &   & 8.5 \\
\hline
$N_{gg}$ & $ 0.343 \pm 0.012$ & $ 1.317 \pm0.084$ & 1.1 \\
\hline 
\end{tabular}
\end{center}
\caption{\label{t_3nlopar} Parameter values
 of the 3NLO-fits of $N_{q\bar q}$ and
  $N_{gg}$ }
\end{table}
The scale-independent values obtained for $r^{(1)}$ already indicate that
there
is no direct sensitivity to the ratio $r^{(2)}$ of the second derivatives of
the multiplicities with respect to energy. However, the parametrisations of
\eref{e_capellang} and \eref{e_capellanq} can be fitted to the data and
derivatives can then be obtained analytically. This procedure has already been
applied in \cite{opalmult} in order to obtain $r^{(2)}$. As already 
observed in \cite{capella}, the use of a common normalisation $k$
for quark and
gluon multiplicity leads to a significant underestimation of the quark
multiplicity and a wrong energy dependence. 
Therefore, the normalisations $k_q$ and $k_g$ are allowed to
vary independently. 
As the quark multiplicities are measured more directly,
$Q_0$ is fixed by the fit of $N_{q\bar q}$. This leaves $k_g$ as the only
free parameter in $N_{gg}$. The values obtained for the parameters are 
given in
\tref{t_3nlopar}, the fitted parametrisations are shown in the upper two plots
of \fref{f_3nlo} as a solid line for $N_{q\bar q}$ and as a dashed line for
$N_{gg}$. The fit of $N_{q\bar q}$ describes the data well, the value
obtained for $\chi^2$ is small and $Q_0=454$ MeV is of 
%a usual order of magnitude. 
of the same order of magnitude as $\Lambda_{\mathrm{QCD}}$.
However, the energy dependence of the gluon multiplicity is not
described well with this value of $Q_0$, the obtained $\chi^2$ is quite
large. Therefore, $Q_0$ is also allowed to vary freely in the fit of
$N_{gg}$ resulting in two different effective energy scales for quarks and
gluons. This apparently implausible finding can be motivated by the fact
that due to the choice of evolution parameters leading to \eref{e_capellang}
and \eref{e_capellanq} both quantities are predicted not in exactly the same
order leading to different effective scales. The variation of $Q_0$ leads
to a good description also of the gluon multiplicities by the prediction
as indicated by the solid line in the upper right plot of \fref{f_3nlo}, the
parameter values obtained for $N_{gg}$ are given in the last line of
\tref{t_3nlopar}. 
\par
The derivatives of these fitted parametrisations can now be 
calculated analytically 
to any order and the ratios of these derivatives can be built.
In the lower plot of \fref{f_3nlo} from bottom to top
the ratios $r$, $r^{(1)}$ and $r^{(2)}$ are
shown in comparison with the measurements of $r$ and $r^{(1)}$. The agreement
between the measurement and the derived ratios is good, especially the
energy dependence of $r$ is well
reproduced. The shaded area
indicates the result obtained for $r^{(2)}$ within the uncertainty given by
the errors of the fit parameters. Interpreting this curve as a
measurement of $r^{(2)}$ the value obtained  e.g.~at 30 GeV would be found as 
\be
r^{(2)}(30~\mathrm{GeV}) = 1.97\pm0.09\quad.
\ee
This value is in good agreement with the value obtained by a similar procedure
in \cite{opalmult}. However, 
theoretical assumptions about the energy
dependence of quark and gluon multiplicity are the basis of the
parametrisations used, 
so that a theoretical bias cannot be avoided in obtaining
this quantity.
\section{Conclusions\label{s_summary}}
The multiplicity of hadronic three-jet events has been measured as a function
of variables depending on
the event topology. A MLLA-prediction of this quantity has been fitted to
the data yielding a measurement of the colour factor ratio 
\begin{displaymath}
\frac{C_A}{C_F}=2.261\pm0.014_{\mathrm{stat.}}
\pm0.036_{\mathrm{exp.}}
\pm0.066_{\mathrm{theo.}}\qquad .
%\pm0.041_{\mathrm{clus.}} 
%\frac{C_A}{C_F}=2.261\pm0.014_{\mathrm{stat.}}
%\pm0.036_{\mathrm{exp.}}
%\pm0.052_{\mathrm{theo.}}
%\pm0.041_{\mathrm{clus.}} \qquad .
\end{displaymath}
With an overall relative uncertainty of 3.4\% this is the most precise
measurement of this quantity so far.
It has been shown that
the formulation \eref{eden_3jet_b} 
of the prediction should be rejected in favour of
the formulation \eref{eden_3jet_a}. 
Using the MLLA-prediction to subtract the
quark-contribution from the three-jet event multiplicity, the multiplicity of
two-gluon colour-singlet systems has been extracted over a wide range of the
effective energy scale. Comparing these multiplicities with the known
multiplicities of quark-antiquark colour-singlet systems, the ratios $r$,
$r^{(1)}$ and $r^{(2)}$ have been studied. It has been found that $r^{(1)}$ is
an observable superior to $r$ as non-perturbative effects affect the energy
development of multiplicity less than the absolute value of
multiplicity. However, the data do not show any
significant sensitivity to the ratio
$r^{(2)}$.
%         Modified on 04-06-1999 by dimartino
%-------------------------------------------------------------------
\subsection*{Acknowledgements}
\vskip 3 mm
 We are greatly indebted to our technical 
collaborators, to the members of the CERN-SL Division for the excellent 
performance of the LEP collider, and to the funding agencies for their
support in building and operating the DELPHI detector.\\
We acknowledge in particular the support of \\
Austrian Federal Ministry of Education, Science and Culture,
GZ 616.364/2-III/2a/98, \\
FNRS--FWO, Flanders Institute to encourage scientific and technological 
research in the industry (IWT), Belgium,  \\
FINEP, CNPq, CAPES, FUJB and FAPERJ, Brazil, \\
Czech Ministry of Industry and Trade, GA CR 202/99/1362,\\
Commission of the European Communities (DG XII), \\
Direction des Sciences de la Mati$\grave{\mbox{\rm e}}$re, CEA, France, \\
Bundesministerium f$\ddot{\mbox{\rm u}}$r Bildung, Wissenschaft, Forschung 
und Technologie, Germany,\\
General Secretariat for Research and Technology, Greece, \\
National Science Foundation (NWO) and Foundation for Research on Matter (FOM),
The Netherlands, \\
Norwegian Research Council,  \\
State Committee for Scientific Research, Poland, SPUB-M/CERN/PO3/DZ296/2000,
SPUB-M/CERN/PO3/DZ297/2000, 2P03B 104 19 and 2P03B 69 23(2002-2004)\\
FCT - Funda\c{c}\~ao para a Ci\^encia e Tecnologia, Portugal, \\
Vedecka grantova agentura MS SR, Slovakia, Nr. 95/5195/134, \\
Ministry of Science and Technology of the Republic of Slovenia, \\
CICYT, Spain, AEN99-0950 and AEN99-0761,  \\
The Swedish Research Council,      \\
Particle Physics and Astronomy Research Council, UK, \\
Department of Energy, USA, DE-FG02-01ER41155. \\
EEC RTN contract HPRN-CT-00292-2002. \\

%=========================================================================%

%\endnumbering
%
%
%
% B E W A R E :: do not remove %lines from thebibliography, they are important
% for the citations
 
%%\end{document}


\begin{thebibliography}{99}
\newcommand{\wwwspires}{http://www.slac.stanford.edu/spires/find/hep/www}
\bibitem{webbermult} B.R.Webber, 
%            {\em Average multiplicities in jets}\\
            Phys.~Lett. {\bf B143} (1984)  501
\bibitem{gaffneymult} J.B.Gaffney and A.H.Mueller, 
%            {\em $\alpha(Q^2)$ corrections to particle multiplicity
%            ratios in gluon and quark jets}\\
            Nucl.~Phys. {\bf B250} (1985) 109
\bibitem{capella}
    A.~Capella et al.,
% I.~M.~Dremin, J.~W.~Gary, V.~A.~Nechitailo and
%    J.~Tran Thanh Van,
%    {\em Evolution of average multiplicities of quark and gluon jets}\\
     Phys.~Rev~{\bf D61} (2000) 074009
%    hep-ph/9910226
%\bibitem{theomult} qqbar mult pred
\bibitem{multmess}
{\sc Mark} II collaboration, 
G.~S.~Abrams et al.,Phys.~Rev.~Lett. {\bf 64} (1990) 1334,\\ 
{\sc Aleph} collaboration, D.~Decamp et al., 
        Phys.~Lett. {\bf B234} (1990) 209,\\ 
{\sc Aleph} collaboration, D.~Decamp et al.,
 Phys.~Lett. {\bf B273} (1991) 181,\\ 
{\sc Aleph} collaboration, D.~Buskulic et al.,
Zeit.~Phys. {\bf C69} (1996) 15,\\ 
{\sc Aleph} collaboration, R.~Barate et al.,
Phys.~Rep. {\bf 294} (1998) 1,\\ 
{\sc Delphi} collaboration, P.~Abreu et al.,
Zeit.~Phys. {\bf C50} (1991) 185,\\ 
{\sc Delphi} collaboration, P.~Abreu et al.,
Zeit.~Phys. {\bf C52} (1991) 271,\\ 
{\sc Delphi} collaboration, P.~Abreu et al.,
Eur.~Phys.J. {\bf C5} (1998) 585 ,\\ 
L3 collaboration, B.~Adeva et al.,
Phys.~Lett. {\bf B259} (1991) 199 ,\\
L3 collaboration, B.~Adeva et al., 
Zeit.~Phys. {\bf C55} (1992) 39,\\ 
{\sc Opal} collaboration, M.~Z.~Akrawy et al.,
Zeit.~Phys. {\bf C47} (1990) 505,\\ 
{\sc Opal} collaboration, P.~D.~Acton et al.,
Phys.~Lett. {\bf B291} (1992) 503,\\ 
{\sc Opal} collaboration, P.~D.~Acton et al.,
Zeit.~Phys. {\bf C53} (1992) 539 ,\\ 
{\sc Opal} collaboration, K.~Ackerstaff et al.,
Eur.~Phys.J. {\bf C7} (1999) 369
\bibitem{restmult}
 {\sc Tasso} collaboration, M.~Althoff et al.,
 Zeit.~Phys. {\bf C22 } (1984) 307,\\
 {\sc Tasso} collaboration, W.~Braunschweig et al.,
 Zeit.~Phys. {\bf C45 } (1989)193 ,\\
 TPC/Two Gamma collaboration, H.Aihara et al.,
 Phys.~Lett. {\bf B184} (1987) 299,\\
 {\sc Mark} II collaboration, P.~C.~Rowson et al.,
 Phys.~Rev.~Lett. {\bf 54} (1985) 2580,\\
 HRS collaboration, M.~Derrick et al.,
 Phys.~Rev. {\bf D34} (1986) 3304,\\
%{\sc Tasso} collaboration, W.~Braunschweig et al., 
% Zeit.~Phys. {\bf C45} (1989) 93,\\
{\sc Amy} collaboration, H.~W.~Zheng et al.,
 Phys.~Rev. {\bf D42} (1990)737,\\
L3 collaboration, M.~Acciarri et al., 
 Phys.~Lett. {\bf B371} (1996) 137,\\
{\sc Delphi} collaboration, P.~Abreu et al., 
 Phys.~Lett. {\bf B372} (1996) 172,\\
{\sc Opal} collaboration, G.~Alexander et al., 
 Zeit.~Phys. {\bf C72} (1996) 191,\\
{\sc Aleph} collaboration, D.~Buskulic et al., 
 Zeit.~Phys. {\bf C73} (1997) 409,\\
{\sc Opal} collaboration, K.~Ackerstaff et al., 
 Zeit.~Phys. {\bf C75} (1997) 193,\\
{\sc Delphi} collaboration, P.~Abreu et al., 
 Phys.~Lett. {\bf B416} (1998) 233,\\
{\sc Opal} collaboration, G.~Abbiendi et al., 
 Eur.~Phys.~J. {\bf C16} (2000) 185,\\
{\sc Delphi} collaboration, P.~Abreu et al., 
 Eur.~Phys.~J. {\bf C18} (2000) 203;\\
  Erratum ibid. {\bf C25} (2002) 493
%\bibitem{expmult} qqbar mult measurements
%\bibitem{theo_r} theorie r
\bibitem{topaz}
{\sc Topaz} collaboration, K.~Nakabayashi et al., 
 Phys.~Lett. {\bf B413} (1997) 447
\bibitem{r1_suggest}
    J.~Fuster, K.~Hamacher, O.~Klapp, P.~Langefeld, S.~Marti and M.~Siebel,
%    {\em The scale dependence of the multiplicity in quark and gluon jets 
%     and a precise determination of $C_A/C_F$}\\
    contributed paper 545 to the 
    International Europhysics Conference on High Energy
    Physics (EPS-HEP'97), Jerusalem 1997, 
    DELPHI 97-81 CONF 67
%    Beitrag Nr.~545
\bibitem{qgstudy} {\sc Delphi} collaboration, P.~Abreu et al., Z.~Phys.~{\bf
    C70} (1996) 179
\bibitem{alephscales} {\sc Aleph} collaboration, R.~Barate et al., 
    Z.~Phys.~{\bf
    C76} (1997) 191
\bibitem{opal_ngg}
    {\sc Opal} collaboration, G.~Abbiendi et al.,
%    {\em Experimental properties of gluon and quark jets from a point
%       source}\\
    Eur.~Phys.~J.~{\bf C11} (1999) 217    
\bibitem{scaling} {\sc Delphi} collaboration, 
    P.~Abreu et al., Eur.~Phys.~J. {\bf C13} (2000)
    573 
\bibitem{opal_recoil} {\sc Opal} collaboration, G.~Abbiendi et al.,
Eur.~Phys.~J. {\bf C37} (2004) 25
% CERN-PH-EP
%  2004-011, submitted to Eur.~Phys.~J.~{\bf C}
\bibitem{coherence} {\sc Delphi} collaboration, P.~Abreu et al.,
Phys. Lett {\bf B605} (2005) 37
%  CERN-PH-EP 2004-018, 
%to be submitted to Phys.~Lett.~{\bf B}
\bibitem{multpaper}
    {\sc Delphi} collaboration, P.~Abreu et al.,
%     {\em The scale dependence of the hadron multiplicity in quark and
%     gluon jets and a precise determination of $C_A/C_F$}\\
     Phys.~Lett.~{\bf B449} (1999) 383
\bibitem{Eden:1998ig}
P.~Eden and G. Gustafson,
%``Energy and virtuality scale dependence in quark and gluon jets,''
JHEP {\bf 9809} (1998) 015,
hep-ph/9805228
%%CITATION = HEP-PH 9805228;%%
%\href{http://www.slac.stanford.edu/spires/find/hep/www?eprint=HEP-PH/9805228}{SPIRES}
\bibitem{Eden:1999vc}
P.~Eden, G.~Gustafson and V.~Khoze,
%``On particle multiplicity distribution in three-jet events,''
Eur.~Phys.~J.  {\bf C11} (1999) 345, 
hep-ph/9904455
%%CITATION = HEP-PH 9904455;%%
%\href{\wwwspires?eprint=HEP-PH/9904455}{SPIRES}
\bibitem{multnote}
    M.~Siebel, K.~Hamacher, P.~Langefeld and O.~Klapp
%    {\em More about the multiplicity in symmetric three-jet events}\\
    contributed paper 640 
    to the International Conference on High Energy Physics, 
    Osaka 2000\\
    CERN OPEN 2000-134
\bibitem{opalmult}    {\sc Opal} collaboration, G.~Abbiendi et al.,
%    {\em Particle multiplicity of unbiased gluon jets from $e^+e^-$ three-jet
%    events}\\
    Eur.~Phys.~J.~{\bf C23} (2002) 597 
\bibitem{det}
%\cite{Aarnio:1990vx}
%\bibitem{Aarnio:1990vx}
{\sc Delphi} collaboration, P.~A.~Aarnio  et al., 
%``The Delphi Detector At Lep,''
Nucl.~Instrum.~Meth.~{\bf A303} (1991) 233
%%CITATION = NUIMA,A303,233;%%
\bibitem{perf}
%\cite{Abreu:1995uz}
%\bibitem{Abreu:1995uz}
{\sc Delphi} collaboration, P.~Abreu et al., 
%``Performance of the DELPHI detector,''
Nucl.~Instrum.~Meth.~{\bf A378} (1996) 57
%%CITATION = NUIMA,A378,57;%%
\bibitem{b-tag2}
     {\sc Delphi} collaboration, J.~Abdallah et al., 
%     {\em b-tagging in {\sc Delphi} at LEP}\\
     Eur.~Phys.~J. {\bf C32} (2004) 185
\bibitem{cambridge}
    Y.~L.~Dokshitzer et al., 
% G.~D.~Leder, S.~Morietti and B.~R.~Webber, 
%    {\em Better jet clustering algorithms}\\ 
    JHEP {\bf 9708} (1997) 001,  \\
    S.~Bentvelsen and I.~Meyer,
%    {\em The Cambridge jet algorithm: Features and applications},\\
    Eur.~Phys.~J. {\bf C4} (1998) 623
\bibitem{durham}
    S.~Catani et al., 
%Yu.~L.~Dokshitzer, M.~Olsson, G.~Turnock and B.~R.~Webber,
%    {\em New clustering algorithm for multi-jet cross-sections in $e^+e^-$
%    annihilation },\\
    Phys.~Lett. {\bf B269} (1991) 432
%\bibitem{cambridge}
\bibitem{luclus}
    T.~Sj\"ostrand,
%    {\em High-energy physics event generation with 
%      Pythia 5.7 and Jetset 7.4},\\
    Computer Physics Comm. {\bf 82} (1994) 74
\bibitem{sphericity} J.~D.~Bjorken and S~.J.~Brodsky, Phys.Rev. {\bf D1} 
(1970) 1416 
\bibitem{mue}
     A.~H.~Mueller,
%    {\em  $\sqrt{\alpha(Q^2)}$ corrections to particle multiplicity ratios in
%    gluon and quark jets}\\
    Nucl.~Phys.~{\bf B241} (1984) 141
%\bibitem{gaffneymult} J.B.Gaffney and A.H.Mueller,
%            {\em $\alpha(Q^2)$ corrections to particle multiplicity
%            ratios in gluon and quark jets}\\
%            Nucl.Phys. {\bf B250} (1985) 109
\bibitem{dremin}
    I.~M.~Dremin and J.~W.~Gary,
%    {\em Energy dependence of mean multiplicities in gluon and quark jets at
%    the next-to-next-to-next-to-leading order}\\
    Phys. Lett. {\bf B459} (1999) 341
%    hep-ph/9905477
%\bibitem{capella}
%    A.~Capella, I.~M.~Dremin, J.~W.~Gary, V.~A.~Nechitailo und
%    J.~Tran Thanh Van,
%    {\em Evolution of average multiplicities of quark and gluon jets}\\
%    hep-ph/9910226
\bibitem{rb}
%  R_b=0.21642\pm 0.00075
%   The LEP collaborations {\sc Aleph}, {\sc Delphi}, L3 and {\sc Opal},
%   the LEP Electroweak Working Group and the
%   SLD Heavy Flavour and Electroweak Groups,
%   {\em A combination of preliminary electroweak measurements and constraints
%   to the standard model}\\
%   CERN-EP-2000-16  
Particle Data Group, S.~Eidelman et al., Phys.~Lett.~{\bf B592} (2004) 1
\bibitem{n0mes_d}
    {\sc Delphi} collaboration, P.~Abreu et al.,
%    {\em Production of charged particles $K^0_s$, $K^\pm$, $p$ and $\Lambda$ 
%      in $Z\to b\bar b$ events and in the decay of $b$ hadrons} \\
    Phys.~Lett.~{\bf B347} (1995) 447
\bibitem{n0mes_o}
    {\sc Opal} collaboration, R.~Akers et al.,
%    {\em A measurement of charged particle multiplicity in $Z^0\to c \bar c$
%     and $Z^0\to b \bar b$} events\\
    Phys.~Lett.~{\bf B352} (1995) 176
\bibitem{mlla_n0const}
    V.A. Khoze and W. Ochs, 
%    {\em Perturbative QCD approach to multiparticle production}\\
%    hep-ph/9701421
    Int.~J.~Mod.~Phys.~{\bf A12} (1997) 2949
\bibitem{kha}
   K.~Hamacher,
%   {\em QCD at $e^+e^-$-experiments}\\
%   Talk given at XXIII. Physics in Collision, Zeuthen 2003, \\
     Physics in Collision Zeuthen 2003 285-300\\
   {\bf eConf C030626:SAAT08, 2003}
\bibitem{multmess_notused}
 {\sc Jade} collaboration, W.~Bartel et al., 
 Zeit.~Phys.~{\bf C20} (1983) 187,\\
 {\sc Pluto} collaboration, C.~Berger et al., 
 Phys.~Lett.~{\bf B95} (1980) 313,\\
 {\sc Jade} collaboration, W.~Bartel et al., 
 Phys.~Lett.~{\bf B88} (1979) 171
\bibitem{jade_reana}
  S.~Kluth,
%  {\em QCD studies with resurrected {\sc Jade} data}\\
  Talk given at the XXXVIII. Rencontres de Moriond, 2003 \\
  hep-ex/0305028
 \bibitem{cleomult}
 {\sc Cleo} collaboration, M.~S.~Alam et al.,  
%% %``Shape studies of quark jets vs. gluon jets at s**(1.2) = 10-GeV,''
 Phys.\ Rev.\  {\bf D46} (1992) 4822
%% %%CITATION = PHRVA,D46,4822;%%
%% %\href{\wwwspires?j=PHRVA\%2cD46\%2c4822}{SPIRES}
\bibitem{edenfehler}  P.Eden, Eur Ph J {\bf C19} (2001) 493  
\bibitem{betafunction}
  {\sc Delphi} collaboration, J.~Abdallah et al., Eur.~Phys.~J. {\bf C29}
  (2003) 285
\bibitem{fourjets}
    {\sc Aleph} collaboration, A.~Heister et al.,
%    {\em Measurements of the strong coupling constant and the QCD colour
%    factors using four-jet observables from hadronic $Z$-decays}\\
    Eur.~Phys.~J.~{\bf C27} (2003) 1 \\   
    {\sc Delphi} collaboration, P.~Abreu et al., Phys.~Lett.~{\bf B414} (1997)
    401 \\
    {\sc Opal} collaboration, G.~Abbiendi et al.,
%    {\em A simultaneous measurement of the QCD colour factors and 
%         the strong coupling}\\
    Eur.~Phys.~J.~{\bf C20} (2001) 601       
\bibitem{drmartin}  M.~Siebel, PhD thesis, University of Wuppertal
    (WUB-DIS 2003-11), \\
%link maintained by the university library of 
%Wuppertal: \\ 
\mbox{http://elpub.bib.uni-wuppertal.de/edocs/dokumente/fbc/physik/diss2003/siebel}
\bibitem{ochs_r} S.~Lupia and W.~Ochs, Phys.~Lett.~{\bf B418} (1998) 214
%% ddddddddddddddddddddddddddddddddddddd
%% \bibitem{pythia} pythia
%% \bibitem{rb}
%% %  R_b=0.21642\pm 0.00075
%%    Die LEP Kollaborationen {\sc Aleph}, {\sc Delphi}, L3 und {\sc Opal},\\
%%    die LEP Electroweak Working Group,\\
%%    und die SLD Heavy Flavour und Electroweak Groups,
%% %   {\em A combination of preliminary electroweak measurements and constraints
%% %   to the standard model}\\
%%    CERN-EP-2000-16  
%% \bibitem{n0mes_d}
%%     Delphi-Kollaboration
%% %    {\em Production of charged particles $K^0_s$, $K^\pm$, $p$ and $\Lambda$ 
%% %      in $Z\to b\bar b$ events and in the decay of $b$ hadrons} \\
%%     Phys.~Lett.~{\bf B347} (1995) 447
%% \bibitem{n0mes_o}
%%     Opal-Kollaboration
%% %    {\em A measurement of charged particle multiplicity in $Z^0\to c \bar c$
%% %     and $Z^0\to b \bar b$} events\\
%%     Phys.~Lett.~{\bf B352} (1995) 176
%% \bibitem{webbermult} B.R.Webber,
%% %            {\em Average multiplicities in jets}\\
%%             Phys.Lett. {\bf 143B} (1984)  501
%% \bibitem{mue}
%%      A.~H.~Mueller,
%% %    {\em  $\sqrt{\alpha(Q^2)}$ corrections to particle multiplicity ratios in
%% %    gluon and quark jets}\\
%%     Nucl.~Phys.~{\bf B241} (1984) 141
%% \bibitem{r1vorschlag}
%%     J.~Fuster, K.~Hamacher, O.~Klapp, P.~Langefeld, S.~Marti und M.~Siebel,
%% %    {\em The scale dependence of the multiplicity in quark and gluon jets 
%% %     and a precise determination of $C_A/C_F$}\\
%%     Beitrag zur International Europhysics Conference on High Energy
%%     Physics (EPS-HEP'97), Jerusalem 1997\\
%%     Beitrag Nr.~545
%% %\bibitem{eden1}
%% %    P.~Ed\'{e}n und G.~Gustafson,
%% %    {\em Energy and virtuality scale dependence in quark and gluon jets}\\
%% %    JHEP {\bf 09} (1998) 015


%% %\cite{Abreu:1999rs}
%% \bibitem{Abreu:1999rs}
%% P.~Abreu {\it et al.}  [DELPHI Collaboration],
%% %``The scale dependence of the hadron multiplicity in quark and gluon jets  and a precise determination of C(A)/C(F),''
%% Phys.\ Lett.\  {\bf B449} (1999) 383
%% [hep-ex/9903073].
%% %%CITATION = HEP-EX 9903073;%%
%% %\href{\wwwspires?eprint=HEP-EX/9903073}{SPIRES}

%% \bibitem{gustafprivcom}
%%  G.~Gustafson, priv. comm.

%% \bibitem{Abreu:1998ve}
%% P.~Abreu {\it et al.}  [DELPHI Collaboration],
%% %``Investigation of the splitting of quark and gluon jets,''
%% Eur.\ Phys.\ J.\  {\bf C4} (1998) 1.
%% %%CITATION = EPHJA,C4,1;%%
%% %\href{\wwwspires?j=EPHJA\%2cC4\%2c1}{SPIRES} 

%% \bibitem{Dremin:1994bj}
%% I.~M.~Dremin and V.~A.~Nechitailo,
%% %``Average multiplicities in gluon and quark jets in higher order perturbative QCD,''
%% Mod.\ Phys.\ Lett.\  {\bf A9} (1994) 1471
%% [hep-ex/9406002].
%% %%CITATION = HEP-EX 9406002;%%
%% %\href{\wwwspires?eprint=HEP-EX/9406002}{SPIRES}



%% \bibitem{Capella:1999ms}
%% A.~Capella, I.~M.~Dremin, J.~W.~Gary, V.~A.~Nechitailo and J.~Tran Thanh Van,
%% %``Evolution of average multiplicities of quark and gluon jets,''
%% hep-ph/9910226.
%% %%CITATION = HEP-PH 9910226;%%
%% %\href{http://www.slac.stanford.edu/spires/find/hep/www?eprint=HEP-PH/9910226}{SPIRES}

%% %\cite{Dokshitzer:1997in}
%% \bibitem{Dokshitzer:1997in}
%% Y.~L.~Dokshitzer, G.~D.~Leder, S.~Moretti and B.~R.~Webber,
%% %``Better jet clustering algorithms,''
%% JHEP {\bf 9708} (1997) 001
%% [hep-ph/9707323].
%% %%CITATION = HEP-PH 9707323;%%
%% %\href{\wwwspires?eprint=HEP-PH/9707323}{SPIRES}
%% %

%% \bibitem{Buskulic:1996sw}
%% D.~Buskulic {\it et al.}  [ALEPH Collaboration],
%% %``Quark and gluon jet properties in symmetric three jet events,''
%% Phys.\ Lett.\  {\bf B384} (1996) 353.
%% %%CITATION = PHLTA,B384,353;%%
%% %\href{\wwwspires?j=PHLTA\%2cB384\%2c353}{SPIRES}

%% %\cite{Abbiendi:1999pi}
%% \bibitem{Abbiendi:1999pi}
%% G.~Abbiendi {\it et al.}  [OPAL Collaboration],
%% %``Experimental properties of gluon and quark jets from a point source,''
%% Eur.\ Phys.\ J.\  {\bf C11} (1999) 217
%% [hep-ex/9903027].
%% %%CITATION = HEP-EX 9903027;%%
%% %\href{\wwwspires?eprint=HEP-EX/9903027}{SPIRES}

%% \bibitem{Abreu:1996na}
%% P.~Abreu {\it et al.}  [DELPHI Collaboration],
%% %``Tuning and test of fragmentation models based on identified particles  and precision event shape data,''
%% Z.\ Phys.\  {\bf C73} (1996) 11.
%% %%CITATION = ZEPYA,C73,11;%%
%% %\href{\wwwspires?j=ZEPYA\%2cC73\%2c11}{SPIRES}

%% \bibitem{Gaffney:1985yd}
%% J.~B.~Gaffney and A.~H.~Mueller,
%% %``Alpha (Q**2) Corrections To Particle Multiplicity Ratios In Gluon And Quark Jets,''
%% Nucl.\ Phys.\  {\bf B250} (1985) 109.
%% %%CITATION = NUPHA,B250,109;%%
%% %\href{\wwwspires?j=NUPHA\%2cB250\%2c109}{SPIRES}

%% \bibitem{Dremin:1999ji}
%% I.~M.~Dremin and J.~W.~Gary,
%% %``Energy dependence of mean multiplicities in gluon and quark jets at the  next-to-next-to-next-to-leading order,''
%% Phys.\ Lett.\  {\bf B459} (1999) 341
%% [hep-ph/9905477].
%% %%CITATION = HEP-PH 9905477;%%
%% %\href{\wwwspires?eprint=HEP-PH/9905477}{SPIRES}

%% \bibitem{jerusalem}
%% J.~Fuster, K.~Hamacher, O.~Klapp, P.~Langefeld, S.~Marti, M.~Siebel
%% [DELPHI Collaboration], 
%% %``The Scale Dependence of the Multiplicity in Quark and Gluon Jets and a
%% %Determination of CA/CF``
%% contrib. paper HEP'97, 545 to the International Europhysics Conference on
%% High Energy Physics, 19-26 August 1997, Jerusalem, Israel.


%% \bibitem{qqbar}
%% %\newcommand{\wwwspires}{http://www.slac.stanford.edu/spires/find/hep/www}
%% %\cite{Braunschweig:1989bp}
%% %\bibitem{Braunschweig:1989bp}
%% W.~Braunschweig {\it et al.}  [TASSO Collaboration],
%% %``Charged Multiplicity Distributions And Correlations In E+ E- Annihilation At Petra Energies,''
%% Z.\ Phys.\  {\bf C45} (1989) 193.
%% %%CITATION = ZEPYA,C45,193;%%
%% %\href{\wwwspires?j=ZEPYA\%2cC45\%2c193}{SPIRES}
%% H.~Aihara {\it et al.}  [TPC/Two Gamma Collaboration],
%% %``Pion And Kaon Multiplicities In Heavy Quark Jets From E+ E- Annihilation At 29-Gev,''
%% Phys.\ Lett.\  {\bf B184} (1987) 299.
%% %%CITATION = PHLTA,B184,299;%%
%% %\href{\wwwspires?j=PHLTA\%2cB184\%2c299}{SPIRES}
%% P.~C.~Rowson {\it et al.},
%% %``Charged Multiplicity Of Hadronic Events Containing Heavy Quark Jets,''
%% Phys.\ Rev.\ Lett.\  {\bf 54} (1985) 2580.
%% %%CITATION = PRLTA,54,2580;%%
%% %\href{\wwwspires?j=PRLTA\%2c54\%2c2580}{SPIRES}
%% M.~Derrick {\it et al.},
%% %``Study Of Quark Fragmentation In E+ E- Annihilation At 29-Gev: Charged Particle Multiplicity And Single Particle Rapidity Distributions,''
%% Phys.\ Rev.\  {\bf D34} (1986) 3304.
%% %%CITATION = PHRVA,D34,3304;%%
%% %\href{\wwwspires?j=PHRVA\%2cD34\%2c3304}{SPIRES}
%% H.~W.~Zheng {\it et al.}  [AMY Collaboration],
%% %``Charged Hadron Multiplicities In E+ E- Annihilations At S**(1/2) = 50-Gev - 61.4-Gev,''
%% Phys.\ Rev.\  {\bf D42} (1990) 737.
%% %%CITATION = PHRVA,D42,737;%%
%% %\href{\wwwspires?j=PHRVA\%2cD42\%2c737}{SPIRES}

%% \bibitem{Webber:1984jp}
%% B.~R.~Webber,
%% B.~R.~Webber,
%% %``Average Multiplicities In Jets,''
%% Phys.\ Lett.\  {\bf B143} (1984) 501.
%% %%CITATION = PHLTA,B143,501;%%
%% %\href{\wwwspires?j=PHLTA\%2cB143\%2c501}{SPIRES}
%% %\cite{Dokshitzer:1992ej}
%% \bibitem{Dokshitzer:1992ej}
%% Y.~L.~Dokshitzer, V.~A.~Khoze and S.~I.~Troian,
%% %``Inclusive particle spectra from QCD cascades,''
%% Int.\ J.\ Mod.\ Phys.\  {\bf A7} (1992) 1875.
%% %%CITATION = IMPAE,A7,1875;%%
%% %\href{\wwwspires?j=IMPAE\%2cA7\%2c1875}{SPIRES}
%% %\newcommand{\wwwspires}{http://www.slac.stanford.edu/spires/find/hep/www}

%% %\cite{Azimov:1985sf}
%% %\bibitem{Azimov:1985sf}
%% %Y.~I.~Azimov, Y.~L.~Dokshitzer, V.~A.~Khoze and S.~I.~Troian,
%% %%``The String Effect And QCD Coherence,''
%% %Phys.\ Lett.\  {\bf B165} (1985) 147.
%% %%%CITATION = PHLTA,B165,147;%%
%% %%\href{\wwwspires?j=PHLTA\%2cB165\%2c147}{SPIRES}
%% %


%% %

%% \bibitem{Rowson:1985xh}
%% P.~C.~Rowson {\it et al.},
%% %``Charged Multiplicity Of Hadronic Events Containing Heavy Quark Jets,''
%% Phys.\ Rev.\ Lett.\  {\bf 54}, 2580 (1985).
%% %%CITATION = PRLTA,54,2580;%%
%% %\href{\wwwspires?j=PRLTA\%2c54\%2c2580}{SPIRES}

%% \bibitem{Abreu:1995pk}
%% P.~Abreu {\it et al.}  [DELPHI Collaboration],
%% %``Production of charged particles, K0(s), K+-, p and Lambda in Z $\to$ b anti-b events and in the decay of b hadrons,''
%% Phys.\ Lett.\  {\bf B347} (1995) 447.
%% %%CITATION = PHLTA,B347,447;%%
%% %\href{\wwwspires?j=PHLTA\%2cB347\%2c447}{SPIRES}

%% \bibitem{Abe:1996zi}
%% K.~Abe {\it et al.}  [SLD Collaboration],
%% %``Measurement of the charged multiplicities in b, c and light quark 
%% events from Z0 decays,''
%% Phys.\ Lett.\  {\bf B386} (1996) 475
%% [hep-ex/9608008].
%% %%CITATION = HEP-EX 9608008;%%
%% %\href{\wwwspires?eprint=HEP-EX/9608008}{SPIRES}
%% %\newcommand{\wwwspires}{http://www.slac.stanford.edu/spires/find/hep/www}
%% %\cite{Akers:1995ww}

%% \bibitem{Akers:1995ww}
%% R.~Akers {\it et al.}  [OPAL Collaboration],
%% %``A Measurement of charged particle multiplicity in Z0 $\to$ c anti-c
%% and Z0 $\to$ b anti-b events,''
%% Phys.\ Lett.\  {\bf B352} (1995) 176.
%% %%CITATION = PHLTA,B352,176;%%
%% %\href{\wwwspires?j=PHLTA\%2cB352\%2c176}{SPIRES}


%% \bibitem{Alam:1997ht}
%% M.~S.~Alam {\it et al.}  [CLEO Collaboration],
%% %``Study of gluon versus quark fragmentation in Upsilon --> g g gamma and  e+ e- --> q anti-q gamma events at s**(1/2) = 10-GeV,''
%% Phys.\ Rev.\  {\bf D56} (1997) 17
%% [hep-ex/9701006].
%% %%CITATION = HEP-EX 9701006;%%
%% %\href{\wwwspires?eprint=HEP-EX/9701006}{SPIRES}
\end{thebibliography}
\end{document}